\documentclass[final,3p,times]{elsarticle}

\usepackage{framed,multirow}

\usepackage{amssymb}
\usepackage{amsmath}
\usepackage{latexsym}

\usepackage{url}
\usepackage[dvipsnames]{xcolor}
\definecolor{newcolor}{rgb}{.8,.349,.1}
\usepackage[utf8]{inputenx}
\usepackage{graphicx}
\usepackage{amsmath,amssymb,mathrsfs}
\usepackage{fancyhdr}
\usepackage{calc}
\usepackage[T1]{fontenc}
\usepackage{mathptmx}
\usepackage{physics}
\usepackage{lineno}
\usepackage{tikz}
\usepackage{pgfplots}

\newcommand{\dif}{\mathrm{d}}
\newcommand{\dxdx}[2]{\frac{\partial #1}{\partial #2}}
\newcommand{\tdxdx}[2]{\frac{\dif #1}{\dif #2}}

\newcommand{\Jd}{\textup J}

\def\du{[\![}
\def\df{]\!]}
\def\ol{\overline}

\newcommand{\bes}{\begin{equation*}}
\newcommand{\ees}{\end{equation*}}
\newcommand{\beq}{\begin{equation}}
\newcommand{\eeq}{\end{equation}}
\newcommand{\bea}{\begin{eqnarray}}
\newcommand{\eea}{\end{eqnarray}}
\newcommand{\beas}{\begin{eqnarray*}}
\newcommand{\eeas}{\end{eqnarray*}}
\newcommand{\ds}{\displaystyle}

\newtheorem{lem}{Lemma}

\newtheorem{sry}{Summary}

\newcommand{\clon}[1]{\textcolor{black}{#1}}
\newcommand{\cltw}[1]{\textcolor{black}{#1}}
\newcommand{\clmy}[1]{\textcolor{black}{#1}}

\begin{document}

\begin{frontmatter}

\title{Analysis of finite-volume discrete adjoint fields for two-dimensional compressible Euler flows}

\author[rvt1]{Jacques Peter\corref{cor1}}
\ead{jacques.peter@onera.fr}
\cortext[cor1]{Corresponding author. Tel.: +33 1 46 73 41 84.}
\address[rvt1]{DAAA, ONERA, Universit\'e Paris Saclay, F-92322 Ch\^atillon, France}

\author[rvt1]{Florent Renac}
\ead{florent.renac@onera.fr}

\author[rvt1]{Cl\'ement Labb\'e}

\begin{abstract} This work deals with a number of questions \cltw{related} to the
  discrete and continuous adjoint fields associated with the compressible Euler equations and classical
   aerodynamic functions.
 The consistency of the discrete adjoint equations
with the corresponding continuous
 adjoint partial differential equations is one of them. It is has been
established or at least discussed only for
 some numerical schemes and a contribution of this article is
to give the adjoint consistency conditions
 for the 2-D Jameson-Schmidt-Turkel scheme in cell-centred finite-volume formulation.
 The consistency issue is also studied here from a new heuristic
point of view by discretizing the continuous adjoint equation for the
discrete flow and adjoint fields. Both points of view prove to provide useful information.
 Besides, it has been
 often noted that discrete or continuous inviscid lift and drag adjoint
exhibit numerical divergence close to
 the wall and stagnation streamline for a wide range of subsonic and
transonic flow conditions. This is analyzed
 here using the physical source term perturbation method introduced in
reference \cite{GilPie_97}. With this point of view,
 the fourth physical source term of \cite{GilPie_97} appears to be the only one
responsible for this behavior. It is also
  demonstrated that the numerical
 divergence of the adjoint variables corresponds to the response of the
flow to the convected increment of stagnation
  pressure and diminution of entropy created at the source and the
resulting change in lift and drag. 
\end{abstract}   


\end{frontmatter}


\section{Introduction}

 Discrete and continuous adjoint methods are well
established methods to efficiently calculate derivatives of aerodynamic
functions with
 respect to numerous design parameters. Adjoint-based derivatives and
adjoint-fields are commonly used
for local shape optimization \cite{FraShu_92,pironneau_74,Jam_88,JamMarPie_98b,CasLozPal_07},
goal-oriented mesh adaptation
 \cite{becker_rannacher_01,VenDar_02,Dwi_08,LosDerAla_10,FidRoe_10,FidDar_11,JPNguTro_12,BelAlaDer_19},
 flow control \cite{lions_68,SarMetSip_15},
 meta-modelling \cite{MorMitYlv_93}, receptivity-sensitivity-stability analyses
 \cite{luchini_bottaro_14}, and data assimilation \cite{talagrand_courtier_87}. These methods are often used
 for the linear analysis of non-linear conservation
laws where the adjoint is defined as the dual to the linearized equations around a given solution to the direct non-linear
problem.
The continuous adjoint method refers to the discretization of the adjoint equations associated with the formal
direct problem, while the discrete adjoint method refers to the adjoint equations to the discrete direct problem.
 \cltw{The advantages and disavantages} of the two approaches have been listed in several articles \cite{GilPie_00,JPDwi_10}.

Concerning the discrete adjoint method, strong efforts
have been devoted to implement (either by hand or using automatic differentiation)
 the exact linearization of intricate schemes for intricate models
 \cite{AndBon_99,NieAnd_99} \cltw{and to} enhance the robustness of the linear solver \cite{NemZin_02,XuRadMey_15}.
 \cltw{However} fundamental questions \cltw{related to the} discrete adjoint
 fields are still open even for two-dimensional (2-D) compressible Euler flows. 
 Although numerical simulations including viscous laminar and turbulent effects are \cltw{currently} common practices in CFD,
 it is desirable to gain \cltw{a further} understanding in the discrete adjoint method
addressing unresolved \cltw{questions} in the framework of 2-D Euler equations.

First, are the discrete adjoint fields consistent with the
continuous equation at the limit of fine meshes?
This subject goes back to a discussion by Giles et al.~in
\cite{GilDutMul_01} where the authors \cltw{discussed} the discrete adjoint
counterpart of a strong
boundary condition at a wall and \cltw{derived} a normal projection of the
momentum adjoint variable that \cltw{did} not correspond to the normal momentum
component of the continuous adjoint solution at the boundary.
Although, of course, the exact discrete adjoint provides the exact output functional
  gradient for a fixed mesh and scheme, this oddness
raised a series of questions: which schemes are adjoint-consistent? What are the consequences of the lack of
dual consistency? How close are continuous adjoints and possibly inconsistent discrete adjoints?
These points have been addressed using numerical comparisons by Nadarajah and Jameson who
  also discussed the influence of the adjoint gradient accuracy on the final optimized
  shapes for airfoil design problems \cite{NadJam_00,NadJam_01}. Neither the discrepancy in
  the gradients, nor the difference in the final shapes they observed on intermediate and coarse meshes were  very
   significant. Nevertheless it is sound to gain \cltw{a more} general understanding in \cltw{this area}.
  The adjoint consistency issue in CFD has also been studied theoretically in some
articles where the authors
have \cltw{to first} precisely describe the numerical method they
consider, including the possibly  specific discretization close
 to the boundary, and also the way the
function of interest is discretized. Lu and Darmofal  \cite{LuDar_04}
considered a discontinuous Galerkin (DG)
scheme and then demonstrated how to formulate the primal boundary
condition and the output functional to obtain dual consistency. The
difference between the
consistent and non consistent formulations is noticed in both the
appearance of spurious oscillations in the adjoint contours close to the function
 support and the convergence rate of the output
of interest w.r.t. the characteristic mesh size. Hartmann \cite{Har_07}
also discussed dual consistency for DG schemes, including interior
penalty DG and
also dealt with Navier-Stokes equations and test-cases. The adjoint consistency
of a high-order correction
procedure via reconstruction formulation for hyperbolic conservation laws was
investigated in \cite{shi_wang_2015}. As concerning
finite-differences and finite-volume (FV), the questions of
adjoint consistency was discussed
by Duivesteijn et al. and Lozano for  quasi-one-dimensional Euler flows
\cite{DuiBijKor_05,Loz_16} and also for an unsteady conservation law by
Liu and Sandu \cite{LiuSan_08}.
In this later article, the authors considered first-order and second-order
upwind schemes and derived the circumstances in which the discrete
adjoint is punctually not consistent (change of convection direction,
inadequate discretization of boundary condition in particular). Hicken
and Zingg
have demonstrated the influence of adjoint consistency for a class of
summation-by-parts difference schemes, illustrating the benefit of
consistency
in terms of regularity of adjoint contours close to boundaries,
functional grid-convergence, and dual-weighted corrected functional mesh
convergence \cite{HicZin_14}.
Recently, St\"uck carried out a corresponding effort for cell-vertex
FV discretization (using median dual grids), deriving the
condition on the output of
interest and scheme residual for dual consistency \cite{Stu_15,Stu_17}. In this
article, the conditions for dual consistency of the classical
Jameson-Schmidt-Turkel (JST)
scheme in 2-D cell-centered FV are derived. Besides, the dual
consistency is also studied here from a heuristic
point of view, discretizing the continuous adjoint equation for the
discrete adjoint fields which adds valuable information
to the theoretical study where the flow or the gradient of the adjoint is discontinuous
and where numerical divergence of the adjoint is observed

 \cltw{In the case} of hyperbolic equations, the validity of the adjoint linearization around discontinuities
 in the direct solution has to be carefully handled because it results in linear equations with discontinuous
 coefficients. The analysis must include the linearization of the jump relations at the discontinuity
 \cite{majda_83} which leads to a so-called interior boundary condition for the adjoint variables
 \cite{ulbrich_02}. The adjoint relations to the Rankine-Hugoniot (RH) relations have been derived
 in \cite{giles_pierce_2001} for the quasi-one-dimensional and in \cite{BaeCasPal_09,Loz_18} for the
 2-D compressible Euler equations and proved continuity of the adjoint variables across the shock.
 In contrast, their derivatives may be discontinuous
\cite{giles_pierce_2001,GilPie_97,BaeCasPal_09}.
  Lozano \cite{Loz_18} also briefly discussed the consistency of the continuous adjoint
  fields but on rather coarse meshes. The conclusions of his work leave the question open whether
these relations are actually satisfied at the limit of  fine discrete
adjoint fields.

Finally, it has been well-documented
\cite{GilPie_97,Loz_17,Loz_18,Loz_19} that, at certain flow conditions,
  the inviscid lift and drag adjoint fields may exhibit increasing
values as the mesh is refined
  close to the stagnation streamline, the wall and, for (at least locally)
  supersonic flows, the Mach lines impacting a shock foot.
  In the case of the stagnation line, Giles and Pierce derived a
$1/\sqrt{d}$ law for lift and drag adjoint \cite{GilPie_97}
 ($d$ being the distance to the line),  but they used the potential flow
theory and this
   $1/\sqrt{d}$ behavior is not always well-observed for compressible
flows \cite{Loz_17,Loz_18}.
 In an effort to analyze the zones of singular behavior, Todarello et
al. \cite{TodVonBou_16} used the characterization
 of the discrete adjoint and looked how the flow is perturbed by a
residual perturbation located in one of the these
 specific areas. This method is used again here but with perturbations
of the residual corresponding to physical
source terms \cite{GilPie_97} which allows fluid mechanics analysis.
In this framework, it appears that the numerical
divergence of the adjoint variables close to the stagnation streamline and the wall is due to the influence of
the fourth of the source term proposed in \cite{GilPie_97},
 the corresponding convected increment of stagnation
 pressure and diminution of  entropy (created at the
\cltw{location of the source}) and the resulting change in lift and drag.
 Finally, some other aspects of 2-D inviscid lift and
 drag adjoint field are studied in details. In particular, in the case of a
   supersonic flow with a detached shock wave, the mathematical expression of the adjoint field upwind
   the shock wave is derived.

The paper is organized as follows. Sections~\ref{sec:adj_cont_glob} and \ref{sec:adj_discr} are devoted to reminders about
 continuous and discrete adjoint solutions for compressible Euler flows. The adjoint consistency of the cell-centered JST scheme
 with structured 2-D meshes \cite{JamSchTur_81} is theoretically discussed in \S~\ref{sec:JST}. 
 The theoretical results about adjoint consistency and adjoint-gradient discontinuity at shock waves are then
 assessed in \S~\ref{sec:num_xp} using lift- and drag-adjoint fields for inviscid flows around the
 NACA0012 airfoil in supersonic (\S~\ref{sec:xp_naca_sup}), transonic (\S~\ref{sec:xp_naca_tra}) and subsonic
 (\S~\ref{sec:xp_naca_sub}) regimes. The dual consistency property is then also practically discussed
    discretizing the continuous adjoint equation for fine grid discrete adjoint fields. This appears to highlight several zones of numerical
   divergence of the adjoint fields and this divergence is analysed in the vicinity of the wall and
   the stagnation streamline. Finally, concluding remarks about this work are given in \S~\ref{sec:concl}.
	
%
%
\section{Continuous adjoint equations for 2-D Euler flows}\label{sec:adj_cont_glob}
%
%
\subsection{Regions of smooth flow}\label{sec:adj_cont}
The continuous adjoint equations for gas dynamics were first derived by Jameson  \cite{Jam_88} in the case of 2-D compressible Euler flows around a profile. He
  considered a body fitted structured grid that was mapped to a conformal rectangle and used the Euler equations in the resulting $(\xi,\eta)$-coordinates.
 A parametrization of the mapping then allowed to vary the airfoil shape in the physical space (without altering the domain of variation of the 
 transformed coordinates) and to define a gradient calculation problem for functional outputs. 

As, in this formulation, the system of transformed coordinates is attached to a structured mesh, the aforementioned equations
 could not be used for unstructured CFD for which a formulation in physical coordinates was
 necessary.  The corresponding system of equations was first derived by Anderson and Venkatakrishnan in \cite{AndVen_99} \cltw{and} then by Hiernaux and
  Essers \cite{HieEss_99,HieEss_99b}. Below we recall shortly the theory from \cite{AndVen_99}.
  
 The quantity of interest is assumed to be the projection of the force applied by
 the fluid onto the solid, projected in the direction $\overline{d}=(-\cos\alpha,\sin\alpha)^T$ for the lift, or $\overline{d}=(\sin\alpha,\cos\alpha)^T$ for drag with $\alpha$ the angle of attack:
\begin{equation}
 J= \int _{\Gamma_w} p (\overline{n} \cdot \overline{d})  ds, \label{eq:obj_func}
\end{equation}

\noindent where $\Gamma_w$ is the boundary of the solid body, $\overline{n}=(n_x,n_y)^T$ is the unit normal vector pointing outward, and $p$ is the static pressure.  

The 2-D steady compressible Euler equations in conservative form read

\beq \frac{\partial F_x(W) }{\partial x} + \frac{\partial F_y(W)}{\partial y}=0 \quad \mbox{in } \Omega, \label{eq:euler2d} \eeq

\noindent $W$ being the vector of conservative variables and $F_x$ and $F_y$ the physical fluxes:

$$ W = \left( \begin{array}{c}
   \rho   \\
    \rho u_x  \\ 
    \rho u_y  \\ 
   \rho E \\
\end{array}
\right), 
\quad
 F_x(W) = \left( \begin{array}{c}
   \rho u_x  \\
   \rho u_x^2+ p  \\ 
    \rho u_x u_y  \\ 
   \rho u_x H \\
\end{array}
\right),
\quad
 F_y(W) = \left( \begin{array}{c}
   \rho u_y \\
   \rho u_x u_y  \\
   \rho u_y^2+ p  \\
   \rho u_y H  \\
\end{array}
\right),
$$  

\noindent with $H=E+p/\rho$ the total specific enthalpy and $\overline{U}=(u_x,u_y)^T$ the velocity vector. We assume an ideal gas law for the static pressure $p = (\gamma-1) \rho e = (\gamma-1) (\rho E -  \tfrac{1}{2} \rho \|\overline{U}\|^2)$ with $\gamma$ the ratio of specific heats. Let $\delta W$ by a perturbation in the steady state flow that is caused be an infinitesimal perturbation of the airfoil shape or flow conditions. As $W$ and $W+\delta W$ are solutions to the steady Euler equations for the initial and perturbed problems, we get by difference
$$ \frac{\partial (A \delta W) }{\partial x} + \frac{\partial (B \delta W)}{\partial y}=0 \quad \mbox{in } \Omega, $$

\noindent with $A$ and $B$ 
 the Jacobians of the fluxes $F_x$ and $F_y$, respectively. The perturbation in $J$ can be augmented by the dot product of the above equation with an arbitrary co-state field $\psi$:
$$ \delta J=  \int _{\Gamma_w} \delta p (\overline{n} \cdot \overline{d})  ds + 
  \int _{\Gamma_w} p (\delta (\overline{n} )\cdot \overline{d}) ds +  \int _{\Gamma_w}  p (\overline{n} \cdot \overline{d}) \delta(ds) +
 \int _\Omega \psi^T \left( \frac{\partial (A \delta W) }{\partial x} + \frac{\partial (B \delta W)}{\partial y} \right) dv, $$

 \noindent where at first order $\delta p=\nabla p\cdot\delta{\bf x}+{\bf p}'({\bf W})\cdot\delta{\bf W}$ represents pressure variations due
 to both geometry variations $\delta{\bf x}=(\delta x, \delta y)^T$ and solution variations $\delta{\bf W}$ due to this change in the geometry.

Assuming smooth direct and adjoint solutions, the last term can be transformed by integration by parts into
\beas
   - \int _\Omega  \left( \frac{\partial \psi^T }{\partial x} A + \frac{\partial \psi^T}{\partial y} B \right) \delta W dv  
  + \int _{\Gamma_w\cup\Gamma_\infty} \psi^T (n_xA + n_yB)\delta W ds,
\eeas

\noindent with $\Gamma_\infty$ the far field boundary, so we have
\bea
\delta J &=&  \int_{\Gamma_w} \big(\delta p (\overline{n} \cdot \overline{d}) + \int _{\Gamma_w} p (\delta (\overline{n} ) \cdot \overline{d})\big) ds
+  p (\overline{n} \cdot \overline{d}) \delta(ds) 
- \int _\Omega  \left( \frac{\partial \psi^T }{\partial x} A + \frac{\partial \psi^T}{\partial x} B \right) \delta W dv  \nonumber\\
  &&+ \int _{\Gamma_w\cup\Gamma_\infty} \psi^T (n_xA + n_yB)\delta W ds
\label{e:i0}
\eea

The adjoint method removes the dependency in the flow perturbation $\delta W$ for the evaluation of the variation of $J$.
 This directly yields the adjoint equation in the fluid domain:
\beq
 -A^T \frac{\partial \psi}{\partial x} - B^T \frac{\partial \psi}{\partial y}=0, \quad \mbox{in }\Omega.
\label{e:adjcdom} 
\eeq

Note that manipulating (\ref{e:adjcdom}) one may deduce 
\beq  \ol{U}\cdot\nabla\psi_1-H\ol{U}\cdot \nabla \psi_4 = 0. \label{e:adjcdom4} \eeq

Besides, using the variation of the boundary condition $ \delta ( \overline{U} \cdot \overline{n})=0$ and expanding
  $\psi^T (n_xA + n_yB)\delta W $ at the wall yields the wall boundary condition \cite{AndVen_99}
\beq
   \overline{n} \cdot \overline{d} + \psi_2 n_x + \psi_3 n_y = 0.
 \label{e:adjcwal}
 \eeq

 In the far field, no variation of the boundary needs to be  considered.
 The Jacobian in direction $\overline{n}$ can be rewritten by using a locally 
 one-dimensional characteristic decomposition to yield

\beq
  \int _{\Gamma_\infty} \psi^T (n_xA + n_yB )\delta W ds = \int _{\Gamma_\infty} (\psi^T P^{-1}  D   P\delta W) ds. 
\label{e:i2}
\eeq
 
It is assumed that $ P\delta W \simeq \delta( P W) $. The variation in these characteristic variables is zero for the components
 corresponding to negative eigenvalues of the Jacobian, $n_xA + n_yB$, in the classical 1D approximate linearization at the boundary
 (information coming from outside of the domain and fixed characteristic value). The far field adjoint BC simply imposes that the other
 components of  $\psi^T P^{-1}$ are zero so that  $\delta W^T(n_xA^T + n_yB^T)\psi$ vanishes for all $\delta W$.
 %
 %
 \subsection{Jump relations for the adjoint derivatives}\label{sec:adj_RH}

 We \cltw{derive here} some useful relations for the jump of the adjoint variables derivatives across the shock.  As reported in the introduction,
 the derivation of the adjoint Rankine-Hugoniot relations for the quasi-one-dimensional \cite{giles_pierce_2001} and 2-D \cite{Loz_18,BaeCasPal_09}
 compressible Euler equations show that the adjoint variables are continuous across the shock, while the adjoint derivatives in the direction normal
 to the shock may be discontinuous. These jump relations for the derivatives have been manipulated in \cite{Loz_18} to provide information on the
 behavior of adjoint variables at shocks. We \cltw{reproduce here} this latter analysis but in a somewhat more general way and numerical experiments on
 very fine meshes in Figs.~\ref{fig:adj_across_shock_sup} and \ref{fig:adj_across_shock_tra} will support our conclusions.

Let consider an isolated discontinuity $\Sigma$ where the direct solution satisfies the RH jump relations 

\beq \du n_xF_x(W)+n_yF_y(W)\df=0, \label{eq:RH} \eeq

\noindent where $\du\cdot\df$ denotes the jump across $\Sigma$ in the unit direction $\ol{n}=(n_x,n_y)^T$ normal to $\Sigma$. Introducing $\ol{t}=(-n_y,n_x)^T$, the adjoint equation (\ref{e:adjcdom}) may be rewritten as

\beq
 -A_n^T \frac{\partial \psi}{\partial n} - A_t^T \frac{\partial \psi}{\partial t}=0 \quad  \mbox{in }\Omega,
\label{eq:adjcdom_n_t}
\eeq

\noindent with

\begin{align*}
 A_n =n_xA+n_yB &=\begin{pmatrix} 0 & n_x & n_y & 0 \\ (\gamma-1)e_cn_x-uv_n & v_n-(\gamma-2)un_x & un_y-(\gamma-1)vn_x & (\gamma-1)n_x \\ (\gamma-1)e_cn_y-vv_n & vn_x-(\gamma-1)un_y & v_n-(\gamma-2)vn_y & (\gamma-1)n_y \\ \big((\gamma-1)e_c-H)v_n & Hn_x-(\gamma-1)uv_n & Hn_y-(\gamma-1)vv_n & \gamma v_n \end{pmatrix}, \\
 A_t =-n_yA+n_xB &=\begin{pmatrix} 0 & -n_y & n_x & 0 \\-(\gamma-1)e_cn_y-uv_t & v_t+(\gamma-2)un_y & un_x+(\gamma-1)vn_y &-(\gamma-1)n_y \\ (\gamma-1)e_cn_x-vv_t &-vn_y-(\gamma-1)un_x & v_t-(\gamma-2)vn_x & (\gamma-1)n_x \\ \big((\gamma-1)e_c-H)v_t &-Hn_y-(\gamma-1)uv_t & Hn_x-(\gamma-1)vv_t & \gamma v_t \end{pmatrix},
\end{align*}

\noindent where $v_n=\ol{n}\cdot\ol{U}$, $v_t=\ol{t}\cdot\ol{U}$, $\partial_n\psi=n_x\partial_x\psi+n_y\partial_y\psi$, and $\partial_t\psi=-n_y\partial_x\psi+n_x\partial_y\psi$. Assuming that $A_n$ is non-singular, the adjoint is continuous across $\Sigma$ \cite{giles_pierce_2001,GilPie_97,Loz_18,BaeCasPal_09}: $\du\psi\df=0$, so $\du\partial_t\psi\df=0$. Likewise we have $\du\rho v_n\df=0$, $\du v_t\df=0$, and $\du H\df = 0$.

Let (\ref{eq:adjcdom_n_t}a-d) denote the four equations in (\ref{eq:adjcdom_n_t}). Then, the operation (\ref{eq:adjcdom_n_t}a)$+\tfrac{u_x}{2}$(\ref{eq:adjcdom_n_t}b)$+\tfrac{u_y}{2}$(\ref{eq:adjcdom_n_t}c) gives
\bes
 \frac{v_n}{2}\big(\partial_n\psi_1-H\partial_n\psi_4\big) + \frac{v_t}{2}\big(\partial_t\psi_1-H\partial_t\psi_4\big) = 0,
\ees

\noindent and applying the jump operator we obtain
\beq \du v_n\partial_n\psi_1\df - H\du v_n\partial_n\psi_4\df = 0.  \label{eq:adj_RH_1}\eeq
Note that for functions of the static pressure and the geometry and uniform far field, one gets from (\ref{e:adjcdom4})
that $\nabla\psi_1=H\nabla\psi_4$ which is actually observed later on in the numerical experiments with lift and drag adjoint (see also equation (\ref{e:lam14}) below).
Likewise, (\ref{eq:adjcdom_n_t}d) gives
\beq  (\gamma-1)\du n_x\partial_n\psi_2+n_y\partial_n\psi_3\df + \gamma\du v_n\partial_n\psi_4\df = 0, \label{eq:adj_RH_2}\eeq
\noindent and the operations $\du n_x$(\ref{eq:adjcdom_n_t}c)$-n_y$(\ref{eq:adjcdom_n_t}b)$\df$ and $\du n_x$(\ref{eq:adjcdom_n_t}b)$+n_y$(\ref{eq:adjcdom_n_t}c)$\df$ result in
\begin{subequations}
\begin{align}
 \du v_n\big(-n_y\partial_n\psi_2 + n_x\partial_n\psi_3 + v_t\partial_n\psi_4 \big)\df + \du v_n\df\big( n_x\partial_t\psi_2 + n_y\partial_t\psi_3\big) &= 0,  \label{eq:adj_RH_3}\\
 \du \partial_n\psi_1\df + \du (u+v_nn_x)\partial_n\psi_2 + (v+v_nn_y)\partial_n\psi_3 + (H+v_n^2)\partial_n\psi_4\df + \du v_n\df v_t\partial_t\psi_4 &= 0.  \label{eq:adj_RH_4}
\end{align}
\end{subequations}

Finally, the adjoint equation to the RH relations (\ref{eq:RH}) reads $\du -n_yF_x(W)+n_xF_y(W)\df\cdot\partial_t\psi=0$ \cite{giles_pierce_2001,Loz_18,BaeCasPal_09} which may be simplified into
\beq v_t\du\rho\df(\partial_t\psi_1+H\partial_t\psi_4) + ( \du p\df + v_t^2\du\rho\df)(-n_y\partial_t\psi_2 + n_x\partial_t\psi_3) = 0. \eeq

%
%
\section{Discrete adjoint equations}\label{sec:adj_discr}
%
%
\subsection{Discrete gradient calculation}
The finite volume scheme of interest defines the steady-state discrete flow vector $W$ (of size $n_W$) as the solution of a set of $n_W$ non-linear equations 
 involving $W$ and the vector of mesh coordinates $X$:
$$ R(W,X) = 0. $$

Let us assume that the mesh is a smooth function of the vector of design parameters $\beta$ (of size $n_\beta$), then the implicit function theorem allows to define $W$ as a function of $X$ and so of $\beta$ \cite{JPDwi_10}. Discrete gradient
 calculation consists of computing the derivatives of the $n_f$ functions
$${\cal J}_k(\beta) = \Jd_k(W(\beta),X(\beta)), \quad  1\leq k \leq n_f,$$

\noindent with respect to the $n_\beta$ design parameters. In external aerodynamic applications, $n_\beta$  is usually much larger than $n_f$ and the most
 efficient way to proceed is to use the discrete adjoint method which requires to solve $n_f$ linear systems
\beq
 \left(\dxdx{R}{W}\right)^T \Lambda_k = -\left(\dxdx{\Jd_k}{W}\right)^T, \quad  1\leq k \leq n_f,
 \label{eq:discr_adj_equ}
\eeq

\noindent and then to calculate
\[
\tdxdx{{\cal J}_k}{\beta} = \dxdx{\Jd_k}{X} \tdxdx{X}{\beta} 
                         + \Lambda_k^T\left(\dxdx{R}{X} \tdxdx{X}{\beta}\right).
\]

The dominant cost is the inversion of the $n_f$ linear systems of size $n_W$ whereas all other classical
methods solve $n_\beta$ linear (or non-linear for finite-differences) systems of size $n_W$.
 The cost of solving $n$ systems with the same matrix and different right-hand sides can
  be mitigated using specific methods like block-GMRES \cite{Gut_06,PinMon_13} or specific CPU
  optimizations but in the common case where $n_f \ll n_\beta$ the
  adjoint method is
 more efficient than any direct approach. 

%
%
\subsection{Numerical characterization of the discrete adjoint}
The adjoint vector  $\Lambda$ associated with one function $\Jd$ is usually identified with the sensitivity of $\Jd$ to a perturbation 
of the residual $R$ followed by re-convergence (see \cite{FidDar_11} and references therein). Following
\cite{GilPie_97, FidDar_11},  we consider here a perturbation $\delta R$ added to the right-hand-side of the discrete flow equations and denote by $W + \delta W$ the converged solution corresponding to the perturbed equation:
$$ R(W + \delta W, X) = \delta R, $$
\noindent or at first order $\tfrac{\partial R}{\partial W} \delta W=\delta R$.
Since  $ \Jd(W+\delta W,X)   \simeq \Jd(W,X) + \tfrac{\partial \Jd}{\partial W}\delta W $  the first-order change in the function of interest $\Jd$ due to $\delta W$ is
$$   \delta \Jd = \frac{\partial \Jd}{\partial W} \Big(\frac{\partial R}{\partial W}\Big)^{-1} \delta R,$$
\noindent or involving the adjoint vector $\Lambda$ defined in (\ref{eq:discr_adj_equ}): $\delta \Jd =  -\Lambda ^T \delta R$.
If only the $a$-th component of $R$ at cell $m$ has been altered by a small quantity $\delta R^{a}_m$, then the previous
equation yields  $  \Lambda^a_m =  - \delta \Jd /\delta R^a_m $, which defines the $a$-th component
of $\Lambda $ at cell $m$ as the limit ratio of the change in  $\Jd$ divided by the  
infinitesimal change in the residual $R$ at the corresponding
cell and component which caused the change in flow solution and function value.	
%
%
\subsection{Physical characterization for 2-D Euler flows}\label{sec:source_term_approach}

We \cltw{recall here} the physical interpretation of the adjoint solution by Giles and Pierce \cite{GilPie_97}. More precisely,
the discussion in \cite{GilPie_97} is based on the continuous adjoint equation and its discrete
 counterpart is presented here.
 Four physical source terms $\delta R$ are defined
 at each individual cell: 1) local  mass source at fixed stagnation pressure and total enthalpy; 2) local normal force; 3) local change in total enthalpy at
 fixed static and total pressure; 4) local change in total pressure at fixed total enthalpy and static pressure:
\beq
 \delta R_m ^1 = \epsilon \left( \begin{array}{c}
   1  \\
   u_x  \\ 
   u_y  \\ 
   H  \\
\end{array}
\right), 
~~
 \delta R_m ^2 = \epsilon \left( \begin{array}{c}
   0  \\
   -\rho u_y  \\
   \rho u_x  \\
   0  \\
\end{array}
\right),
~~
 \delta R_m ^3 = \epsilon \left( \begin{array}{c}
   -\frac{1}{2H} \\
   0  \\
   0  \\
   \frac{1}{2}  \\ 
\end{array}
\right),
~~
 \delta R_m ^4 = \epsilon \left( \begin{array}{c}
   \frac{1}{p_0} \left( \frac{\gamma - 1}{\gamma} + \frac{1}{\gamma M^2} \right)  \\
   \frac{u_x}{p_0} \left( \frac{\gamma - 1}{\gamma} + \frac{2}{\gamma M^2} \right)  \\
   \frac{u_y}{p_0} \left( \frac{\gamma - 1}{\gamma} + \frac{2}{\gamma M^2} \right)  \\
   \frac{H}{p_0} \left( \frac{\gamma - 1}{\gamma} + \frac{1}{\gamma M^2} \right)  \\ 
\end{array}
\right),
\label{e:deltaR} 
\eeq
%
\noindent with $0<\epsilon\ll1$ and where $H$ denote the total enthalpy and $M$ is the Mach number. 
At first order, the corresponding function
variations  $\delta \Jd^1_m$, $\delta \Jd^2_m$, $\delta \Jd^3_m$, $\delta \Jd^4_m$ 
 due to the source terms in cell $m$, (\ref{e:deltaR}), read
\bea
    (\delta \Jd^1_m, \delta \Jd^2_m, \delta \Jd^3_m, \delta \Jd^4_m) = - (\Lambda^1_{m}, \Lambda^2_{m}, \Lambda^3_{m}, \Lambda^4_{m}) \times
\left(
\begin{array}{cccc}
  \delta R^1_{1,m}   & \delta R^2_{1,m}   &   \delta R^3_{1,m}   &   \delta R^4_{1,m} \\
  \delta R^1_{2,m}   & \delta R^2_{2,m}   &   \delta R^3_{2,m}   &   \delta R^4_{2,m} \\
  \delta R^1_{3,m}   & \delta R^2_{3,m}   &   \delta R^3_{3,m}   &   \delta R^4_{3,m} \\
  \delta R^1_{4,m}   & \delta R^2_{4,m}   &   \delta R^3_{4,m}   &   \delta R^4_{4,m} 
\end{array}
\right),
 \label{e:physpre}  
\eea

\noindent where $\delta \Jd^d_m$ corresponds to the change in $\Jd$ due to $d$-th change of $R$ in cell $m$. Converging again the perturbed solution $W+\delta W^d$ leads to the change $\delta \Jd^d_m$ in the function of interest. As the four changes in $R$ are linearly independent,
Giles and Pierce define the adjoint vector at cell $m$ as the solution to (\ref{e:physpre}) so, using the perturbation terms in (\ref{e:deltaR}), we obtain
\bea 
     (\Lambda^1_m, \Lambda^2_m, \Lambda^3_m, \Lambda^4_m) &=&  - (\delta \Jd_m^{1}, \delta \Jd_m^{2},\delta \Jd_m^{3}, \delta \Jd_m^{4}) \times \nonumber \\
    &&\ds\frac{1}{\epsilon} \left(
\begin{array}{cccc}
  1+\frac{(\gamma-1)}{2}M^2  & -\frac{(1+(\gamma-1)M^2)u_x}{\|\overline{U}\|^2} & -\frac{(1+(\gamma-1)M^2)u_y}{\|\overline{U}\|^2} & \frac{1+\frac{(\gamma-1)}{2}M^2}{H}\\
  0                          & -\frac{u_y}{\rho \|\overline{U}\|^2}             & \frac{u_x}{\rho \|\overline{U}\|^2}              &    0\\
  -H                         &     0                           &            0                    &    1\\
  -\frac{\gamma p_0 M^2}{2}   & \frac{\gamma p_0 u_x}{c^2}         & \frac{\gamma  p_0 u_y}{c^2}          & -\frac{\gamma  p_0 M^2}{2H} 
\end{array}
\right).
 \label{e:physp3}  
\eea

 The residual perturbation $\delta R$ has been physically defined
 thus giving a physical meaning to the local discrete adjoint vector. \clon{The adjoint vector is intrinsic to the system of equations for which a similar
 demonstration can be done, and we can expect similar solutions from different discretizations on the same mesh}.

Finally, Giles and Pierce \cite{GilPie_97} have noted for continuous adjoint that the third perturbation 
 does not alter the pressure field for inviscid flows and hence leaves the drag and lift unchanged.
 From the expression of this perturbation, it is then straightforward to prove \cltw{that 
 the} lift and drag continuous adjoint fields satisfy all over the fluid domain the following equation: 
\beq
 \psi_1 = H \psi_4.
\label{e:lam14}
\eeq

This property may be compared to (\ref{e:adjcdom4}) and is seen to be  well satisfied in the numerical adjoint fields (see \S~\ref{sec:xp_naca_tra_adj_sol}).
The fourth perturbation  $\delta R^4$ also preserves the static pressure at \cltw{the location of the source term} but the authors of \cite{GilPie_97} \cltw{proved} that,
 contrary to $\delta R^3$, for arbitrary flow conditions, it \cltw{does not preserve the static pressure field  nor} the streamtube structure of
  the base flows  (in particular, the mass flux varies
  downstream the source term in the corresponding streamtube of the original flow).
	
%
%
\section{Adjoint consistency of the cell-centered JST scheme}\label{sec:JST} 
 The dual consistency of the JST scheme is discussed here for 2-D FV cell-centred simulations on structured meshes.
 We use the classical method
of equivalent differential equation which consists of introducing Taylor series expansions for all
 state variables and adjoint
values in order to analyze the truncation error of the scheme \cite[ch.~9]{Hir_07}.
 Let us stress that this approach remains valid in
the case of discontinuous solutions \cite{GOODMAN1985336}.

Lozano studied the adjoint consistency of the discrete adjoint of the JST scheme for both cell-centered and
cell-vertex discretizations \cite{Loz_16} of the quasi-one-dimensional compressible Euler equations. Boundary
conditions using artificial dissipation (AD) were applied which lead to inconsistent discrete adjoint scheme at
the boundaries. Such boundary conditions are not suitable for simulations of internal flows that require
 conservation and, in this work, no artificial dissipation is added to the boundary fluxes.
 However, several classical AD formulas may still be used for the second to last faces close to a physical boundary
  \cite{SwaTur_87,SwaTur_92}.
  In the following we will first consider the default discretization of our code (see equation (\ref{e:fjst2}) below)
  and then consider in \S\ref{sec:adj_cons_Fb_Fc} 
  two alternative formulas (equations (\ref{e:fjst2b}) and (\ref{e:fjst2c})) with different properties with respect to
  dual consistency. 
  Finally, this analysis is completed in \S~\ref{sec:xp_naca_sup} to \ref{sec:xp_naca_sub}
  by numerical and  visual examination of the residuals
of the continuous adjoint equation calculated for the discrete lift and drag adjoint fields.

The reference continuous equations will be the adjoint of Euler equations in transformed coordinates.
Jameson presented these equations in 2-D and 3-D \cite{Jam_88}.
 Later on, Giles and Pierce  derived the corresponding direct differentiation equation
 in 2-D \cite{GilPie_99}. In this framework, the physical space (typically about an airfoil)
 is seen as the image of  the unit domain $[0,1]\times[0,1]$ with coordinates $(\xi,\eta)$.
 The transformed 2-D Euler equations in the $(\xi,\eta)$-coordinates are well-known and,
 for functions of the form (\ref{eq:obj_func}), the adjoint equations read
\beq
- \dxdx{\Lambda^T}{\xi}  \left( \dxdx{y}{\eta} A(W) - \dxdx{x}{\eta} B(W) \right) -  \dxdx{\Lambda^T}{\eta} \left(-A(W)\dxdx{y}{\xi}  + B(W)\dxdx{x}{\xi}  \right) = 0,
\label{e:conttc}
\eeq

\noindent \cltw{where $A(w)$ and $B(w)$ are} the Jacobians of Euler flux in the physical $x$ and $y$ directions. Let us assume that
 the wall corresponds to the $\xi=0$ boundary and that the function of
 interest is the force applied by the fluid on the solid in direction $\overline{d}=(d_x,d_y)$,
 then the corresponding boundary condition reads: 
 $$ - \Lambda^T \left( A(W)\dxdx{y}{\eta} - B(W)\dxdx{x}{\eta} \right) - (\dxdx{y}{\eta}d_x- \dxdx{x}{\eta} d_y)\dxdx{p}{W} = 0~~~~ \textrm{at}~~ \xi=0. $$

 \noindent It can be simplified by benefiting from the specific form of the Jacobian at a wall yielding
\beq
 \Lambda_2 \dxdx{y}{\eta} -\Lambda_3  \dxdx{x}{\eta} +(\dxdx{y}{\eta}d_x- \dxdx{x}{\eta} d_y)=0.
 \label{e:contbc}
\eeq
 
The adjoint consistency of the FV scheme of interest is discussed hereafter referring to
the formal equations (\ref{e:conttc}) and  (\ref{e:contbc}).
Using a structured mesh, the JST scheme in the $i$-direction reads (the $j$ subscripts have been
dropped for the sake of readability):
\bea
 F^{JST}_{i+1/2} = \tfrac{1}{2} ({\bf F}(W_i)+{\bf F}(W_{i+1}))\cdot S_{i+1/2}
 &-&k^2 \nu_{i+1/2} \kappa_{i+1/2} (W_{i+1}-W_i)   \nonumber \\
 &+& \overline{k^4}_{i+1/2} \kappa_{i+1/2} (W_{i+2}-3 W_{i+1}+3 W_{i} - W_{i-1})   \label{e:fjst1}
\vspace{1mm} \\
 \nu_i = \frac{|p_{i+1}-2 p_i + p_{i-1}|}{(p_{i+1}+2 p_i + p_{i-1})}~~&&~~~ \nu_{i+1/2} = \max (\nu_i,\nu_{i+1})~~~~~
\vspace{1mm}  \label{e:fjstsa} \\
\overline{k^4}_{i+1/2}= \max (0,k^4-\nu_{i+1/2} k^2)~~~&&~~  \kappa_{i+1/2}= |\overline{U}_{i+1/2}.S_{i+1/2}|+c_{i+1/2} \| S_{i+1/2} \|, \label{e:fjstsb}
\eea

\noindent where ${\bf F}=(F_x,F_y)$. \cltw{The fluid velocity $\overline{U}$ and the speed of sound $c$ at the interface $i+1/2$ are} evaluated from
one of the classical conserved variable means. Dual consistency close to the boundary
 depends on the specific flux formulas at the boundary and second to last interfaces:
\bea
 F^{JST}_{1/2}  &=&    {\bf F}(W_b)\cdot S_{1/2}    \label{e:fjst3} \\
 F^{JST_a}_{3/2} &=& \tfrac{1}{2} ({\bf F}(W_1)+{\bf F}(W_{2}))\cdot S_{3/2} -k^2 \nu_{3/2} \kappa_{3/2} (W_{2}-W_1) + \overline{k^4}_{3/2} \kappa_{3/2} (W_{3}-3 W_{2}+2 W_{1}) \label{e:fjst2} 
 \eea

 \noindent where $p_0=2p_b-p_1$ is used in the evaluation of $\nu_1$, and $W_b$ is a boundary state satisfying all Dirichlet-like boundary conditions.
 For the sake of comparison, two alternative formulas \cite{SwaTur_87,SwaTur_92} will be considered:
 \bea
 F^{JST_b}_{3/2} &=& \tfrac{1}{2} ({\bf F}(W_1)+{\bf F}(W_{2}))\cdot S_{3/2} -k^2 \nu_{3/2} \kappa_{3/2} (W_{2}-W_1) + \overline{k^4}_{3/2} \kappa_{3/2} (W_{3}-2 W_{2}+ W_{1}), \label{e:fjst2b} \\
 F^{JST_c}_{3/2} &=& \tfrac{1}{2} ({\bf F}(W_1)+{\bf F}(W_{2}))\cdot S_{3/2} -k^2 \nu_{3/2} \kappa_{3/2} (W_{2}-W_1) + \overline{k^4}_{3/2} \kappa_{3/2} (W_{3}-3 W_{2}+3 W_{1}-W_g), \label{e:fjst2c} 
 \eea
 where $W_g$ is a mirror state of $W_1$ w.r.t. the boundary.

Note that at specific faces  where either $\overline{U}\cdot S=0$, or $k^4-\nu k^2=0$, or $\nu_i=\nu_{i+1}$, the flux is not differentiable. This issue is discussed in \cite{Ngu_14} for a simulation about a symmetric airfoil, with a symmetric mesh and zero angle of attack leading to two complete lines of faces
where  $|\overline{U}\cdot S|$ is extremely close to machine-level-zero. The small irregularities of adjoint quantities that appear close to these lines are cured by regularization
but no such problem is observed for more complex cases and, most generally, this question is discarded \cite{DwiBre_06} considering that the occurrence of such equalities is marginal in practice.

  Under the classical assumption that $\Delta \xi$ and $\Delta \eta$ vary smoothly and have the same order of
  magnitude, the consistency of the discrete adjoint of equations (\ref{e:fjst1}) to (\ref{e:fjst3}) completed by
   (\ref{e:fjst2}), (\ref{e:fjst2b}) or (\ref{e:fjst2c}),    
  is studied. The main result of this section is the following.	

\begin{lem}
  For a wall integral, the discrete  discrete adjoint of the finite-volume  JST scheme is:
	\begin{itemize}
    \item consistent with the continuous adjoint equations (\ref{e:conttc}) in interior cells;
    \item consistent with the continuous adjoint equations (\ref{e:conttc}) in a penultimate cell w.r.t. a boundary using $F^{JST_b}_{3/2}$  whereas it is inconsistent using $F^{JST_a}_{3/2}$ or $F^{JST_c}_{3/2}$;
    \item consistent with the wall boundary condition (\ref{e:contbc}) in a cell adjacent to a wall.
	\end{itemize}
The first and third results are independent of the choice of the $F^{JST}_{3/2}$ formula.
\end{lem}

	Unfortunately, the choice of $F^{JST_b}_{3/2}$ induces significant robustness issues in the simulations on
    very fine meshes and we finally discuss in \S4.5 how to slightly alter the exact Jacobian obtained
    with $F^{JST_a}_{3/2}$
    and  $F^{JST_c}_{3/2}$ to get adjoint consistency.      
%
%
\subsection{Consistency of discrete adjoint inside the domain}\label{sec:JST_int_domain} 

The discrete adjoint equation at a current cell $(i,j)$ reads
\bea
     (\Lambda_{(i-2,j)}^T-\Lambda_{(i-1,j)}^T) \frac{\partial F_{i-3/2,j}}{\partial W_{(i,j)}} 
    +(\Lambda_{(i-1,j)}^T-\Lambda_{(i  ,j)}^T) \frac{\partial F_{i-1/2,j}}{\partial W_{(i,j)}} \label{ep:i1} && \\
    +(\Lambda_{(i  ,j)}^T-\Lambda_{(i+1,j)}^T) \frac{\partial F_{i+1/2,j}}{\partial W_{(i,j)}} 
    +(\Lambda_{(i+1,j)}^T-\Lambda_{(i+2,j)}^T) \frac{\partial F_{i+3/2,j}}{\partial W_{(i,j)}} \label{ep:i2} &&  \\ 
    +(\Lambda_{(i,j-2)}^T-\Lambda_{(i,j-1)}^T) \frac{\partial F_{i,j-3/2}}{\partial W_{(i,j)}} 
    +(\Lambda_{(i,j-1)}^T-\Lambda_{(i  ,j)}^T) \frac{\partial F_{i,j-1/2}}{\partial W_{(i,j)}} \label{ep:j1} && \\
    +(\Lambda_{(i , j)}^T-\Lambda_{(i,j+1)}^T) \frac{\partial F_{i,j+1/2}}{\partial W_{(i,j)}}
    +(\Lambda_{(i,j+1)}^T-\Lambda_{(i,j+2)}^T) \frac{\partial F_{i,j+3/2}}{\partial W_{(i,j)}}  && = 0.   \label{ep:j2}
\eea

Only fluxes in the $i$-direction appear in (\ref{ep:i1})-(\ref{ep:i2}), while only fluxes in the $j$-direction appear in (\ref{ep:j1})-(\ref{ep:j2})
 and we will analyze them separately. This is done below in the $i$-direction (again dropping the $j$ subscripts).
  The contribution  of the centered flux in (\ref{ep:i1})-(\ref{ep:i2}) is
\beq
  \frac{1}{2} \Big( (\Lambda_{i-1}^T-\Lambda_{i }  ^T)  \frac{\partial {\bf F}}{\partial W_i} S_{i-1/2} 
  +  (\Lambda_{i  }^T-\Lambda_{i+1}^T)  \frac{\partial {\bf F}}{\partial W_i}  S_{i+1/2} \Big).
  \label{e:adjc1}
\eeq  

The $i$ and $j$ subscripts and $\xi$ and $\eta$ coordinates are linked \cltw{by the} simple affine transformations
$$ \xi(i) = \frac{i-1}{i_{max}-1} = (i-1) \Delta \xi, ~~~~~~~~~  \eta(j) = \frac{j-1}{j_{max}-1} =  (j-1) \Delta \eta,$$ 

\noindent and the expression of the surface vectors is easily derived from those of the coordinates
$$ S_{i+1/2,j} = \left( \begin{array}{c}
   y(\xi(i+1/2),\eta(j+1/2))- y(\xi(i+1/2),\eta(j-1/2)) \\
  -x(\xi(i+1/2),\eta(j+1/2))+ x(\xi(i+1/2),\eta(j-1/2)) \\ 
\end{array}
\right).
$$

The two terms in (\ref{e:adjc1}) are hence consistent with
$$- \frac{1}{2} \dxdx{\Lambda^T}{\xi} \left( \dxdx{y}{\eta} A  - \dxdx{x}{\eta} B \right)  \Delta \xi \Delta \eta $$

\noindent respectively at points $(i-1/2,j)$ and $(i+1/2,j)$, with $\Delta \xi \Delta \eta$ the volume of a cell, and the sum of both contributions is a consistent approximation  of
 $$ - \dxdx{\Lambda^T}{\xi} \left( \dxdx{y}{\eta} A  - \dxdx{x}{\xi} B \right)  \Delta\xi \Delta \eta $$

\noindent at point $(i,j)$. Moreover it is second-order accurate as it uses symmetric means and differences w.r.t. $(\xi(i),\eta(j))$.
Carrying out the corresponding expansions for the linearized fluxes in the $j$-direction,
  the second term of equation (\ref{e:conttc}) times  $\Delta \xi \Delta \eta $ is recovered.
  The first part of the discrete adjoint equation, obtained by isolating the differentiation
  of the centered flux, is hence a consistent discretization of equation (\ref{e:conttc}) up to a multiplicative  $\Delta \xi \Delta \eta$ factor.
  
  After this first step, the discussion of the interior cells adjoint consistency of
 the  JST scheme comes back to the examination of all the other terms
in (\ref{ep:i1})-(\ref{ep:j2}) and the question whether other second-order terms
in $\Delta \xi$ and $\Delta \eta$ appear (inconsistency) or only higher-order terms appear (consistency).
  The terms in (\ref{ep:i1})-(\ref{ep:i2}) that involve the derivatives of the sensor, read
\bea
     k^2 \bigg(~ (\Lambda_{i-2}^T-\Lambda_{i-1}^T)  \frac{\partial \nu_{i-3/2}}{\partial W_{(i,j)}} \kappa_{i-3/2} \Big(- (W_{i-2}-W_{i-1})- 1_{[\overline{k^4}_{i-3/2}>0]} (W_{i}-3 W_{i-1}+3 W_{i-2} - W_{i-3})  \Big)  \nonumber  \\
    +(\Lambda_{i-1}^T-\Lambda_{i}^T)   \frac{\partial \nu_{i-1/2}}{\partial W_{(i,j)}} \kappa_{i-1/2} \Big(- (W_{i-1}-W_{i})   -  1_{[\overline{k^4}_{i-1/2}>0]}(W_{i+1}-3 W_{i}+3 W_{i-1} - W_{i-2})  \Big)  \nonumber \\
    +(\Lambda_{i}^T-\Lambda_{i+1}^T)  \frac{\partial \nu_{i+1/2}}{\partial W_{(i,j)}}\kappa_{i+1/2} \Big(- (W_{i}-W_{i+1})   -  1_{[\overline{k^4}_{i+1/2}>0]}(W_{i+2}-3 W_{i+1}+3 W_{i} - W_{i-1})  \Big)  \nonumber \\
    +(\Lambda_{i+1}^T-\Lambda_{i+2}^T)\frac{\partial \nu_{i+3/2}}{\partial W_{(i,j)}}\kappa_{i+3/2} \Big(- (W_{i+1}-W_{i+2})  -  1_{[\overline{k^4}_{i+3/2}>0]} (W_{i+3}-3 W_{i+2}+3 W_{i+1} - W_{i})  \Big) \bigg),
    \label{e:adjc2}
\eea

\noindent where $1_{[\overline{k^4}_{i+1/2}>0]}$ stands for 1 if  $\overline{k^4}_{i+1/2} $ is strictly positive and 0 if not. As the spectral radius $\kappa$ is $O(\Delta \eta)$, all the terms in (\ref{e:adjc2}) are at least $O(\Delta \xi^2 \Delta \eta)$. The terms of (\ref{ep:i1})-(\ref{ep:i2}) that involve the derivatives of
the spectral radius are
\bea
    (\Lambda_{i-1}^T-\Lambda_{i}^T)  \frac{\partial \kappa_{i-1/2} }{\partial W_{(i,j)}} \Big(-k^2  \nu_{i-1/2} (W_{i-1}-W_{i})   +\overline{k^4}_{i-1/2} (W_{i+1}-3 W_{i}+3 W_{i-1} - W_{i-2})  \Big)   \nonumber \\
    (\Lambda_{i}^T-\Lambda_{i+1}^T) \frac{\partial \kappa_{i+1/2}}{\partial W_{(i,j)}} \Big(-k^2 \nu_{i+1/2} (W_{i}-W_{i+1})   + \overline{k^4}_{i+1/2}(W_{i+1}-3 W_{i}+3 W_{i} - W_{i-1})  \Big).
\eea

As $\kappa$ and its derivatives w.r.t. $W$ are $O(\Delta \eta)$ and as $\nu$ is $O(\Delta \xi ^2)$ these terms are $O(\Delta \xi ^4 \Delta \eta)$.
 The terms arising when differentiating the first-difference and the third-difference are  
\beas
     && (\Lambda_{i-2}^T-\Lambda_{i-1}^T)  ( \kappa_{i-3/2} ~\overline{k^4}_{i-3/2}  )  
    + (\Lambda_{i-1}^T-\Lambda_{i}^T)  ( -k^2 \nu_{i-1/2} ~\kappa_{i-1/2}   - 3 \overline{k^4}_{i-1/2} \kappa_{i-1/2}  ) \nonumber  \vspace{1mm} \\
    +&& (\Lambda_{i}^T-\Lambda_{i+1}^T)  ( k^2 \nu_{i+1/2} ~\kappa_{i+1/2}    + 3 \overline{k^4}_{i+1/2} \kappa_{i+1/2}  )  
   + (\Lambda_{i+1}^T-\Lambda_{i+2}^T) ( - \kappa_{i+3/2} ~ \overline{k^4}_{i+3/2} ),
\eeas

\noindent and may be rewritten as 
\beas
     && (\Lambda_{i-2}^T-\Lambda_{i-1}^T)  \kappa_{i-3/2} ~\overline{k^4}_{i-3/2}   
    - 3 (\Lambda_{i-1}^T-\Lambda_{i}^T)  \kappa_{i-1/2}   ~\overline{k^4}_{i-1/2} \\ 
    +&& 3  (\Lambda_{i}^T-\Lambda_{i+1}^T)  \kappa_{i+1/2}  ~\overline{k^4}_{i+1/2}
    -    (\Lambda_{i+1}^T-\Lambda_{i+2}^T) \kappa_{i+3/2} ~ \overline{k^4}_{i+3/2}  \\    
    -&& k^2 (  (\Lambda_{i-1}^T-\Lambda_{i}^T)  \nu_{i-1/2} ~\kappa_{i-1/2}  
           - (\Lambda_{i}^T-\Lambda_{i+1}^T)  \nu_{i+1/2} ~\kappa_{i+1/2} ).
\eeas

The last two terms that stem from first-difference dissipation fluxes, are both $O(\Delta\xi^3 \Delta\eta)$ 
as the sensor in $\xi$ ($I$-mesh) direction is $O(\Delta \xi^2)$.
The possibly problematic terms are the four first terms stemming from the third-difference dissipation fluxes.
These are individually $O(\Delta \xi \Delta \eta)$ terms and the question is whether their linear combination
is a higher-order term.
 If the algebraic expression of $\overline{k^4}$ as function of the state variables is the same
for all four values, then a third-order difference is identified and linear combination of these four terms
is  $O(\Delta \xi^4 \Delta \eta)$.
If some of the $\nu_{l+1/2}$ are equal to the left value $\nu_{l}$ and some to the 
 right value $\nu_{l+1}$ but  none of the $\bar{k^4}$ is zero, the terms stemming from the differentiation of 
 third-order differences may be rewritten as
\beas
     && (\Lambda_{i-2}^T-\Lambda_{i-1}^T)  \kappa_{i-3/2} ~k^4   
    - 3 (\Lambda_{i-1}^T-\Lambda_{i}^T)    \kappa_{i-1/2}  ~k^4 
    + 3  (\Lambda_{i}^T-\Lambda_{i+1}^T)  \kappa_{i+1/2}  ~k^4
    -    (\Lambda_{i+1}^T-\Lambda_{i+2}^T) \kappa_{i+3/2} ~ k^4 \\
 &&    -(\Lambda_{i-2}^T-\Lambda_{i-1}^T)  \kappa_{i-3/2} ~k^2 \nu_{i-3/2}    
    + 3 (\Lambda_{i-1}^T-\Lambda_{i}^T)    \kappa_{i-1/2}  ~k^2 \nu_{i-1/2} \\  
 &&   - 3  (\Lambda_{i}^T-\Lambda_{i+1}^T)  \kappa_{i+1/2}  ~k^2\nu_{i+1/2}  
    +    (\Lambda_{i+1}^T-\Lambda_{i+2}^T) \kappa_{i+3/2} ~ k^2  \nu_{i+3/2}.  
\eeas

 In this case, the sum of the terms involving $k^4$ is $O(\Delta \xi^4 \Delta \eta)$ and the terms in $k^2$
 may be individually identified as  $O(\Delta \xi^3 \Delta \eta)$ terms. For inconsistency to be observed,
 it is hence necessary that: (a) due to the  $\max$ function in equation (\ref{e:fjstsa}), the actual formulas of the
 involved $\overline{k^4}$ are not the same up to a shift of subscripts; (b) some of the $\overline{k^4}$ are zero.
 (Note that these conditions are expected to appear at a marginal number of cells in particular as
  $\nu$ is very close to zero except in the vicinity of shock waves. Also note that these conditions are not
 even sufficient for dual inconsistency -- if $\overline{k^4}_{i-3/2}= \overline{k^4}_{i+3/2}=0$ and  
 $\overline{k^4}_{i+1/2} \neq0$, $\overline{k^4}_{i+3/2}\neq0$ or crisscross, dual consistency is observed).
 Such conditions for inconsistency thus correspond to local non-differentiability of the JST scheme.
 This may be compared with the results of reference \cite{LiuSan_08} and the local lack of dual consistency
 of the studied schemes where upwinding changes.

%
%
\subsection{Consistency of the discrete adjoint at a penultimate cell w.r.t. a boundary}\label{sec:adj_cons_penult}
In a penultimate cell of indices $(2,j)$, the previous equations are slightly altered as numerical
fluxes $F_{1/2,j}$ and $F_{3/2,j}$  have
specific definitions involving a boundary state, $W_{b_{1/2,j}}$. Actually $W_{b_{1/2,j}}$
depends on the adjacent conservative variables $W_{1,j}$ and the $S_{1/2,j}$ surface vector only;
$W_b$ satisfies all  Dirichlet like relations to be imposed at the boundary.
The numerical flux at the i=1/2 boundary interface is given by (\ref{e:fjst3}).
   The numerical flux at the i=3/2 boundary interface
  is assumed in this subsection to be defined by (\ref{e:fjst2}).
  This formula is satisfactory from a robustness point of view
  but it induces an inconsistent scheme in the last two cells adjacent to a boundary. This issue
  will be fixed by using formulas (\ref{e:fjst2b}) and (\ref{e:fjst2c}) in \S \ref{sec:adj_cons_Fb_Fc}
  \footnote{As the factors multiplying $\overline{k^4}_{3/2} \kappa_{3/2}$ in (\ref{e:fjst2}) (\ref{e:fjst2b})
 and (\ref{e:fjst2c}) are respectively first-, second-, and third-order terms, this is the simplest way to discuss
  dual consistency for these three $F^{JST}_{3/2}$ formulas.}.

\begin {figure}[htbp]
  \begin{center}
	  \includegraphics[width=0.5\linewidth]{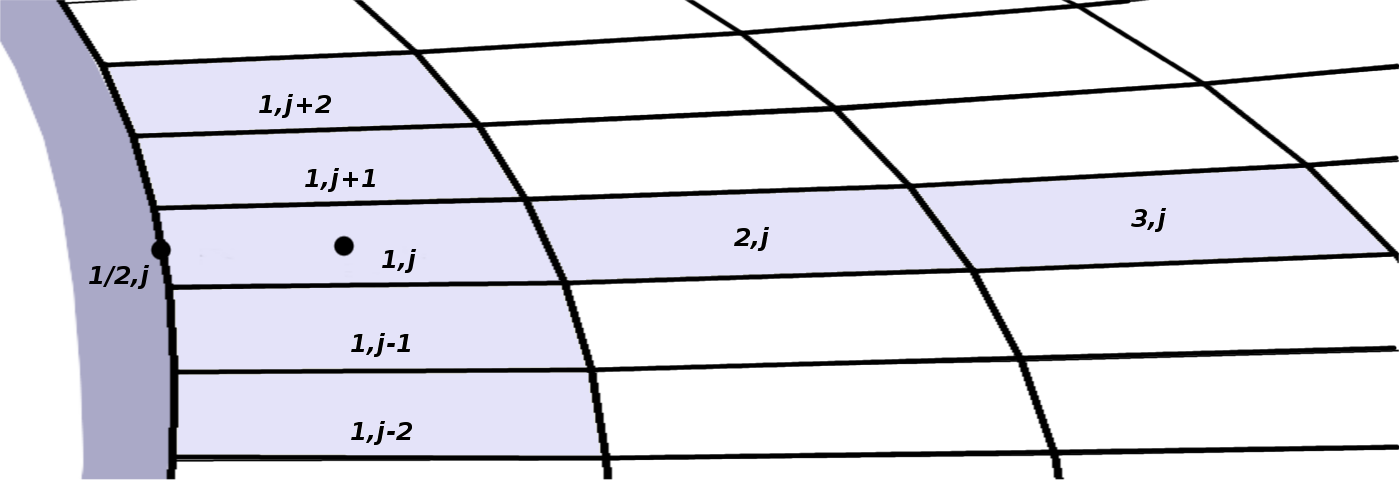}
	\caption{Location of the state variable $W$ and the corresponding boundary variable $W_b$ for the $(1,j)$ cell. Stencil of the JST scheme for this cell.}
  \label{f:meshWWb}
  \end{center}
\end {figure}

 We now consider the discrete adjoint equation in the penultimate cell $(2,j)$ (see Fig.~\ref{f:meshWWb}). Note first that the contributions of
the centered flux are standard for both mesh directions since $F^{JST}_{1/2,j}$ does not depend on $W_{2,j}$.
The terms of lines (\ref{ep:j1})-(\ref{ep:j2}) are the same as in a current cell and do not need to be investigated again.
 Those of lines (\ref{ep:i1})-(\ref{ep:i2}) read
\beq
(\Lambda_1^T-\Lambda_2^T) \frac{\partial F_{3/2}}{\partial W_2}+ (\Lambda_2^T-\Lambda_3^T)\frac{\partial F_{5/2}}{\partial W_2}+ (\Lambda_3^T-\Lambda_4^T) \frac{\partial F_{7/2}}{\partial W_2} =0
\label{ep:repi12}
\eeq

\noindent (dropping $j$ indices) as $F_{1/2}$ does not depend on $W_2$.
  The current face contributions stemming from the differentiation of 
  the sensor and the spectral radius
 have been bounded individually in the previous subsection. We hence
only need to look at the specific contributions of the $3/2$ face that depends on the boundary condition
 through $\nu_1$ in $\nu_{3/2}$. Differentiating the sensor yields
   $$ k^2 (\Lambda_{1}^T-\Lambda_{2}^T)\frac{\partial \nu_{3/2}}{\partial W_{2}}\kappa_{3/2} \Big(-(W_{2}-W_{1})- 1_{[\overline{k^4}_{3/2}>0]} (W_{3}-3 W_{2}+2 W_{1})\Big) $$

\noindent that is $O(\Delta \xi^2 \Delta \eta)$.
When differentiating the spectral radius, the only specific term depending on the boundary condition,
is also the one stemming from  $F^{JST}_{3/2}$,
$$    (\Lambda_{1}^T-\Lambda_{2}^T) \frac{\partial \kappa_{3/2}}{\partial W_{2}} \Big(-k^2 \nu_{3/2} (W_{2}-W_{1})+ \overline{k^4}_{3/2}(W_{3}-3 W_{2}+2 W_{1})\Big). $$

The first part of this expression is  $O(\Delta \xi^3 \Delta \eta)$  (a more precise assertion depends whether $\nu_{3/2}=\nu_1 $ that is
 $O(\Delta \xi)$ or $\nu_{3/2}=\nu_2$ that is $O(\Delta \xi^2)$). The second part of this expression is  $O(\Delta \xi^2 \Delta \eta)$.
  None of them hence prevents dual consistency.
Finally, the terms arising from the differentiation of the differences terms in (\ref{ep:repi12}) read
$$  (\Lambda_{1}^T-\Lambda_{2}^T)  ( -k^2 \nu_{3/2} ~\kappa_{3/2}    - 3 \overline{k^4}_{3/2} \kappa_{3/2}  )
    + (\Lambda_{2}^T-\Lambda_{3}^T)  (  k^2 \nu_{5/2} ~\kappa_{5/2}    + 3 \overline{k^4}_{5/2} \kappa_{5/2}  )  
   + (\Lambda_{3}^T-\Lambda_{4}^T) ( - \kappa_{7/2} ~ \overline{k^4}_{7/2} ).    
$$

They may be rewritten as 
\bea
&&  -3 (\Lambda_{1}^T-\Lambda_{2}^T) \overline{k^4}_{3/2} \kappa_{3/2} + 3 (\Lambda_{2}^T-\Lambda_{3}^T) \overline{k^4}_{5/2} \kappa_{5/2}
   - (\Lambda_{3}^T-\Lambda_{4}^T) \kappa_{7/2} ~ \overline{k^4}_{7/2} \label{e:constpb}\\
&&  - (\Lambda_{1}^T-\Lambda_{2}^T) k^2 \nu_{3/2} ~\kappa_{3/2} + (\Lambda_{2}^T-\Lambda_{3}^T) k^2 \nu_{5/2} ~\kappa_{5/2}. \nonumber
\eea

The two terms  of the last line are at least $O(\Delta \xi^2 \Delta \eta)$ (a more precise assertion depends
 whether $\nu_{3/2}$ is equal to $\nu_1$ or $\nu_2$).
The sum of the terms of the first line is only $O(\Delta \xi \Delta \eta)$
and introduces an inconsistency. (Compared to the current-point equation, due to the specific $F_{1/2,j}$ formula, a
$(\Lambda_{0}^T-\Lambda_{1}^T)\overline{k^4}_{1/2} \kappa_{1/2}$ is missing to have a third-order difference to appear.)

Whether  $F^{JST_b}_{3/2}$ or $F^{JST_c}_{3/2}$ leads to a dual-consistent scheme is discussed in \S~\ref{sec:adj_cons_Fb_Fc}.
How the exact linearization of $F^{JST_a}_{3/2}$ (and also $F^{JST_c}_{3/2}$) can be slightly modified to obtain dual consistency 
is discussed in \S~\ref{sec:adj_cons_BC}.

%
%
\subsection{Consistency of discrete adjoint at a wall boundary cell}
 The discrete counterpart of the output functional (\ref{eq:obj_func}) reads
\beq
 J =  \sum _j   p({\tiny W_{b_{(1/2,j)}}}) S_{(1/2,j)}.\overline{d},
\label{e:funj}
\eeq
\noindent where $(1/2,j)$ are the wall indices. We stress that the static pressure $p$ is evaluated for the boundary
 state $W_b $ that also appears in the boundary flux (\ref{e:fjst3}).
The discrete adjoint equation in a cell adjacent to a wall-boundary
 (see cell $(1,j)$ in Fig.~\ref{f:meshWWb}) then reads
\beq
  \sum _k  \Lambda_k^T \frac{\partial R_k}{\partial W_{(1,j)}} = -\frac{\partial J}{\partial W_{b_{(1/2,j)}}} \frac{\partial W_{b_{(1/2,j)}}}{\partial W_{(1,j)}}. 
\label{e:dab1} 
\eeq

Due to the stencil of the JST scheme in (\ref{e:fjst1}), (\ref{e:fjst2}) and (\ref{e:fjst3}), the $k$ subscript varies in $ \{ (1,j-2),(1,j-1),(1,j),(1,j+1),(1,j+2),(2,j),(3,j) \}$ as illustrated in Fig.~\ref{f:meshWWb}.
The expanded expression of (\ref{e:dab1}) reads
\bea
     (\Lambda_{(1,j-2)}^T-\Lambda_{(1,j-1)}^T) \frac{\partial F_{1,j-3/2}}{\partial W_{(1,j)}} 
    +(\Lambda_{(1,j-1)}^T-\Lambda_{(1,j  )}^T) \frac{\partial F_{1,j-1/2}}{\partial W_{(1,j)}} && \label{e:fin1} \\
    +(\Lambda_{(1,j )}^T-\Lambda_{(1,j+1)}^T)  \frac{\partial F_{1,j+1/2}}{\partial W_{(1,j)}} 
    +(\Lambda_{(1,j+1)}^T-\Lambda_{(1,j+2)}^T) \frac{\partial F_{1,j+3/2}}{\partial W_{(1,j)}} && \label{e:fin2} \\ 
    +(\Lambda_{( 2,j)}^T-\Lambda_{(3,j)}^T)    \frac{\partial F_{5/2,j}}{\partial W_{(1,j)}} 
    +(\Lambda_{(1,j)}^T-\Lambda_{(2,j)}^T)    \frac{\partial F_{3/2,j}}{\partial W_{(1,j)}} && \label{e:fin3} \\
     - \Lambda_{(1,j)}^T  \frac{\partial F_{1/2,j}}{\partial W_{(j,1)}} 
  &=& -\frac{\partial J}{\partial W_{b_{(i,1/2)}}} \frac{\partial W_{b_{(i,1/2)}}}{\partial W_{(1,j)}}.
\label{e:fin4} 
\eea
\noindent As noted by Nadarajah et al. \cite{NadJam_00}, this equation mixes O($\Delta \eta$) terms (in (\ref{e:fin4}) only) and higher order terms (in all three
   first lines), and as the mesh is refined the continuous adjoint wall boundary condition is recovered from the first-order terms.
   Keeping only the $O(\Delta \eta)$ terms, the previous equations yield
$$ - \Lambda_{(i ,1)}^T  \left( \begin{array}{c}
    0 \\
    Sx_{(1/2,j)}  \\
   Sy_{(1/2,j)}  \\
    0  \\ 
\end{array}
\right)
\frac{\partial p_{b_{(1/2,j)}}}{\partial W_{b_{(1/2,j)}}}   \frac{\partial W_{b_{(1/2,j)}}}{\partial W_{(1,j)}}  
   \simeq -\frac{\partial ( (-\overline{S_{(1/2,j)}}\cdot\overline{d}) p_{b_{(1/2,j)}}) }{\partial W_{b_{(1/2,j)}}} \frac{\partial W_{b_{(1/2,j)}}}{\partial W_{(1,j)}},
 $$ 
 
\noindent which may be simplified into
\beq   
   \Lambda^2_{(i ,1)} Sx_{(1/2,j)} + \Lambda^3_{(i ,1)} Sy_{(1/2,j)} \simeq  -  \overline{S_{(1/2,j)}}\cdot\overline{d}. 
\label{e:dab3} 
\eeq 

Note that it involves metric terms at the wall and the adjoint at adjacent cells
 and also that it has been derived by equalizing lower-order (i.e., first-order) terms in space. 
 How accurately equation  (\ref{e:dab3}) is satisfied is discussed in sections~\ref{sec:xp_naca_sup} to \ref{sec:xp_naca_sub}. 
%
%
%
%
 %
 %
\subsection{Alternative $F_{3/2}$ formulas ~~$F^{JST_b}_{3/2}$ and $F^{JST_c}_{3/2}$ ~~and adjoint consistency}\label{sec:adj_cons_Fb_Fc}
The discrete  adjoint of the cell-centered JST scheme (\ref{e:fjst1}) to (\ref{e:fjst2}) 
is thus adjoint-consistent except at penultimate cells w.r.t. a boundary. Possible changes in $F_{3/2}$
to get adjoint consistency are discussed here in relation with classical publications \cite{SwaTur_87,SwaTur_92}
that  presented specific formulas for JST AD close to the boundaries of a structured mesh for steady state flow
 simulations.

 A satisfactory  modification of $F^{JST}_{3/2}$ for dual consistency, would substitute
$$
F^{JST_b}_{3/2} = \tfrac{1}{2} ({\bf F}(W_1)+{\bf F}(W_{2}))\cdot S_{3/2} -k^2 \nu_{3/2} \kappa_{3/2} (W_{2}-W_1) + \overline{k^4}_{3/2} \kappa_{3/2} (W_{3}-2 W_{2}+ W_{1})
$$
to $F^{JST_a}_{3/2}$. To check this, first note that $(W_{3}-2 W_{2}+ W_{1})$ in (\ref{e:fjst2b})
 is a second-order difference whereas $(W_{3}-3 W_{2}+ 2W_{1})$ in (\ref{e:fjst2})
 is a first-order term. Consequently, the  terms of  $(\partial F_{3/2}/\partial W_1)$ and
 $(\partial F_{3/2}/\partial W_2)$ stemming from the differentiation of the spectral radius or the sensor
 may not be of lower order than their counterpart in the calculations of \S 4.1 to \S 4.3. When differentiating the
 $(W_{3}-2 W_{2}+ W_{1})$ factor, the contribution in  $(\partial F_{3/2}/\partial W_3)$ is unchanged;
 the contribution in $(\partial F_{3/2}/\partial W_2)$ precisely cures the issue that appears in line
 (\ref{e:constpb});  the contribution in  $(\partial F_{3/2}/\partial W_1)$ is a $O(\Delta \xi \Delta \eta)$
 term in line (\ref{e:fin3}) dominated by the $O(\Delta \eta)$ terms of line  (\ref{e:fin4}).

 Moreover, this $(W_{3}-2 W_{2}+ W_{1})$ factor is one of the adapted discretizations of the third-order difference
 that is recommended close to boundaries in \cite[B1 formula]{SwaTur_87} and \cite{SwaTur_92}. Compared to the corresponding factor in (\ref{e:fjst2}), it doesn't introduce inconsistent terms in
 the discretization of the Euler equations in the last two rows of cells close to a physical boundary. 

 The numerical flux (\ref{e:fjst2b}) has been implemented and proved to be more accurate than (\ref{e:fjst2}). For example, for our 2-D
 subsonic test case \clmy{(\S5.4)}, using $F^{JST_b}_{3/2}$, the spurious drag values
 calculated on a hierarchy of embedded meshes are  30.1, 
 8.7, 3.2 and 1.8 drag counts on the $128^2$, $256^2$, $512^2$ and $1024^2$-cell meshes,
 to be compared to 37.9, 11.3, 4.0 and 2.1 drag counts using $F^{JST_a}_{3/2}$.
 Besides, other accuracy indicators like total enthalpy losses also support the superiority of $F^{JST_b}_{3/2}$ over $F^{JST_a}_{3/2}$.\\
  This discretization, unfortunately,
 significantly reduces the robustness of the JST scheme on very fine meshes
 and, when selecting it,  we could not get converged flows for our transonic (\S 5.3) and subsonic (\S 5.4)
  2-D test cases and our finest meshes (with $2048^2$ and $4096^2$ cells). As there is no artificial dissipation in $ F_{1/2}$,
 the $ \overline{k^4}_{3/2} \kappa_{3/2} (W_{3}-2 W_{2}+ W_{1})$ term
 of this $ F_{3/2}$ flux introduces an anti-dissipative term in the equivalent equation of the cells adjacent
 to the boundary. Most probably, this is the reason of this loss of robustness.
 \\
    
 Once again, formulas involving AD at boundaries are not taken into consideration here in order
 to keep the same discretizations for internal and external flows. On the contrary, regarding the second to last faces,
 formulas recommended by Swanson and Turkel in \cite{SwaTur_92} are recalled and studied
 with respect to the inconsistency issue that appeared in \S 4.2.
 $W_g$ being a mirror state of $W_1$ w.r.t. the boundary, the final formula they recommend reads
$$
F^{JST_c}_{3/2} = \tfrac{1}{2} ({\bf F}(W_1)+{\bf F}(W_{2}))\cdot S_{3/2} -k^2 \nu_{3/2} \kappa_{3/2} (W_{2}-W_1) 
  + \overline{k^4}_{3/2} \kappa_{3/2} (W_{3}-3 W_{2}+ 3W_{1}-W_g).
$$

 It is simply a standard discretization leading to a standard fourth-order dissipative term in the penultimate
 cells w.r.t. to the boundary \cite[Eq.~(2.17)]{SwaTur_92}. This formula
  is one of those available in our code \cite{CamHeiPlo_13} with $W_g = 2 Wb - W_1 $.
  Let us note first that  $(W_{3}-3 W_{2}+3 W_{1}-W_g)$ is a third-order term in space as for formula $F^{JST_b}_{3/2}$.
 The same arguments as above are relevant and the adjoint (in)consistency discussion in cell $(2,j)$ is again reduced to the 
 examination of the terms stemming from the differentiation of the $(W_{3}-3 W_{2}+ 3W_{1}- W_g)$ factor
 in $(\partial F_{3/2}/\partial W_3)$,
 $(\partial F_{3/2}/\partial W_2)$,  $(\partial F_{3/2}/\partial W_1)$ compared to those of the default discretization.
 As in the discussion just above, $(\partial F_{3/2}/\partial W_3)$ is unchanged and
  $(\partial F_{3/2}/\partial W_1)$ brings $O(\Delta \xi \Delta \eta)$ terms in
 line (\ref{e:fin3}) that are dominated by the $O(\Delta \eta)$ terms of line  (\ref{e:fin4}). Finally
 $(\partial F_{3/2}/\partial W_2)$ contribution in (\ref{e:constpb}) is a
  $-3 (\Lambda_{1}^T-\Lambda_{2}^T) \overline{k^4}_{3/2} \kappa_{3/2}$ term just as in the default
  $F^{JST_a}_{3/2}$ discretization and 
 inconsistency is also noticed for the penultimate cells w.r.t. a boundary. %

 %
%
%
%
%
%
\subsection{Modified linearization of $F^{JST_a}_{3/2}$ and $F^{JST_c}_{3/2}$ for adjoint consistency}\label{sec:adj_cons_BC}
Adjoint consistency may be recovered with a minor and local approximation in the Jacobian of the scheme.
Following the analysis of \S~\ref{sec:adj_cons_penult},
 the needed change in the differentiation of the $F^{JST_a}_{3/2,j}$ flux is
the replacement of
$$ -3 (\Lambda_{1}^T-\Lambda_{2}^T) \overline{k^4}_{3/2} \kappa_{3/2}~~~~ \textrm{ by }~~~~  -2 (\Lambda_{1}^T-\Lambda_{2}^T) \overline{k^4}_{3/2} \kappa_{3/2} ~~~~  \textrm{in}~~~~
 (\partial F_{3/2,j}/ \partial W_2), $$

\noindent when differentiating the third-order difference. The same modification in the linearization of $F^{JST_c}_{3/2,j}$ is required to ensure
dual consistency as shown in \S~\ref{sec:adj_cons_Fb_Fc}.

Whereas it was simpler to start the dual consistency analysis for $F^{JST_a}_{3/2}$ and 
then adapt it for $F^{JST_b}_{3/2}$ and $F^{JST_c}_{3/2}$, only  $F^{JST_c}_{3/2}$ is considered from now on.
The robustness of the numerical simulations and the final level of the steady state residuals on very fine meshes
are almost as satisfactory as with  $F^{JST_a}_{3/2}$. Besides, $F^{JST_c}_{3/2}$ does not
introduce any inconsistency or reduction
of the order of accuracy for the cells adjacent to the boundaries and their first neighbors. Concerning the adjoint fields,
the modified linearization is used when discussing adjoint consistency; the exact linearization and possibly also
 the modified one are used in the other sections.

\begin {figure}[htbp]
  \begin{center}
	  \includegraphics[width=0.9\linewidth]{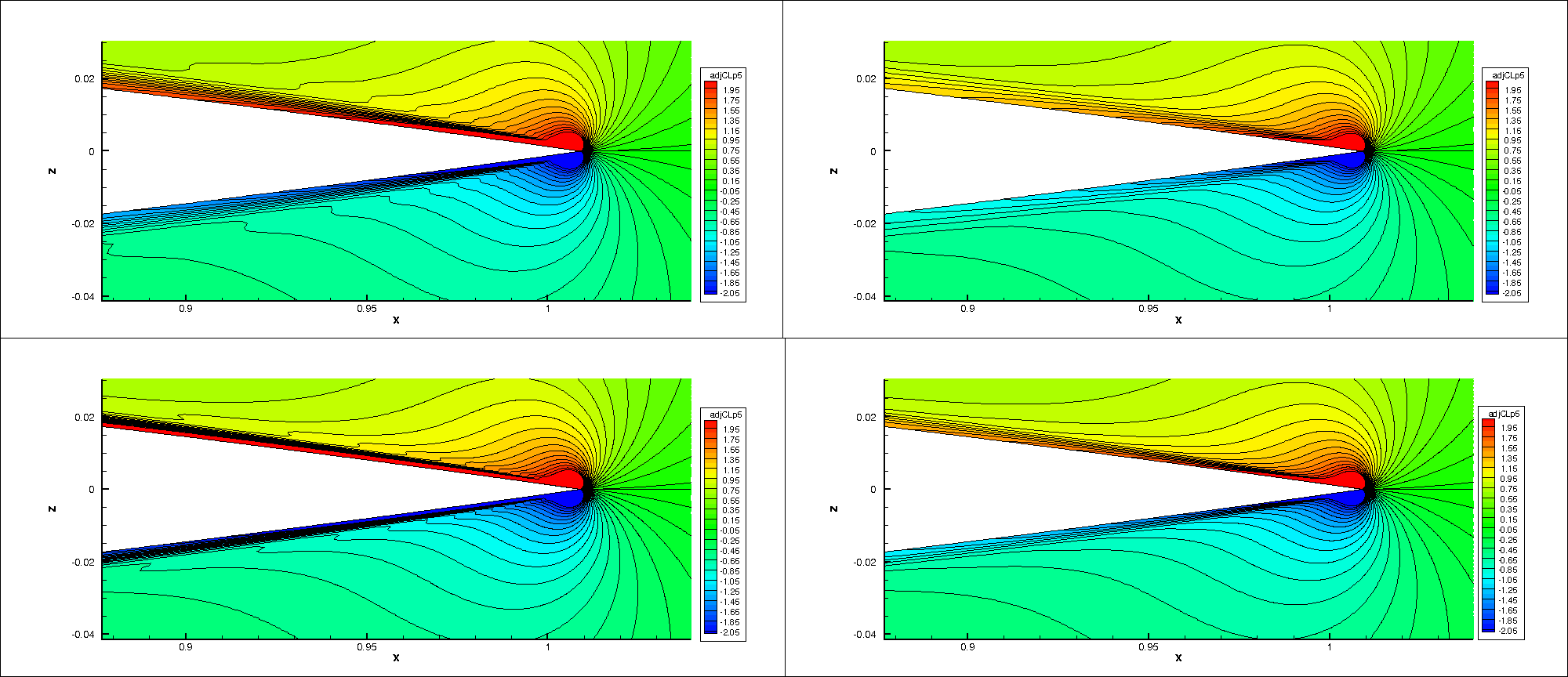}
	  \caption{($M_\infty=0.4$, $\alpha=5^o$) Energy component of the lift adjoint.
      Top 2049$\times$2049 mesh, bottom 4097$\times$4097 mesh
      --  Left, inconsistent discrete adjoint, right consistent discrete adjoint.}
  \label{f:consist}
  \end{center}
\end {figure}

 These linearizations are compared by using the energy component of the lift adjoint in Fig.~\ref{f:consist}.
  The dual inconsistency is illustrated by spurious oscillations
in the iso-lines in the vicinity of the two problematic cells. The amplitude of these oscillations is damped away from the wall and is reduced when using finer meshes. As expected, the oscillations vanish when the consistent adjoint linearization is used.

%
%

%
\section{Numerical experiments}\label{sec:num_xp}

We \cltw{present here the} discrete direct and adjoint solutions to inviscid flows around the NACA0012 airfoil in different regimes.
Steady-state solutions are computed  with the $elsA$ code \cite{CamHeiPlo_13} on a series of six structured meshes 
ranging from 129$\times$129 to $4097\times4097$ nodes
used and described in \cite{VasJam_10}.
The far field boundary is set at about $150$ chords. The computations have been converged
with a second-order accurate FV method using the JST scheme (\ref{e:fjst1})-(\ref{e:fjst3})
and (\ref{e:fjst2c}) with
$k^2=0.5$ and $k^4=0.032$, except for the subsonic flow
for which we used $k^2=0$ and $k^4=0.032$.
The derivatives of the lift and the drag w.r.t. the flow field variables in the discrete adjoint
equation (\ref{eq:discr_adj_equ})
were  calculated by using the post-processing tool from \cite{Des_03}
 before the adjoint equations were solved by the adjoint module of $elsA$ \clmy{\cite{JPRenDum_15,JadBloMar_20}} for the two functions
 and the complete set of meshes. 
 
\subsection{Main results}
\begin{sry}
The dual consistency of the linearization of the JST scheme  (\ref{e:fjst1}) to (\ref{e:fjst3}) and \clmy{(\ref{e:fjst2c})}, involving the minor
  correction of \clmy{$dF^{JST_c}_{3/2}$} introduced in \S~\ref{sec:adj_cons_penult},
  is confirmed by discretizing the
  continuous adjoint equations for consistent and inconsistent discrete adjoint fields on a hierarchy of grids
   and computing the corresponding residuals. The residuals highlight two types of zones were their norm keeps large values as the mesh is refined: (i) zones of discontinuity of the adjoint field;
  (ii) zones of numerical divergence of the adjoint field. In both types of location, a Taylor expansion cannot be applied and dual consistency is not discussed.

 Among the zones where numerical divergence is observed and mathematical divergence is suspected (see below)
   we focus on the vicinity of the wall and the stagnation streamline:
  in the framework of the physical source terms approach \cite{GilPie_97}, it appears that the fourth
  source term $\delta R^4$ in (\ref{e:deltaR}) is the one responsible for the numerical divergence at the wall. The way it disturbs the flow
  is described. The numerical divergence of the adjoint components in the vicinity of the wall and stagnation streamline appears to
  be numerically more complex than the algebraic law derived in  \cite{GilPie_97} under simplifying assumptions.

 Completing the dual consistency discussions, complementary analysis is provided in the case of the supersonic flow providing insight on
 the structure of the theoretical continuous adjoint field.
 Finally, the jump relations on the adjoint derivatives derived in \S~\ref{sec:adj_RH} are numerically checked on the finest grid.

\end{sry}
Let us stress, first, that an adjoint discontinuity is typically observed at
the trailing edge of a supersonic flow.
 The zones where numerical divergence of lift and drag inviscid adjoints is observed, close to the airfoil,
are also recalled in the
case of a lifting airfoil (avoiding marginal cases like symmetric airfoil and flow):\\
 -- for a subsonic flow, $\Lambda_{CD_p}$ does not exhibit numerical divergence  whereas $\Lambda_{CL_p}$ does at
 the stagnation streamline and at the wall;\\
 -- for a transonic flow for which a shock wave foot is located upstream the trailing edge, $\Lambda_{CD_p}$ and
 $\Lambda_{CL_p}$ exhibit numerical divergence at the stagnation streamline and at the wall;\\
 -- for a transonic flow for which all shock wave feet are located at the trailing edge, $\Lambda_{CD_p}$ and
$\Lambda_{CL_p}$ do not exhibit numerical divergence;\\
 -- for a supersonic flow with a detached shock wave and shocks based on the trailing edge, $\Lambda_{CD_p}$ and
$\Lambda_{CL_p}$ do not exhibit numerical divergence.
%
%
\subsection{Supersonic regime}\label{sec:xp_naca_sup}

We first consider a supersonic flow around the NACA0012 airfoil with a free-stream Mach number $M_\infty=1.5$ and an angle of attack $\alpha=1^o$.   

\begin {figure}[htbp]
  \begin{center}
	  \includegraphics[width=0.45\linewidth]{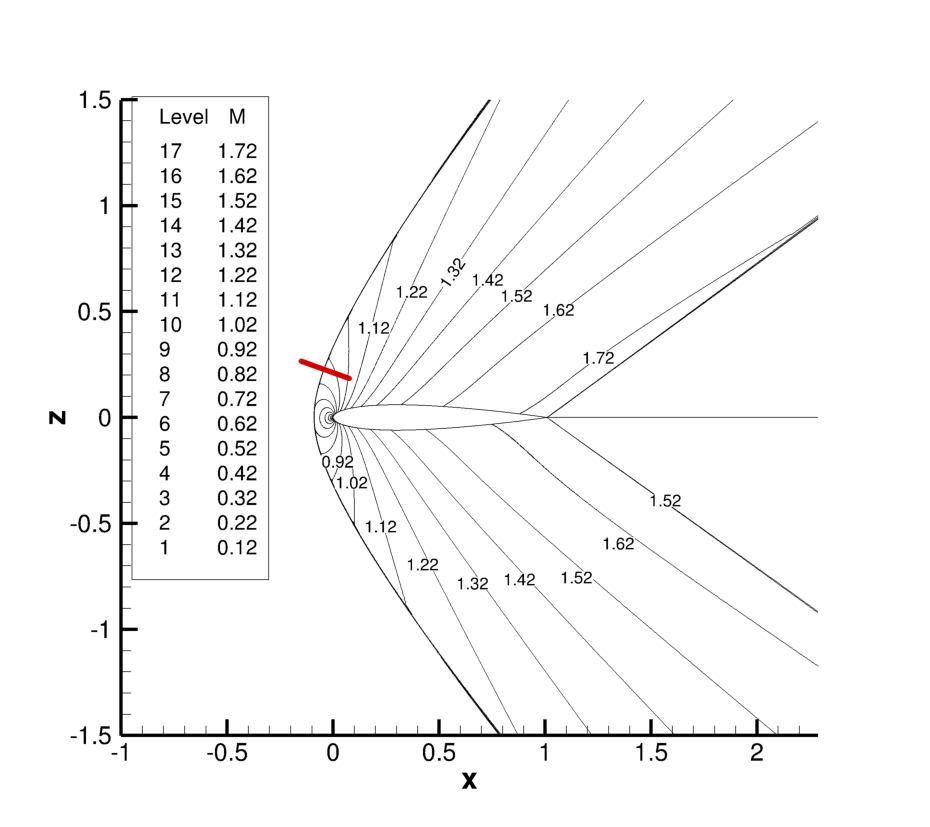}
	  \includegraphics[width=0.45\linewidth]{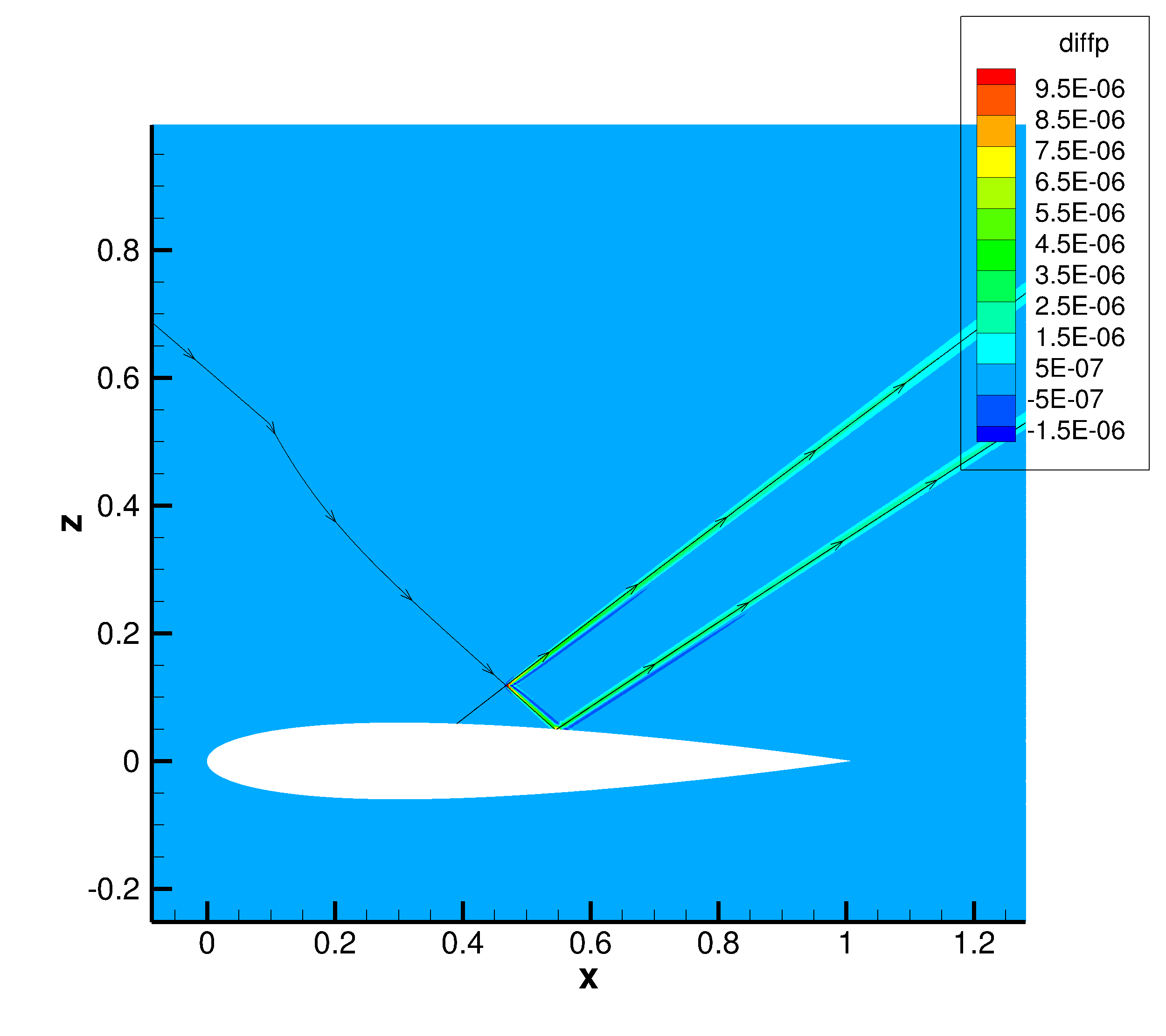}
	\caption{($M_\infty=1.50$, $\alpha=1^o$, 2049$\times$2049 mesh) Left: contours of Mach number. Right: Change in pressure due to $\delta R^1$ source term  ($\epsilon$=$6.10^{-6}$) at point $(x=0.4702,y=0.1184)$.}
  \label{f:isoMach_M150}
  \end{center}
\end {figure}

The direct solution is displayed in Fig.~\ref{f:isoMach_M150}. The flow is supersonic and constant
 up to a detached shock wave. Downstream the shock wave, the flow is subsonic in a small bubble close to the airfoil leading edge
 and supersonic elsewhere. It accelerates along the airfoil up to a fishtail shock wave from the trailing edge.
 Downstream this second shock wave, the flow is still supersonic with a Mach number close to the upwind far field
 Mach number.

%
%
\subsubsection{Lift and drag adjoint solutions}\label{sec:heur_consist}  
\begin {figure}[htbp]
  \begin{center}
	  \includegraphics[width=0.9\linewidth]{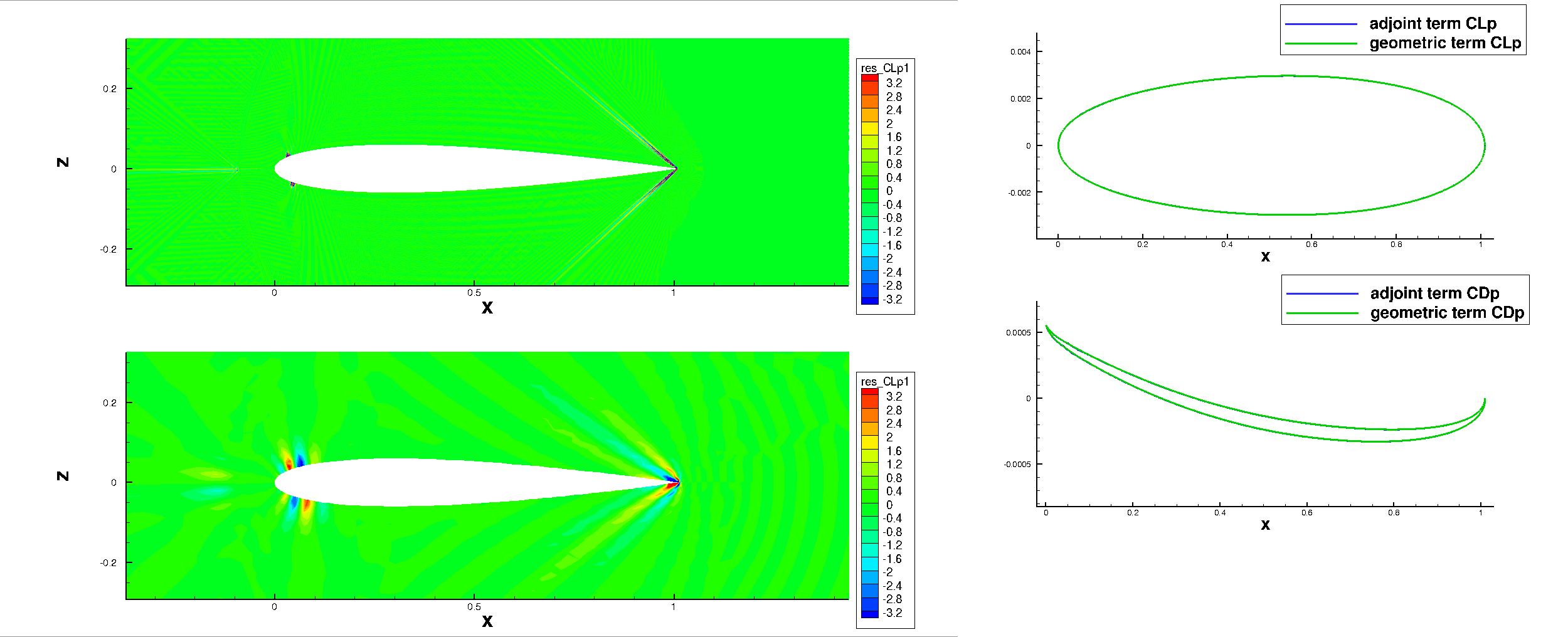}
	\caption{($M_\infty=1.5$, $\alpha=1^o$) Left: Residual of the continuous equation (first component)
  evaluated with the lift discrete adjoint fields  (bottom: 129$\times$129 mesh, top: 2049$\times$2049 mesh), 
 right: geometric and adjoint terms of equation (\ref{e:dab3}) for lift and drag (2049$\times$2049 mesh)}
  \label{f:contdisc_M150}
  \end{center}
\end {figure}

 Figure \ref{f:g_M150} presents the contours of the four components of the adjoint field.
The gradient of the discrete adjoint is then evaluated at cell centers using a Green formula and  the residual in cell $(i,j)$ of the continuous adjoint equation,
\beq
 res_{ij} = -A^T_{ij} \left(\frac{\partial \Lambda}{\partial x}\right)_{ij} - B^T_{ij} \left(\frac{\partial \Lambda}{\partial y}\right)_{ij}, 
\label{e:resadj}
\eeq

\noindent is evaluated for both functions. In the above equation, $A_{ij}$ and  $B_{ij}$ are the Jacobians evaluated with cell-center values. For the sake of validation, both terms of the residual 
and their sum are plotted separately for the lift, on the 2049$\times$2049 mesh in Fig.~\ref{f:bands} (left) and then added 
 in Fig. \ref{f:contdisc_M150} (top left). The cancellation of almost opposite terms is obvious.
 The percentage of cells with $|res|$ below a threshold is \clon{increasing} as the mesh is refined.
 \\
 Nevertheless significant (and increasing)  $|res|$ values are observed in Fig.~\ref{f:contdisc_M150} on the finest mesh at the trailing edge and along
 the Mach lines passing through the trailing edge.
 \clon{ In the vicinity of the trailing edge, the exact adjoint is zero downstream the
   backward characteristics emanating from the trailing edge because  no perturbation downstream those
   lines can affect the pressure on the airfoil profile and, thus, the lift and drag. Upstream those lines, perturbations do affect lift and drag, and thus
   the adjoints take non-zero values. The corresponding simple discontinuity is more and more accurately captured by any consistent discrete adjoint but
   the discrete $res_{ij}$, that is independant of the numerical scheme $R$, exhibits high values along
   the characteristics passing through the trailing edge}.
 \\
 The dual consistency improvement is also checked quantitatively calculating the mean norm of the $res$ components 
 inside a fixed region close to the profile (interior of the 
  $ ((x-0.5)/0.55)^2+(z/0.1)^2 = 1$ ellipse and $0.1c$ upwind the two discontinuous lines issued from the trailing edge
 or their contribution appears to be overwhelming). These means  are then  summed using 
 ponderations by the far field velocity to account for the dimension of the components and the resulting
 all-components residual, denoted
 $RES$, appears to be decreased
 by a few percent by the modification of the $F^{JST_c}_{3/2}$ differential for adjoint consistency. More precisely
 the decrease of $RES$ goes form 1.5\% (coarsest mesh)  to 3.3\% (finest mesh) for the lift whereas it goes from
 1.4\% (coarsest mesh)  to 3.8\% (finest mesh) for the drag.
Concerning far field boundary conditions, the normal Mach number appears to be supersonic at the intersection of the characteristic geometrical
 strips (see  Fig.~\ref{f:bands}) and the far field boundary. No continuous adjoint boundary conditions is hence to be applied there
 and no check of discrete versus continuous adjoint is required. Equation (\ref{e:dab3}) is the discrete counterpart of the continuous boundary
 condition at the wall in (\ref{e:adjcwal}) and its two terms are plotted in the right part of Fig.~\ref{f:contdisc_M150}. They appear to be  superimposed, so (\ref{e:dab3})
 is actually well satisfied.

%
%
\subsubsection{Adjoint fields upstream of the shock}
 The flow is supersonic and constant upstream  the detached shock wave. The Jacobians $A$ and $B$
  in the continuous adjoint equation (\ref{e:adjcdom}) are hence constant matrices in this part of the fluid domain and
  specific analysis
  of the continuous adjoint solution is possible in this region. These calculations,
partly presented in \cite{TodVonBou_16}, are here
 further developed and assessed. 
In a supersonic regime, the direct and adjoint equations both exhibit specific directions of 
 propagation corresponding to simple waves solutions to (\ref{e:adjcdom}):
$$ \psi(x,y) = \phi (x \sin(\zeta) - y \cos(\zeta) ) \lambda_0, $$

\noindent where $\zeta$ is the angle made by the direction of propagation with the x-axis, $\lambda_0$ is a vector
 representing the convected information and $\phi$ is a scalar function.
 Injecting this expression into (\ref{e:adjcdom}) yields
$$ \phi' (x \sin(\zeta) - y \cos(\zeta)) \times (\sin(\zeta) A^T - \cos(\zeta) B^T)\lambda_0 = 0.$$

This equation admits a non-trivial solution $\lambda_0$ if and only if 
$ \det (\sin(\zeta) A^T - \cos(\zeta) B^T) = 0 $ and simple algebraic arguments \cite{TodVonBou_16} allow to prove that
\beas
 \zeta &=& \alpha ~~~~ \textrm{or} ~~~~ \alpha + \pi \\
 \zeta &=& \alpha - \mu  ~~~~ \textrm{or} ~~~~  \zeta = \alpha - \mu + \pi \\
 \zeta &=& \alpha + \mu  ~~~~ \textrm{or}  ~~~~ \zeta = \alpha + \mu + \pi, 
\eeas

\noindent  where $\alpha$ denotes the angle of attack (so $u_x = \|\overline{U}\| \cos(\alpha) $ and
  $u_y = \|\overline{U}\| \sin(\alpha)$) and $\mu = \arcsin (1/M_\infty)$ is the Mach angle. Hence, the information propagates along the three privileged directions
 $\zeta = \alpha$, $\zeta = \alpha - \mu$ et $\zeta = \alpha + \mu$. Evidently, the domains of influence and dependency
 are inverse one to another for the state and adjoint variables.

The $\lambda_0$ eigenvectors of interest are those of $ \sin(\zeta) A^T - \cos(\zeta) B^T $. These are also the left eigenvectors that
appear in the more classical diagonalization of the inviscid flux Jacobian. Following \cite{Hir_07}, the (left)
eigenvectors associated with
  $ \zeta = \alpha \pm \mu  $ (null eigenvalues $u_x \sin(\zeta) - u_y \cos(\zeta) \mp c$) read
$$
\lambda_0^{\alpha-\mu} = \begin{pmatrix}
   \frac{c}{\rho} \left( \frac{-u_xn_x -u_yn_y}{c} + \frac{(\gamma - 1)}{2} M^2 \right)\\
   \frac{1}{\rho} \left( n_x - (\gamma - 1)\frac{u_x}{c} \right)\\
   \frac{1}{\rho} \left( n_y - (\gamma - 1)\frac{u_y}{c} \right) \\
   \frac{\gamma - 1}{\rho c}
\end{pmatrix}, \quad
\lambda_0^{\alpha+\mu} = \begin{pmatrix}
   \frac{c}{\rho} \left( \frac{u_xn_x + u_yn_y}{c} + \frac{(\gamma - 1)}{2}M^2 \right)\\
   -\frac{1}{\rho} \left( n_x + (\gamma - 1)\frac{u_x}{c} \right)\\
   -\frac{1}{\rho} \left( n_y + (\gamma - 1)\frac{u_y}{c} \right)\\
   \frac{\gamma - 1}{\rho c}
\end{pmatrix}, 
 $$

\noindent for a general direction $(n_x,n_y)^T$, which may be replaced by $(\sin(\zeta),-\cos(\zeta))^T$ to recover the solution as a function of the simple wave angles. 
 Checking that the simple waves solution satisfy relation (\ref{e:lam14}) uses the nullity of the eigenvalue ($u_xn_x +u_yn_y+c=0$ for
 the first vector and $u_xn_x +u_yn_y-c=0$ for the second eigenvector).

The dimension of the eigenspace associated with $ \zeta = \alpha $ is two but the classically exhibited left eigenvectors \cite{Hir_07} 
 do not satisfy the Giles and  Pierce relation for pressure-based outputs (\ref{e:lam14}).
 The vectors of this 2-D space that satisfy this relation is the 1D space generated by 
$$
\lambda_0^{\alpha} =  \begin{pmatrix}
   -1 - \frac{(\gamma - 1)}{2} M^2\\
   \frac{(\gamma - 1)u_x}{c^2} + \frac{2n_y}{n_yu_x-n_xu_y}\\
   \frac{(\gamma - 1)u_y}{c^2} - \frac{2n_x}{n_yu_x-n_xu_y}\\
   -\frac{\gamma - 1}{c^2}.
\end{pmatrix}.
$$

\begin {figure}[htbp]
  \begin{center}    
	  \includegraphics[width=0.45\linewidth]{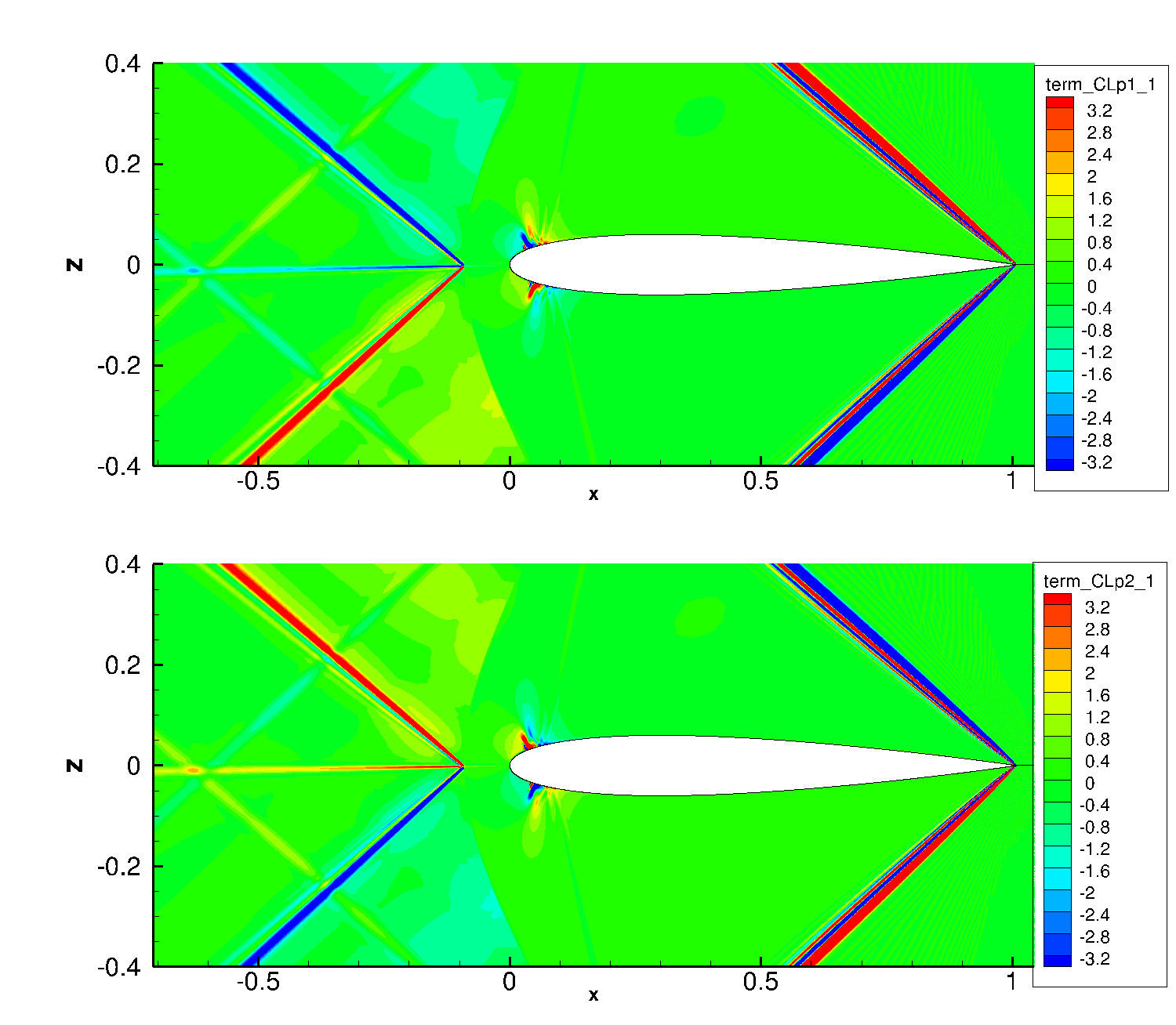}
	  \includegraphics[width=0.45\linewidth]{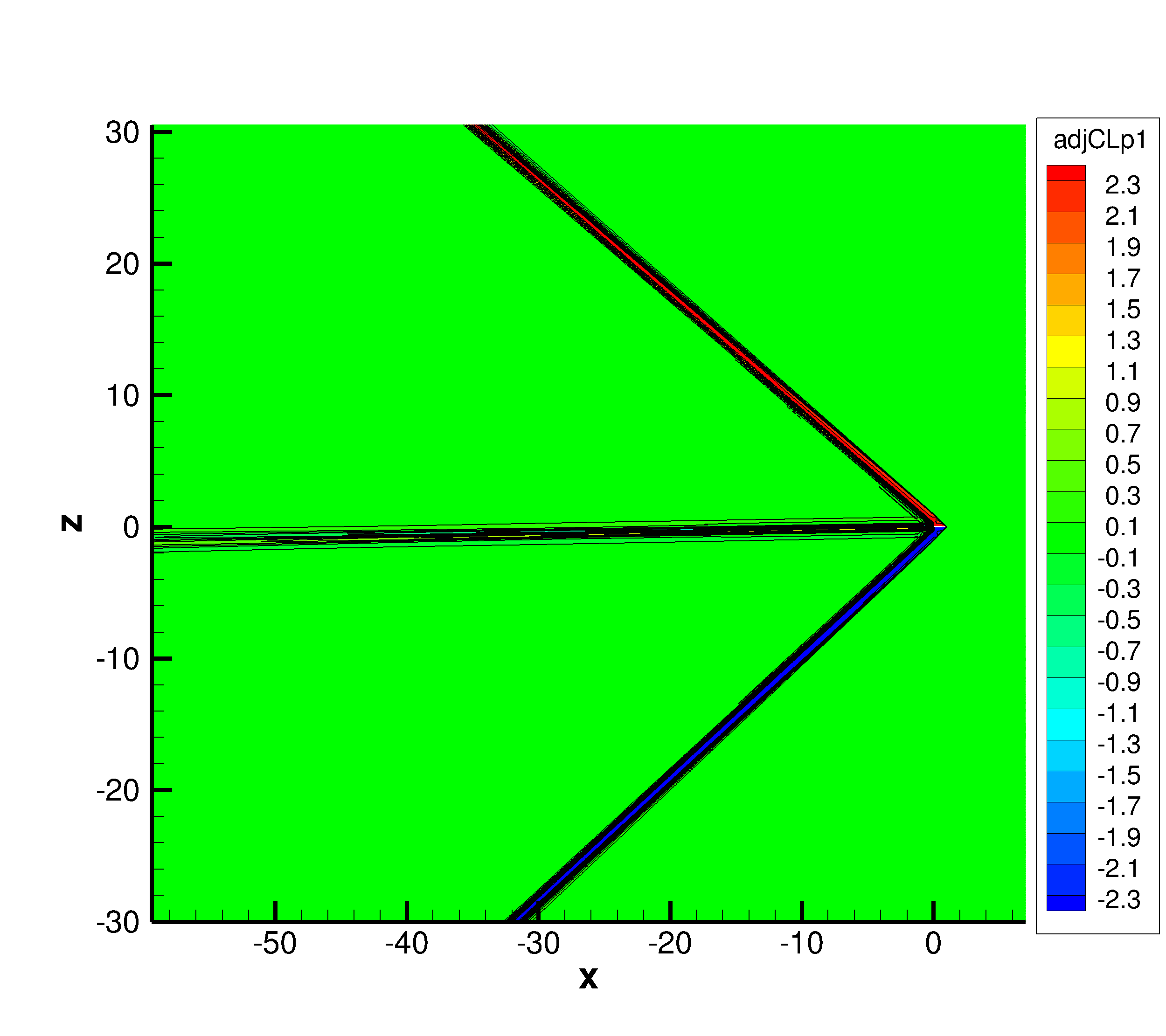}
	\caption{($M_\infty=1.5$, $\alpha=1^o$) Top left: $-A^T (\partial \Lambda_{CL_p} / \partial x)$ discretized with discrete adjoint field (first component). Bottom left:  $-B^T (\partial \Lambda_{CL_p} / \partial y)$ discretized with discrete adjoint field (idem) ($2049\times 2049$ mesh), right: Far field view of first component of $CL_p$ adjoint ($4097\times 4097$ mesh).}
  \label{f:bands}
  \end{center}
\end{figure}

Figure~\ref{f:bands} (right) highlights the directions of propagation in the first component discrete adjoint field.
 The ratio of the adjoint components  inside all three 
 geometrical strips is found equal to the ratio of the corresponding components of the associated $\lambda_0$ vector. The final demonstration of the consistency
 between the discrete adjoint fields and the theoretical form of the continuous adjoint solution upwind the shock wave, is obtained by performing a line extraction of the three terms in
\begin{equation} 
  \psi(x,y) =  \phi_{\alpha} (x \sin(\alpha) - y \cos(\alpha) ) \lambda^{\alpha}_0 + \sum_{\nu=\pm\mu}\phi_{\alpha+\nu} (x \sin(\alpha+\nu) - y \cos(\alpha+\nu) ) \lambda^{\alpha+\nu}_0,
  \label{e:exadjsup}
\end{equation}

\noindent and propagating these values in the complete zone
  upwind the detached  shock wave. More precisely, the values of these terms are extracted from a fine discrete adjoint field at a distance $2.3c$ from the leading edge 
 (where the three bands are separated). They are then interpolated everywhere upwind the shock wave according
 to the local values of $x\sin(\zeta)-y\cos(\zeta)$, with $\zeta\in\{\alpha,\alpha\pm\mu\}$. For the finer meshes, the resulting field appears to be identical to the  actual discrete adjoint field, upstream of the shock wave, where the bands are superimposed (results not shown here).  
 
\subsubsection{Residual perturbation analysis} \label{sec:dRsup}
 When the residual perturbation is located in a supersonic zone, its influence on the flow close to the source 
 appears to be restricted  to the downstream part of the  streamline and of the two
 characteristic curves passing by the \cltw{location of the source}. Moreover, assuming a constant flow $W$,
 by the same arguments as in previous section, its
 perturbation $\delta W$ can be expressed as
\bea
  \delta W(x,y) =  \Phi_{\alpha} (x \sin(\alpha) - y \cos(\alpha) ) r^{\alpha} + \sum_{\nu=\pm\mu}  \Phi_{\alpha+\nu} (x \sin(\alpha+\nu) - y \cos(\alpha+\nu) ) r^{\alpha+\nu},  
\label{e:3bdir}
\eea

\noindent where  $\psi_\zeta$, with $\zeta\in\{\alpha,\alpha\pm\mu\}$, stands for a small scalar perturbation and each $r^\zeta$ vector
is a right eigenvector of the Jacobian $\sin(\zeta) A - \cos(\zeta) B$.
The consistency  of the actual numerical $\delta W$
with this analytical model can be simply assessed,  upwind the detached shock,
 for $\delta R^1$ and $\delta R^2$ in (\ref{e:deltaR}), since they activate
 the Mach-lines part of the solution,
$$\Phi_{\alpha-\mu} (x \sin(\alpha-\mu) - y \cos(\alpha-\mu) ) r^{\alpha-\mu} +
 \Phi_{\alpha+\mu} (x \sin(\alpha+\mu) - y \cos(\alpha+\mu) )  r^{\alpha+\mu}, $$

\noindent with one-dimensional vector spaces $\mbox{span}\{r^{\alpha-\mu}\}$ and $\mbox{span}\{r^{\alpha-\mu}\}$: it was successfully
 verified that the ratio of the components of the discrete $\delta W^1$ and $\delta W^2$
 is the one of the corresponding right eigenvector (results not shown here).
 
On the contrary, if a \clon{$\delta R^1$, $\delta R^2$ or $\delta R^4$} source term is located in the subsonic bubble or if one of the three curves
 enters the subsonic bubble, then the support of the flow perturbation includes this complete subsonic area 
 in agreement with the elliptic nature of the equations \cite[chap.~11]{And_03}. \clon{A $\delta R^3$ perturbation, conversely,
  affects the region along the streamline downstream the location of the source, even if located in a subsonic zone.}
 
 More details are now given for the case where the source term is located in the supersonic zone. Let consider the decomposition (\ref{e:3bdir}) of the perturbation.
 If the entropy or the total enthalpy
 (that are convected in steady state Euler flows) are perturbed ($\delta R^3$ or $\delta R^4$),
  close to the source,
 the perturbation is located along the trajectory downstream the source and possibly
 along shock waves for $\delta R^4$.
  If these quantities are not altered by the perturbation  ($\delta R^1$ or $\delta R^2$),
  the response of the steady state flow to the perturbation, close to the source, is located along
  the two characteristic lines downstream the source and
  the usual relative-variations along trajectories of thermodynamic and kinetic variables are valid between nominal and locally  perturbed
   flow.
  More precisely \footnote{all variations are given for a positive $\epsilon$ in equation (\ref{e:deltaR}) and positive components of velocity
 for the nominal flow}: 
\begin{itemize}
\item $\delta R^1$ induces no change in the stagnation quantities and entropy.
 The Mach number decreases along the two characteristic lines
 starting from the \cltw{location of the perturbation}. The static pressure, temperature and density increase and velocity components are also perturbed along these two lines 
 with a reduction of the velocity magnitude;
\item $\delta R^2$  induces no change in the stagnation quantities and entropy.
 Along the upper characteristic curve
 starting from the perturbation point, the static pressure, density, temperature increase, whereas the Mach number, velocity magnitude, and
 x-component of velocity decrease. Opposite variations are observed along the lower characteristic curve. The y-component of  the velocity increases
  along both characteristic curves;
 \item $\delta R^3$  induces no variation of stagnation pressure, static pressure and Mach number. An increase of total enthalpy (stagnation temperature),
 temperature, both components of velocity and
 entropy is observed along the trajectory starting from the \cltw{location of the perturbation}. The density is reduced;
\item $\delta R^4$ induces no local variation of the static pressure except along shock waves
 and no variation of the 
 total enthalpy. The Mach number, density, stagnation pressure, stagnation density, and both velocity components
 increase along
 the trajectory starting from the \cltw{location of the perturbation}. Entropy and temperature decrease along the trajectory.
 As mentioned end of \S~3.3 this notion of perturbation downstream the source and along a streamtube of the initial
 flow is only a description at first approximation
 as the streamtubes structure of a non-constant flow is perturbed by the $\delta R^4$ source.
\end{itemize}
 
These zones of influence and dependency of the lift and the drag explain in particular the aspect of the discrete adjoint fields
 close to the trailing edge, that are bounded by the two characteristics curves passing through the trailing edge
 as seen in Fig.~\ref{f:discadj_M150}.
 They are also illustrated in Fig.~\ref{f:isoMach_M150}(right) presenting the $\delta W^1$ perturbation on the pressure field plotted together
 with the characteristic curves passing by the \cltw{location of the source}.

\subsubsection{Adjoint fields across the shock}
\begin {figure}[htbp]
  \begin{center}
	  \includegraphics[width=0.8\linewidth]{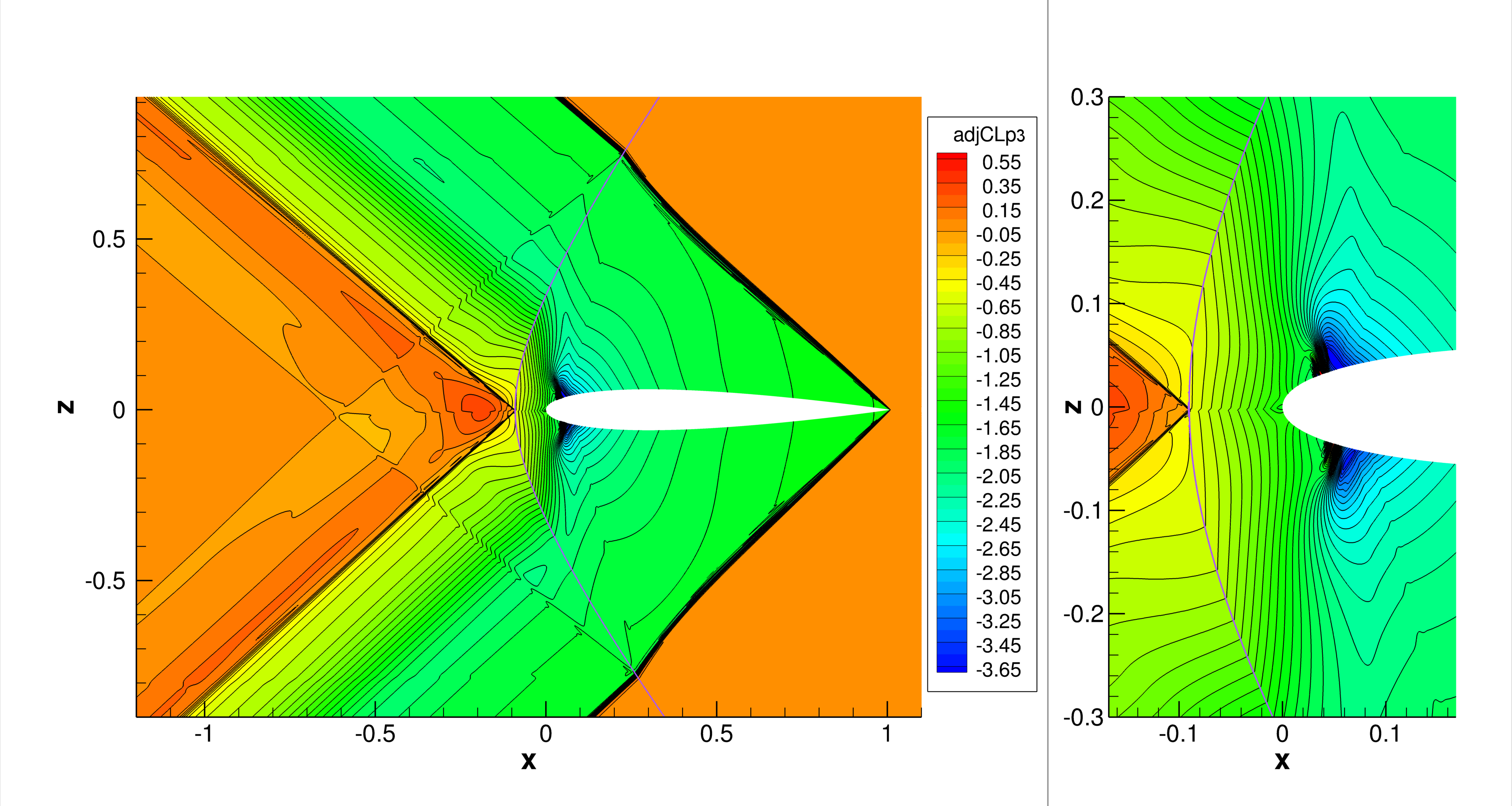}
	  \caption{($M_\infty=1.5$, $\alpha=1^o$, $4097\times4097$ mesh) Contours of the adjoint component associated with
      $z$-coordinate momentum equation. The position of the shock is indicated by a pink line.}
  \label{f:discadj_M150}
  \end{center}
\end {figure}

The adjoint variables are continuous across shocks, whereas their derivatives may be discontinuous \cite{giles_pierce_2001,GilPie_97,Loz_18}. 
 In Fig.~\ref{f:discadj_M150}, a clear discontinuity of the adjoint-drag or the adjoint-lift gradient is observed 
 for the adjoint component associated with $z$-coordinate momentum equation. 

\begin{figure}
\centering
\begin{tikzpicture}
\begin{axis}[ width=6.0cm, 
xlabel=$s$, legend style={font=\footnotesize}, legend pos=south west,
xmin=0, xmax=0.2,
ymin=-15,ymax=10]
\addplot[color=black, thick] table[x=s,y=DnPsi1] {SHOCK/data_across_supersonic_shock_drag_DnDpsi.dat};
\addplot[color=red  , thick] table[x=s,y=DnPsi2] {SHOCK/data_across_supersonic_shock_drag_DnDpsi.dat};
\addplot[color=green, thick] table[x=s,y=DnPsi4] {SHOCK/data_across_supersonic_shock_drag_DnDpsi.dat};
\addplot[color=blue , thick] table[x=s,y=DnPsi5] {SHOCK/data_across_supersonic_shock_drag_DnDpsi.dat};
\addplot[color=black, thick,densely dashdotted] table[x=s,y=DnPsi1] {SHOCK/data_across_supersonic_shock_lift.dat};
\addplot[color=red  , thick,densely dashdotted] table[x=s,y=DnPsi2] {SHOCK/data_across_supersonic_shock_lift.dat};
\addplot[color=green, thick,densely dashdotted] table[x=s,y=DnPsi4] {SHOCK/data_across_supersonic_shock_lift.dat};
\addplot[color=blue , thick,densely dashdotted] table[x=s,y=DnPsi5] {SHOCK/data_across_supersonic_shock_lift.dat};
\legend{$\partial_n\Lambda_1$,$\partial_n\Lambda_2$,$\partial_n\Lambda_3$,$\partial_n\Lambda_4$}
\addplot[dashed] coordinates {(0.114,-15) (0.114,10)};
\end{axis}
\end{tikzpicture}\hspace{-0.25cm}
\begin{tikzpicture}
\begin{axis}[ width=6.0cm, 
xlabel=$s$, legend style={font=\footnotesize}, 
xmin=0, xmax=0.2,
ymin=-2,ymax=5 ]
\addplot[color=black, thick] table[x=s,y=vnDnPsi1-H*vnDnPsi5] {SHOCK/data_across_supersonic_shock_drag.dat};
\addplot[color=red, thick] table[x=s,y=gam1*(nx*DnPsi2+nz*DnPsi4)+gam*vn*DnPsi5] {SHOCK/data_across_supersonic_shock_drag.dat};
\addplot[color=green, thick]   table[x=s,y=vn(-nz*DnPsi2+nz*DnPsi4+vt*DnPsi5)+u*DtPsi2+w*DtPsi4] {SHOCK/data_across_supersonic_shock_drag.dat};
\addplot[color=blue, thick] table[x=s,y=DnPsi1+(u+vn*nx)*DnPsi2+(w+vn*nz)*DnPsi4+(H+vn^2)*DnPsi5+vn*vt*DtPsi5] {SHOCK/data_across_supersonic_shock_drag.dat};
\legend{(\ref{eq:adj_RH_1}),(\ref{eq:adj_RH_2}),(\ref{eq:adj_RH_3}),(\ref{eq:adj_RH_4})} 
\addplot[dashed] coordinates {(0.114,-2) (0.114,5)};
\end{axis}
\end{tikzpicture}\hspace{-0.25cm}
\begin{tikzpicture}
\begin{axis}[ width=6.0cm, 
xlabel=$s$, legend style={font=\footnotesize},
xmin=0, xmax=0.2,
ymin=-25,ymax=10 ]
\addplot[color=black, thick,densely dashdotted] table[x=s,y=vnDnPsi1-H*vnDnPsi5] {SHOCK/data_across_supersonic_shock_lift.dat};
\addplot[color=red, thick,densely dashdotted] table[x=s,y=gam1*(nx*DnPsi2+nz*DnPsi4)+gam*vn*DnPsi5] {SHOCK/data_across_supersonic_shock_lift.dat};
\addplot[color=green, thick,densely dashdotted]   table[x=s,y=vn(-nz*DnPsi2+nz*DnPsi4+vt*DnPsi5)+u*DtPsi2+w*DtPsi4] {SHOCK/data_across_supersonic_shock_lift.dat};
\addplot[color=blue, thick,densely dashdotted] table[x=s,y=DnPsi1+(u+vn*nx)*DnPsi2+(w+vn*nz)*DnPsi4+(H+vn^2)*DnPsi5+vn*vt*DtPsi5] {SHOCK/data_across_supersonic_shock_lift.dat};
\legend{(\ref{eq:adj_RH_1}),(\ref{eq:adj_RH_2}),(\ref{eq:adj_RH_3}),(\ref{eq:adj_RH_4})} 
\addplot[dashed] coordinates {(0.114,-25) (0.114,10)};
\end{axis}
\end{tikzpicture}
\caption{($M_\infty=1.5$, $\alpha=1^o$, $4097\times4097$ mesh) Evolutions of drag (continuous lines) and lift (dash-dotted lines) adjoint quantities
  along a straight line across and normal to the upstream shock between points $(-0.151,0.266)$ and $(0.077,0.184)$ as a function of the local
  coordinate $s$ (the line is indicated in red in the Mach number contours of Fig.~\ref{f:isoMach_M150}). An equation number refers to the arguments between brackets $\du\cdot\df$ in the corresponding \clmy{equation, that} should be continuous across the shock, e.g., $v_n(\partial_n\Lambda_1 - H\partial_n\Lambda_4)$ has been plotted for (\ref{eq:adj_RH_1}).}
\label{fig:adj_across_shock_sup}
\end{figure}
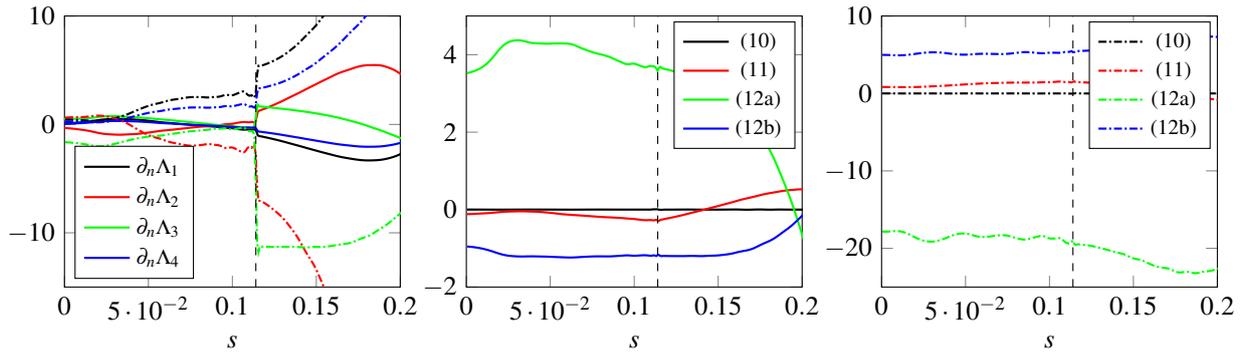

Figure~\ref{fig:adj_across_shock_sup} validates the jump relations on the adjoint derivatives (\ref{eq:adj_RH_1})
to (\ref{eq:adj_RH_4}) derived in \S~\ref{sec:adj_RH} at the discrete level for both lift and drag adjoints. We \cltw{display here}
 the evolution of the arguments between the jump operator $\du\cdot\df$ in the jump relations along a line perpendicular
  to the shock (the line is indicated in red in Fig.~\ref{f:isoMach_M150}) which should be continuous across the shock. The normal and
  tangential derivatives in (\ref{eq:adj_RH_1}) to (\ref{eq:adj_RH_4}) are here given in a frame attached to the shock and evaluated from
  the discrete adjoint field $\Lambda$. The shock location is about $s=0.11$ and we observe that the jump relations are well
satisfied, while the gradients normal to the shock of the adjoint variables, $\partial_n\Lambda$, display large discontinuities at the shock position. 

%
%

%
\subsection{Transonic regime}\label{sec:xp_naca_tra}
\begin {figure}[htbp]
  \begin{center}
	  \includegraphics[width=0.45\linewidth]{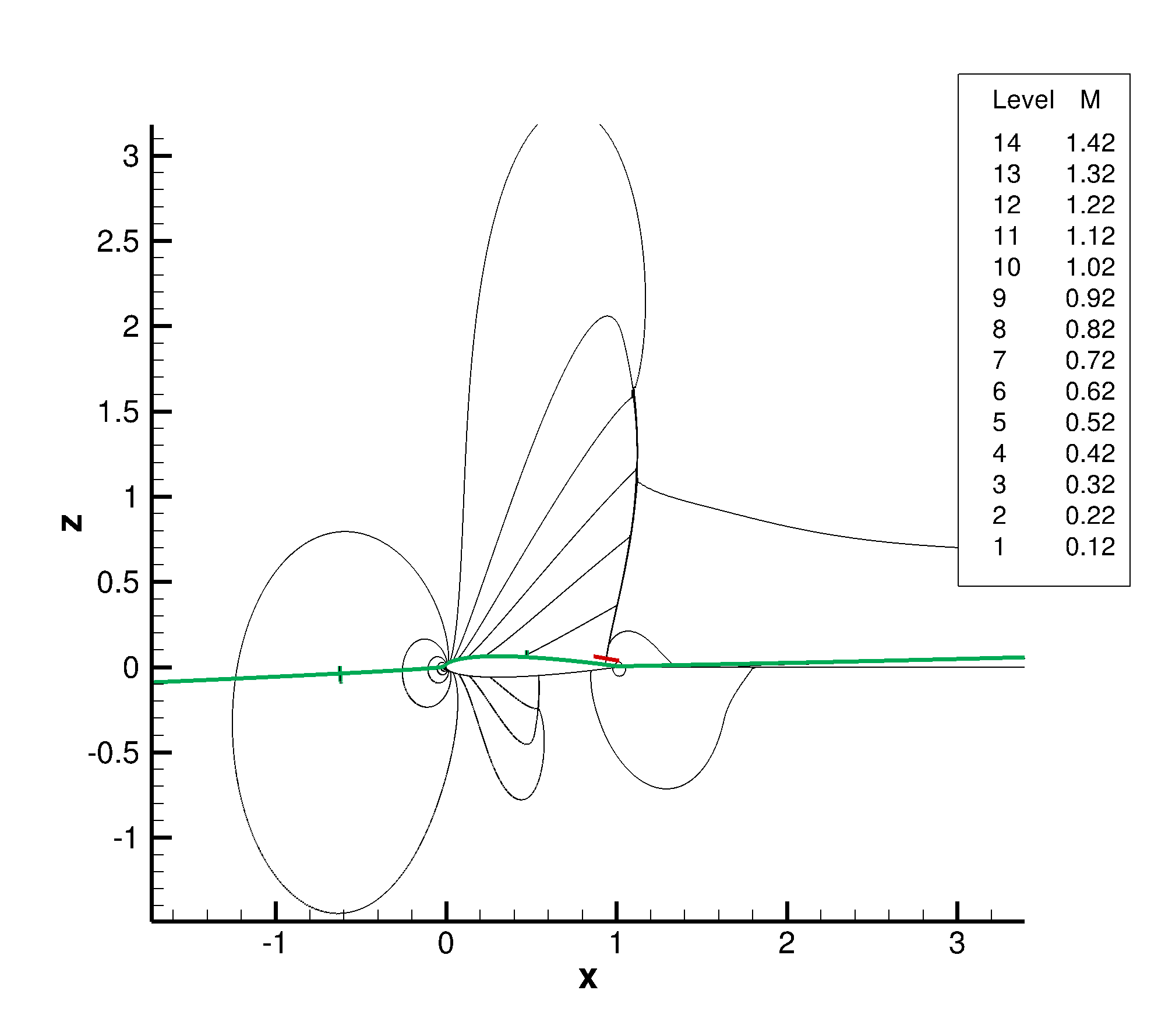}
	  \includegraphics[width=0.45\linewidth]{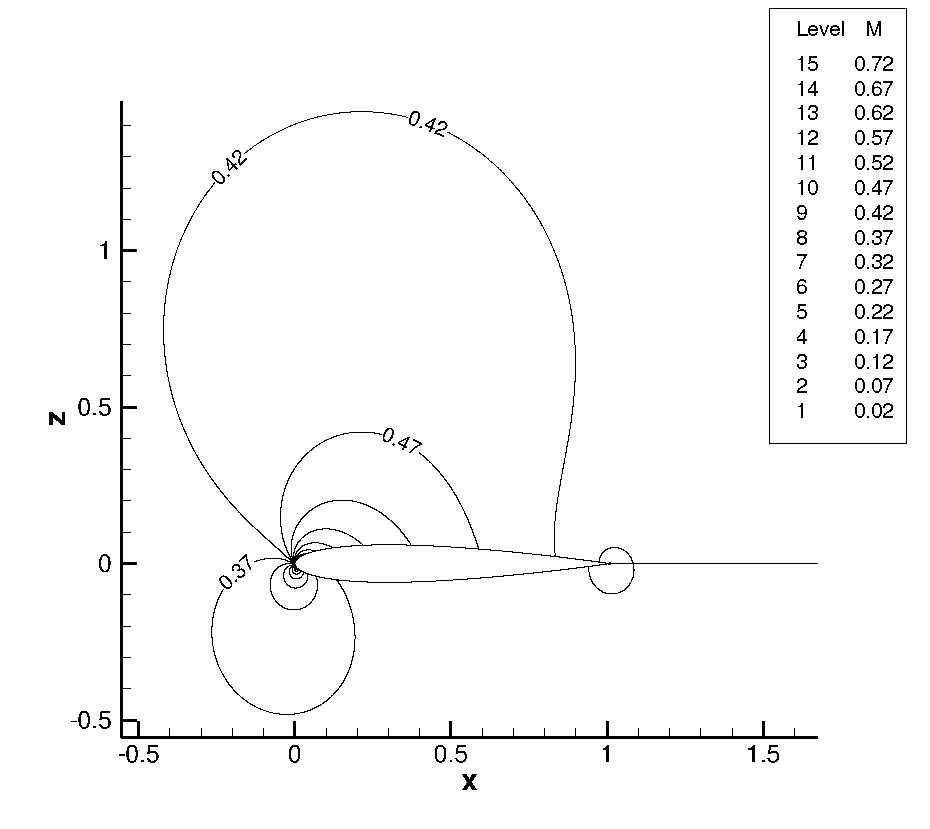}
	\caption{ (4097$\times$4097 mesh) Contours of Mach number. Left: $M_\infty=0.85$, $\alpha=2^o$, right: $M_\infty=0.4$, $\alpha=5^o$.}
  \label{f:isoM_M085_M040}
  \end{center}
\end {figure}

We now consider a transonic flow around the NACA0012 airfoil with conditions $M_\infty=0.85$ and $\alpha=2^o$. A strong shock wave develops on the suction side and a weaker shock on the pressure side (see Fig.~\ref{f:isoM_M085_M040} left).

%
%
\subsubsection{Lift and drag adjoint solutions}\label{sec:xp_naca_tra_adj_sol}
 Figure~\ref{f:g_M085} presents the contours
 of all four components of the drag and lift adjoint fields. As indicated before, the choice of the scale for the contours
 of first and last adjoint components in Figs.~\ref{f:g_M150}, \ref{f:g_M085} and \ref{f:g_M040} allows to visually check
 that (\ref{e:lam14}) is well satisfied by the discrete adjoint fields. 
 Both fields exhibit strong values and gradients close to the wall, the stagnation streamline,
 the two characteristics impinging the upper side and lower shock foot and two other characteristic curves that form a circumflex with the previous two.
 These features have been observed previously \cite{VenDar_02,TodVonBou_16,Loz_17,Loz_18}. 

 The physical point of view of Giles and Pierce \cite{GilPie_97} recalled in \S~\ref{sec:source_term_approach}
 is used again and the $\delta CL_p^d$ terms corresponding to the four source terms in
 (\ref{e:deltaR}) are plotted in Fig.  \ref{f:dCLp124_M085}.
 In the sense of equations (\ref{e:physpre})-(\ref{e:physp3}) the strong values and gradients  of $\delta R^1$ and $\delta R^2$ 
   along the characteristic curves that form the two circumflex correspond to those of $\Lambda_{CLp}$. 
 Accordingly, the high values and gradients of  $\delta R^4$ located close to the wall or to the stagnation streamline correspond to those of $\Lambda_{CLp}$.
 The asset of this approach is obvious when studying the singular behavior of the adjoint fields:
   the $\delta J^{1\leq d\leq4}$ responses to the physical source terms exhibit only part of the singular zones of
   the classical adjoint fields and the high values/gradients of one $\delta J^d$ are transferred to all or to several of the usual
    adjoint components according to equation (\ref{e:physp3}).

\begin {figure}[htbp]
  \begin{center}
	  \includegraphics[width=0.9\linewidth]{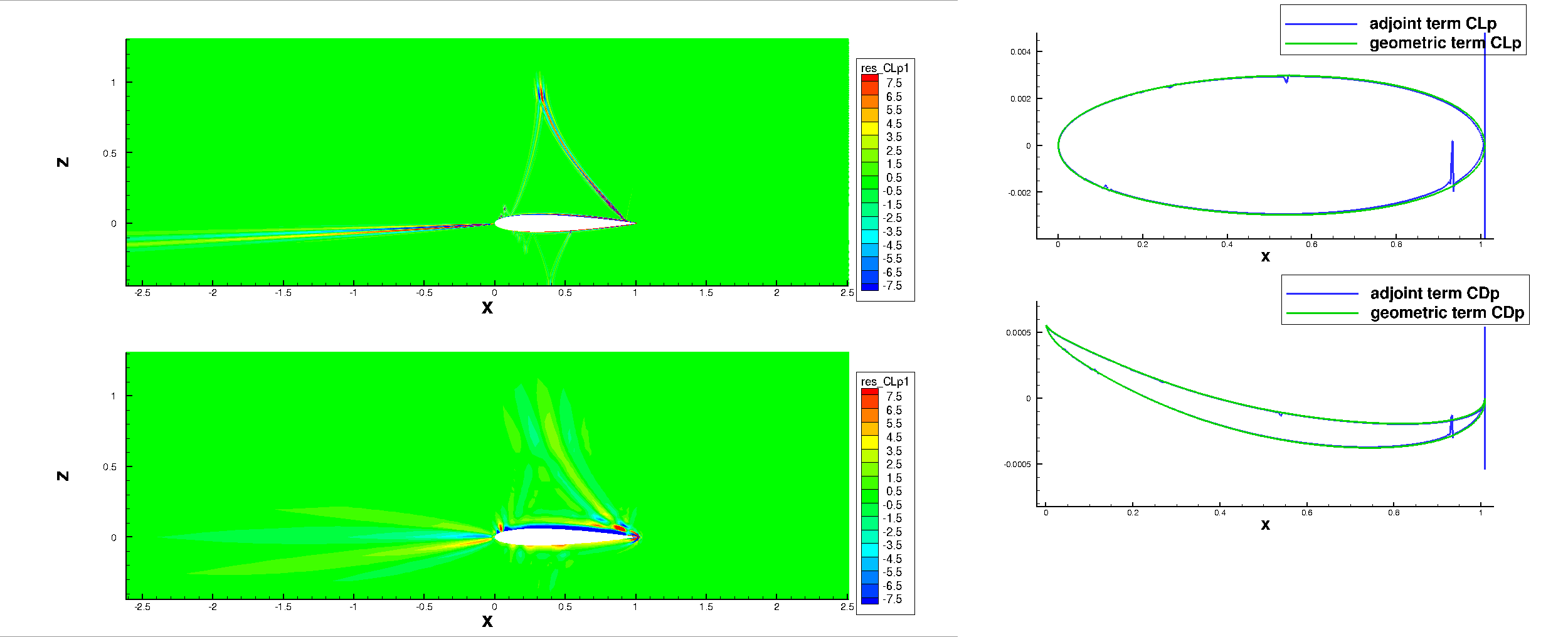}
	\caption{($M_\infty=0.85$, $\alpha=2^o$) Left: residual of the continuous equation  (first component)
 evaluated with the lift discrete adjoint fields  (bottom: 129$\times$129 mesh, top: 2049$\times$2049 mesh).
   Right: geometric and adjoint terms of (\ref{e:dab3}) for lift and drag (2049$\times$2049 mesh).}
  \label{f:contdisc_M085}
  \end{center}
\end {figure}

 The residuals of the continuous adjoint equation are evaluated by using the method introduced in \S~\ref{sec:heur_consist}. 
 As the mesh is refined, the zones with significant residuals have a decreasing area but remain visible.
 Increasing values are also observed close to the wall,
 the stagnation streamline
 and the characteristic line that impacts the upper side shock foot,
 that are known to be zones of large values and large gradients for
 the lift and drag adjoints \cite{TodVonBou_16,Loz_17,Loz_18} as highlighted in the left part of Fig.~\ref{f:contdisc_M085}.
 The far field adjoint boundary condition 
 is satisfied by the discrete adjoint fields as the adjoint field is almost zero at the far field boundary. 
 Right part of Fig.~\ref{f:contdisc_M085} illustrates the verification of continuous-like adjoint wall boundary condition.
 It is observed that (\ref{e:dab3})
 is satisfied except at the shock feet and the trailing edge. At the shocks feet, the first differences in
   neighboring $\Lambda$  along the wall in (\ref{e:fin1}) and (\ref{e:fin2})  may be too large 
   for a simple lower-order spatial analysis. At the trailing edge, the divergence of the discrete adjoint momentum components
   and the discontinuity of $\Lambda_{CDp_2}$ and $\Lambda_{CLp_2}$ between upper side and lower side
   (see Fig. \ref{f:g_M040}) may explain the observed behavior.
%
%
\subsubsection{Numerical assessment of exact discrete adjoint and modified consistent adjoint}\label{sec:numassgrad}
  The transonic test case  is retained for the assessment of the  exact discrete adjoint in our code
  \cite{CamHeiPlo_13,JPRenDum_15}.
  Also tested is the accuracy of the gradients when the Jacobian modification for consistency is applied.\\
   First, a series of ten symmetric
   bumps with  $0.4 c$ span is considered. The wall mesh is deformed normally according to
    $\exp(-0.25 (s_{bmax}-s_{bmin})^2/((s-s_{bmin})(s_{bmax}-s))) $. In the fluid domain, the mesh deformation follows
   the normal vector at the closest point of the wall and the displacement amplitude 
   is smoothed from maximum value at the wall, to zero at the distance $0.4 c$.
 The ten bumps are applied with small positive and negative increments to the (513$\times$513) mesh
   and steady state simulations are run to produce reference lift and drag derivatives. 
   The lift and drag gradients are calculated besides using the exact discrete adjoint, the modified discrete adjoint
   for consistency and the discrete adjoint without linearization of the spectral radius and the sensor \cite{DwiBre_06}.
   All gradient values are gathered in table \ref{tb:gradbumps}. Considering lift gradients, the mean error for exact
   linearization is less than 0.1\% (and calibration of the finite-difference step may improve this accuracy) whereas
   it is 11.5\% for
   the modified dual consistent version. This 11.5\% error
   is close to the 10.6\% error obtained with the frozen sensor and frozen
   spectral radius approximation, and indicates that this gradient may be involved in actual adjoint
   based shape optimization
    although probably not with best efficiency. Similar conclusions may be drawn from the drag gradients.
\\
   Besides, the agreement between the linear and the non-linear source term approaches is  checked.
  Let us recall that
  the linear approach just calculates $ (\delta \Jd^1_m, \delta \Jd^2_m, \delta \Jd^3_m, \delta \Jd^4_m)$ from
  the standard adjoint fields using  equation (\ref{e:physpre}) whereas the non-linear approach
   consists of adding one of the source
   terms $\delta R^l $  -- eq. (\ref{e:deltaR}) --  to the scheme flux balance, for a specific cell $m$,
   running a steady state simulation and calculating  $\delta \Jd^l_m$ as the difference of
    the resulting $\Jd^l$  with its
    nominal evalutation. Note that, as above with the calculation of the shape optimization gradients,
    any error in the scheme Jacobian $\partial R/\partial W$ would become apparent with
    this second type of comparisons.
    The non linear approach is obviously expensive and can only be tested for a limited
    number of cells. It has been applied for the fourth source term $\delta R^4$,
    for two series of cells, upper-side close to the
    wall at $x \simeq  0.5$ and close to the stagnation streamline at $x \simeq -0.6$.
   \clon{ The value of $\epsilon$ and the number of iterations have been chosen such that the
     first four significant numbers of the functions of interest do not change and the next four or five
     significant numbers are known upon reaching the steady-state, at the end of the perturbed calculation.}
     The comparison between the linear (red curve) and non linear (green squares) values is presented
     in Fig. \ref{f:dCLp4lnl}. The agreement is very good and allows to use
     the mechanical perturbations induced by the source terms (\ref{e:deltaR}) to discuss properties of
     $ (\delta \Jd^1_m, \delta \Jd^2_m, \delta \Jd^3_m, \delta \Jd^4_m)$ and then move to properties
     of the standard adjoint fields. This is done hereafter only for $\delta R^4$ / $ \delta \Jd^4$.  
 
%
\subsubsection{Influence of a $\delta R^4$ perturbation in the vicinity of the wall and the stagnation streamline}\label{sec:dR4_pert_ana_tra}
As mentioned in \S~\ref{sec:xp_naca_tra_adj_sol}, considering the physical source terms first --  equation (\ref{e:physp3}) --
$\delta R^4$ is the only physical source ``responsible''
for high values of lift/drag-adjoint and gradient
of adjoint close to the wall and the stagnation streamline. The mechanism by which  $\delta R^4$ modifies the flow field and
the forces values is discussed here for the vicinity of the stagnation streamline and the wall (upper side) using
 the non-linear approach which accuracy has just been discussed. 

 In the first series of considered perturbed flows, the $\delta R^4$ source terms are located 
 on the upper side in the supersonic area,
 along a mesh line perpendicular to the wall and close to $x=0.5$.
 In a supersonic zone, the perturbation of the flow caused by a  $\delta R^4$ source term along the corresponding
 streamline has been described in \S~\ref{sec:dRsup}: no local variation of the static pressure (except along shock waves
   -- see below)
   and no variation of the total enthalpy are observed whereas the Mach number, the density, the stagnation pressure, stagnation
   density and the velocity  are increased along the trajectory starting from the \cltw{location of the perturbation}.
  Hence, for the almost normal upperside shock, the classical equation for the ratio of downwind
  over upwind static pressure  as a function of the upwind Mach number cannot be satisfied
  where the perturbed streamline hits the shock with an increased Mach number.
   In fact in the perturbed flow, the upper side shock is moved upstream and the shifted shock goes with a
   decreasing upwind static pressure and an increasing downwind static pressure.
  It is observed that the smaller the distance of the source  and perturbed streamline to the wall, the larger
  the displacement of the upper side shock foot which explains the growth of $\delta CLp_4$ in the vicinity of
   the wall. Observing the static pressure changes at smaller scales, it is noted,
   as expected, that the perturbation of the static pressure 
   in the subsonic area is not   restricted to the continuation
  of the streamline but extends to the whole subsonic area close to the profile downstream the shocks. 
 \cltw{Regarding} the static pressure close to the lower side, at these
  finer scales, an increase is observed upwind the shock and below the trailing edge
  whereas a decrease is observed downwind the shock; this goes with a backward displacement of the lower
   shock. Both changes of the static  pressure at the wall
  correspond to an increase of the 
  lift that is consistent with the local values of $\delta CLp^4$.
  To complement this description, the fraction
    of the forces variation due to the two shocks displacement is calculated
 for all source points of the series (Fig. \ref{f:dCLp4lnl} left).
   The pressure difference w.r.t. the  nominal pressure
 is summed along 36 faces only among the 2048 of the wall mesh in the vicinity of the shock waves
  ($x\in$ [0.535,0555] lower side and $x\in$ [0.925,0945] upperside) and is then divided by the total
 change in the function estimate. 
 For the lift, theses ratios are included in [0.806, 0.921] with a mean value of 0.880. For the drag,
 the corresponding interval is [0.925,1.038] and the mean value is 0.985. This confirms the main effect
  of the displacement of the two shocks.\\
   If the $\delta R^4$ source is located close to
  the wall but downwind the upper side shock, similar shock displacements and lower side fine
  scale perturbations of the pressure are observed. On the contrary, the flow perturbations created downwind
  the upper side shock are very different for sources located on the same streamline either sides of the shock.
  \\
\begin {figure}[htbp]
  \begin{center}
	  \includegraphics[width=0.95\linewidth]{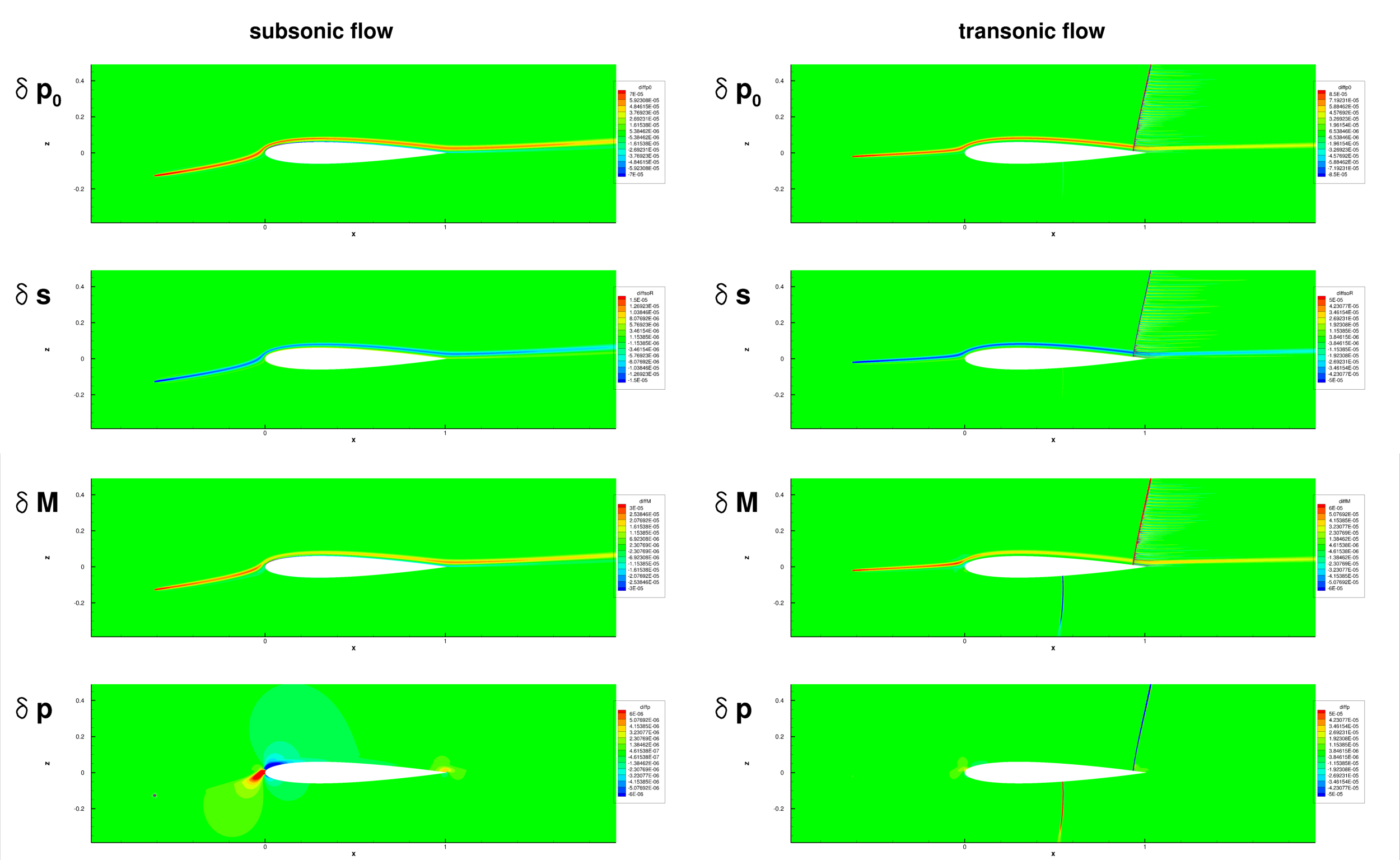}
	  \caption{(2049$\times$2049 mesh) Difference in stagnation pressure, entropy
      (divided by the adiabatic exponent), Mach number and static pressure w.r.t. nominal flow due to $\delta R^4$. Left: $M_\infty=0.4$, $\alpha=5^o$, source at point (-0.612,-0.127). Right: $M_\infty=0.85$, $\alpha=2^o$, source in (-0.623,-0.020).} 
  \label{f:diff_p0sp}
  \end{center}
\end{figure}
  
  Concerning the points located close to the stagnation streamline, the mechanism that produces 
   the lift perturbation $\delta CLp^4$  due to the $\delta R^4$ source also seems to be correlated with
   convection since the iso-$\delta CLp^4$ roughly follow the streamlines
   (see Fig.~\ref{f:dCLp124_M085} right).
   Of course, the increment of stagnation  pressure and stagnation density, and the decrease of entropy are convected
     from the \cltw{location of the source} where they have been created (see Fig. \ref{f:diff_p0sp}). Surprisingly, although the source
     is now located in a subsonic zone, the main change in the velocity and Mach number fields also occur
      along the trajectory. These increments being propagated up to
  the upper side shock,
 the mechanism through which both shocks are moved is then the one described before.  
  This description is complemented with
  the ratios of the function variations that result from the changes in the pressure
 in the vicinity of the shocks only ($x\in$ [0.535,0555] lower
  side and $x\in$ [0.925,0945] upperside)
  divided by the corresponding total variation. 
  The considered points are those appearing in the right part of Fig. \ref{f:dCLp4lnl} ($x\simeq$ -0.6
  $y\simeq$-0.04) at a distance lower
 than $0.025 c$ to the stagnation streamline.
   For the lift, theses ratios are included in [0.793, 1.245] with a mean value of 0.882. For the drag
   the interval is [0.762,0.907] and the mean value is 0.840. Again, the dominant influence of shock displacement 
 in the lift and drag variation is confirmed.
\\
   Finally, we would like to investigate whether the influence of $\delta R^4$ is evolving smoothly from the vicinity
  of the stagnation streamline to the vicinity of the wall. We have already mentioned the pros (the main process
  is the convection of the source perturbation) and cons (the streamline structure
  of the original flow is not strictly preserved) of a positive
  answer to this question. It is investigated numerically extracting $\delta CLp^4$ and $\delta CDp^4$ along a
  streamline close to the stagnation streamline and the upperside wall (see Fig. \ref{f:isoM_M085_M040} then Fig. \ref{f:dCLp4lnltr} -- the gaps
  between the curve and the non-linear verification calculations come from the fact that not all the selected cells
  are exactly crossed at their center by the streamline). \clon{The  $\delta CLp^4$ curve of Fig. \ref{f:dCLp4lnltr}
    is completed in Fig.
     \ref{f:3streamtr} with its counterparts
     on a finer and a coarser mesh and it is checked that the general shape of these curves is not
      affected by the mesh size.
  The levels of  $\delta CLp^4$ seen in Fig. \ref{f:dCLp4lnltr} upwind the airfoil
  at x $\simeq$ -0.1 are observed along the wall but a downwards oscillation just upwind the profile makes it difficult
 to give a sure conclusion. }
%
%
\subsubsection{Asymptotic behavior of lift and drag adjoint at the stagnation streamline and the wall}\label{sec:asym_staglwall}

 In \cite{GilPie_97}, the authors consider a 2-D inviscid flow about an airfoil and an output $J$ 
 defined as the integral along the wall of the pressure times a local factor.  With the additional approximation
 of potential flow, they
 derive the asymptotic behavior of  $\delta J^4_m$, the variation of $J$ due to $\delta R^4$,
 when the source term is located in the vicinity of the stagnation streamline:
 under this potential flow assumption, they  prove that $\delta J^4 \simeq \pm d^{-1/2}$
 where $d$ is the distance of the source $m$ to the stagnation streamline
 and the sign is changed when crossing the streamline.
 This relation is not always numerically well satisfied  by numerical lift / drag
 adjoint fields of compressible Euler flows \cite{Loz_17,Loz_18}.
 \\
  For the transonic test-case of interest,  $\delta CLp^4_m$ and  $\delta CDp^4_m$
  seem to result of similar convective perturbations of the flow for points $m$ located close to the stagnation
  streamline and close to the wall upwind the shock.
  Using the structured and regular meshes from \cite{VasJam_10},
  one may evaluate the $\varphi$ exponent such that $\delta J^4 \simeq d^\varphi$ in the vicinity of the wall.
  This exponent appears to depend on the $x$ location
  (rather than on  $CD$ vs $CL$, lower side vs
 upper side). It is decreasing with increasing $x$, from  0. at the leading edge to about $-0.1$ at $x=0.8$
 and then more rapidly when moving closer to the trailing edge. (\clon{Curves not shown here because 
   changing the mesh density, the $F^{JST}_{3/2}$ formula or selecting its adjoint-consistent
   linearization instead of its exact linearization, 
    significantly modifies these curves although not
   the decreasing trend from a null value.)}
\\
  Finally the values of $\delta CLp^4$ are extracted along the wall
  on the finest used meshes (Fig. \ref{f:finewalladj}).  They appear to grow as the cells adjacent to the wall get thiner
    coherently with the similar behavior observed for the lift or drag adjoint components in other publications
    (\cite{Loz_18} fig. 12, \cite{JP_20} fig. 1.1).
    As the wall boundary condition that involes a linear boundary combination of $\Lambda_2$ and $\Lambda_3$, is numerically well satisfied (equation
    (\ref{e:dab3}) and fig. \ref{f:contdisc_M085} right),
    this is surprising although not impossible: it should be noted that in the vicinity of a wall, with local normal vector $(n_x,n_y)$,
    it is worthwhile to calculate the linear combination of the second and third columns of equation (\ref{e:physp3}) by $(n_x, n_y)$ to show that the term involving
    $\delta CLp_4$ in $\Lambda_2 n_x + \Lambda_3 n_y$  is $-\gamma p_0 /c^2 (u_x n_x + u_y n_y) \delta CLp_4 $ with $(u_x n_x + u_y n_y)\rightarrow 0 $.
    We have checked also that this behavior close to the wall is not an artefact
    due to numerical dissipation by performing calculations with different $k^4$. The influence of the cell-height
    at the wall always dominates
    the one of the artificial dissipation and, for a given mesh density, the largest values of $\delta CLp^4$ are
    obtained for the lowest $k^4$.
%
%
%
\subsubsection{Adjoint fields across the shock} 

\begin {figure}[htbp]
  \begin{center}
	  \includegraphics[width=0.9\linewidth]{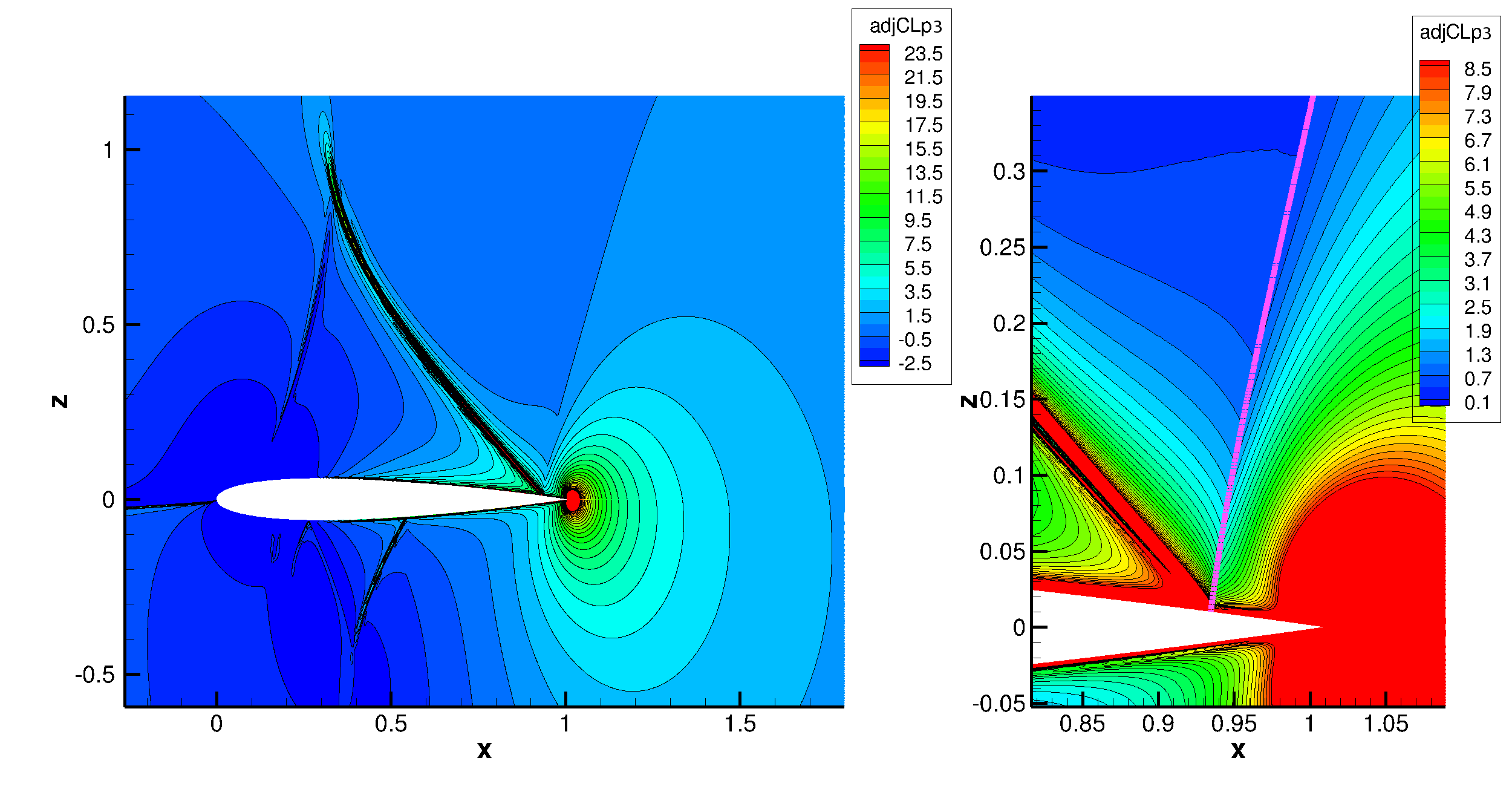}
	\caption{($M_\infty=0.85$, $\alpha=2^o$, $4097\times4097$ mesh) Contours of adjoint component associated with the $z$-coordinate momentum
 equation (sonic line in pink).}
  \label{f:discadj_M085}
  \end{center}
\end {figure}

\begin{figure}
\centering
\begin{tikzpicture}
\begin{axis}[ width=6.0cm, 
xlabel=$s$, legend style={font=\footnotesize}, legend pos=north west,
xmin=0, xmax=0.2,
ymin=-20,ymax=60]
\addplot[color=black, thick] table[x=s,y=DnPsi1] {SHOCK/data_across_transonic_shock_drag.dat};
\addplot[color=red  , thick] table[x=s,y=DnPsi2] {SHOCK/data_across_transonic_shock_drag.dat};
\addplot[color=green, thick] table[x=s,y=DnPsi4] {SHOCK/data_across_transonic_shock_drag.dat};
\addplot[color=blue , thick] table[x=s,y=DnPsi5] {SHOCK/data_across_transonic_shock_drag.dat};
\addplot[color=black, thick,densely dashdotted] table[x=s,y=DnPsi1] {SHOCK/data_across_transonic_shock_lift.dat};
\addplot[color=red  , thick,densely dashdotted] table[x=s,y=DnPsi2] {SHOCK/data_across_transonic_shock_lift.dat};
\addplot[color=green, thick,densely dashdotted] table[x=s,y=DnPsi4] {SHOCK/data_across_transonic_shock_lift.dat};
\addplot[color=blue , thick,densely dashdotted] table[x=s,y=DnPsi5] {SHOCK/data_across_transonic_shock_lift.dat};
\legend{$\partial_n\Lambda_1$,$\partial_n\Lambda_2$,$\partial_n\Lambda_3$,$\partial_n\Lambda_4$}
\addplot[dashed] coordinates {(0.116,-20) (0.116,60)};
\end{axis}
\end{tikzpicture}\hspace{-0.25cm}
\begin{tikzpicture}
\begin{axis}[ width=6.0cm, 
xlabel=$s$, legend style={font=\footnotesize}, 
xmin=0, xmax=0.2,
ymin=-2,ymax=12]
\addplot[color=black, thick] table[x=s,y=vnDnPsi1-H*vnDnPsi5] {SHOCK/data_across_transonic_shock_drag.dat};
\addplot[color=red, thick] table[x=s,y=gam1*(nx*DnPsi2+nz*DnPsi4)+gam*vn*DnPsi5] {SHOCK/data_across_transonic_shock_drag.dat};
\addplot[color=green, thick]   table[x=s,y=vn(-nz*DnPsi2+nz*DnPsi4+vt*DnPsi5)+u*DtPsi2+w*DtPsi4] {SHOCK/data_across_transonic_shock_drag.dat};
\addplot[color=blue, thick] table[x=s,y=DnPsi1+(u+vn*nx)*DnPsi2+(w+vn*nz)*DnPsi4+(H+vn^2)*DnPsi5+vn*vt*DtPsi5] {SHOCK/data_across_transonic_shock_drag.dat};
\legend{(\ref{eq:adj_RH_1}),(\ref{eq:adj_RH_2}),(\ref{eq:adj_RH_3}),(\ref{eq:adj_RH_4})} 
\addplot[dashed] coordinates {(0.116,-2) (0.116,12)};
\end{axis}
\end{tikzpicture}\hspace{-0.25cm}
\begin{tikzpicture}
\begin{axis}[ width=6.0cm, 
xlabel=$s$, legend style={font=\footnotesize},
xmin=0, xmax=0.2,
ymin=-20,ymax=120 ]
\addplot[color=black, thick,densely dashdotted] table[x=s,y=vnDnPsi1-H*vnDnPsi5] {SHOCK/data_across_transonic_shock_lift.dat};
\addplot[color=red, thick,densely dashdotted] table[x=s,y=gam1*(nx*DnPsi2+nz*DnPsi4)+gam*vn*DnPsi5] {SHOCK/data_across_transonic_shock_lift.dat};
\addplot[color=green, thick,densely dashdotted]   table[x=s,y=vn(-nz*DnPsi2+nz*DnPsi4+vt*DnPsi5)+u*DtPsi2+w*DtPsi4] {SHOCK/data_across_transonic_shock_lift.dat};
\addplot[color=blue, thick,densely dashdotted] table[x=s,y=DnPsi1+(u+vn*nx)*DnPsi2+(w+vn*nz)*DnPsi4+(H+vn^2)*DnPsi5+vn*vt*DtPsi5] {SHOCK/data_across_transonic_shock_lift.dat};
\legend{(\ref{eq:adj_RH_1}),(\ref{eq:adj_RH_2}),(\ref{eq:adj_RH_3}),(\ref{eq:adj_RH_4})} 
\addplot[dashed] coordinates {(0.116,-20) (0.116,120)};
\end{axis}
\end{tikzpicture}
\caption{($M_\infty=0.85$, $\alpha=2^o$, $4097\times4097$ mesh) Evolutions of drag (continuous lines) and lift (dash-dotted lines)
  adjoint quantities along a straight line across and normal to the upper shock between points $(0.865,0.064)$ and $(1.016,0.038)$
  as a function of the local coordinate $s$ (the line is indicated in red in the Mach number contours of Fig.~\ref{f:isoM_M085_M040}).
  An equation number refers to the arguments between brackets $\du\cdot\df$ in the corresponding \clmy{equation, that} should be continuous across
  the shock, e.g., $v_n(\partial_n\Lambda_1 - H\partial_n\Lambda_4)$ has been plotted for (\ref{eq:adj_RH_1}).}
\label{fig:adj_across_shock_tra}
\end{figure}
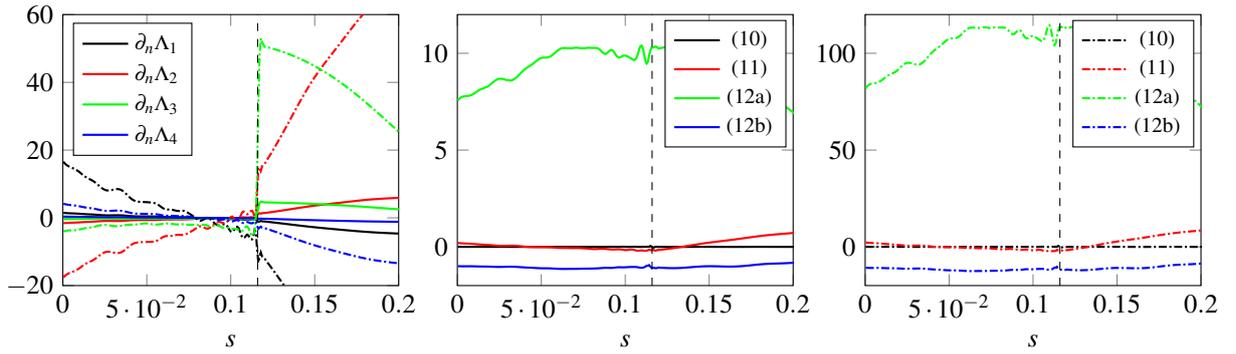

Figure~\ref{f:discadj_M085} highlights a clear discontinuity across the shock of the gradient of
the drag or the lift adjoint component associated with the $z$-coordinate momentum. This is further
analyzed in Fig.~\ref{fig:adj_across_shock_tra} where the arguments \cltw{in the} brackets in (\ref{eq:adj_RH_1}) to (\ref{eq:adj_RH_4}) are plotted
along the line displayed in Fig.~\ref{f:isoM_M085_M040} (left) that crosses the shock. Again, it is observed that the jump relations (\ref{eq:adj_RH_1})
to (\ref{eq:adj_RH_4}) are well \cltw{satisfied while} the adjoint derivatives  in the direction normal to the shock, $\partial_n\Lambda$, are discontinuous across the shock.

%
%
\subsection{Subsonic regime}\label{sec:xp_naca_sub}
Finally, we consider a smooth subsonic flow with free-stream conditions $M_\infty=0.4$ and $\alpha=5^o$. The Mach number contours are presented in Fig.~\ref{f:isoM_M085_M040} where we observe the stagnation point located in the lower part of the leading edge as well as a strong acceleration of the flow on the upper side due to the profile incidence.

%
%
\subsubsection{Lift and drag adjoint solutions}
Figure~\ref{f:g_M040} presents the contours of all four components of the adjoint fields. The lift adjoint components exhibit strong values and gradients close to the wall and
 the stagnation streamline. The amplitudes of these values increase as the mesh is
 refined. Conversely, all components of the drag adjoint appear to be bounded and to converge as the mesh is refined.
 For this flow without \cltw{shocks}, these different behaviors may be attributed to the facts that the adjoint field quantifies the sensitivity of
   the quantity of interest w.r.t. the residuals,
 and that  the limit inviscid drag is zero for a subsonic flow 
 whereas the limit lift is not.

\begin {figure}[htbp]
  \begin{center}
	  \includegraphics[width=0.9\linewidth]{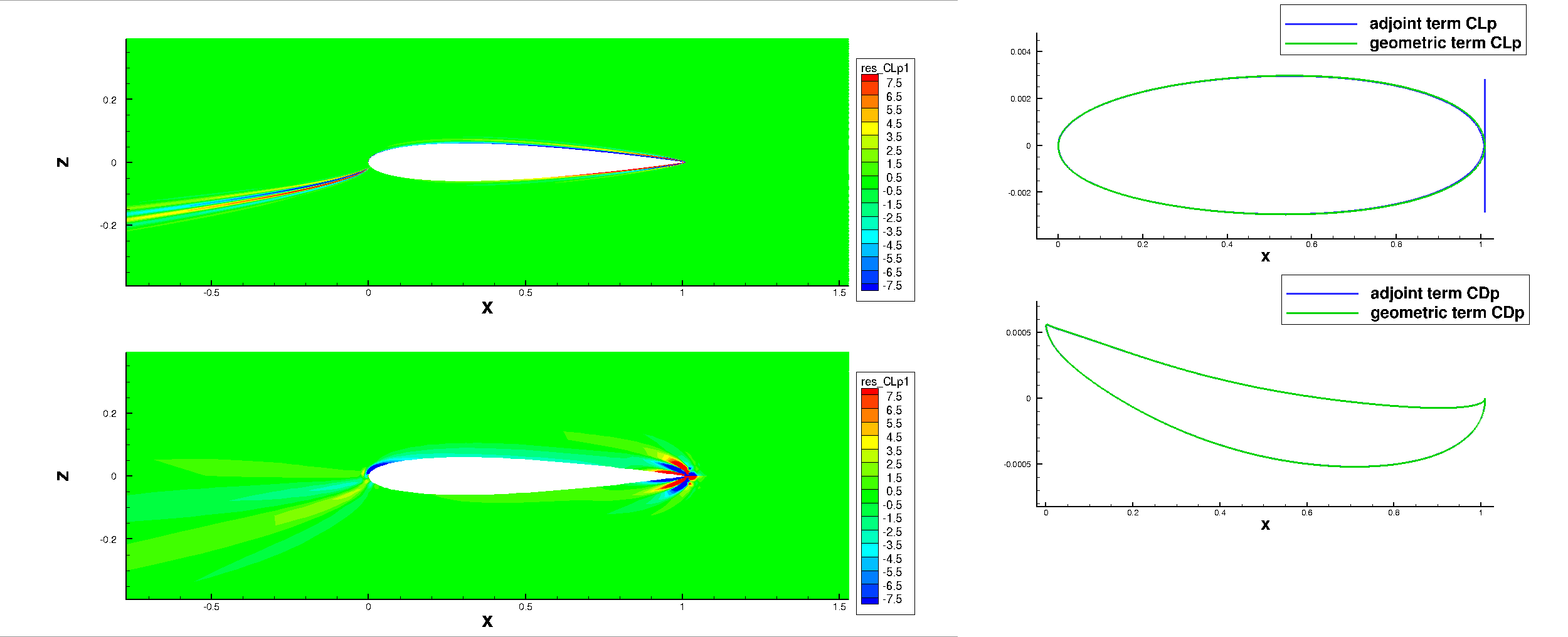}
	\caption{($M_\infty=0.4$, $\alpha=5^o$) Left: residual of the continuous equation (first component)
  evaluated with the lift discrete adjoint fields (bottom: 129$\times$129 mesh, top: 2049$\times$2049 mesh).
  Right: geometric and adjoint terms of equation (\ref{e:dab3}) for lift and drag (2049$\times$2049 mesh).}
  \label{f:contdisc_M040}
  \end{center}
\end {figure}

 The residuals of the continuous adjoint equations have been evaluated as detailed in \S~\ref{sec:heur_consist}.  
 As the mesh is refined, the zones with significant residuals for the lift-adjoint have a decreasing support but exhibit
  increasing values close to the wall and to the stagnation streamline (see left part of Fig.~\ref{f:contdisc_M040}).
 The far field adjoint boundary condition 
 is satisfied by the discrete adjoint fields as the adjoint field is almost zero at the far field boundary. 
 Right part of Fig.~\ref{f:contdisc_M040} shows that the adjoint wall boundary condition (\ref{e:dab3}) is well satisfied for both lift and drag.\\
  The approximate linearization  for consistency of $F^{JST_c}_{3/2}$ proposed in \S4.5 is
 assessed quantitatively for $CDp$ ($CDp$ only as the adjoint of $CLp$ is numerically diverging)
 calculating the mean norm of the $res$ components 
 inside a fixed region close to the profile (interior of 
  $ ((x-0.5)/0.55)^2+(z/0.1)^2=1$ ellipse excluding a $0.005c$ circle about the trailing edge 
 or the contribution of the trailing edge vicinity appears to be dominant). These means  are then  summed in 
 the aggregate residual denoted
 $RES$. As in the supersonic case, it appears \cltw{to decrease}
 be a few percent by the modification of the $F^{JST_c}_{3/2}$ differential for adjoint
 consistency. More precisely, $RES$ is decreased by 2.6\% (coarsest mesh) to 5.0\% (finest mesh). 
 
 The physical point of view of Giles and Pierce \cite{GilPie_97} is then considered to gain understanding in
 the behavior of the lift-adjoint.
The $\delta CLp^1$,  $\delta CLp^2$, $\delta CLp^3$,
and $\delta CLp^4$ responses have been calculated and plotted in Fig.~\ref{f:dCLp124_M040}. We observe that $\delta R^4$  is the only source responsible for the singular
 behavior of the adjoint-lift close to the wall and stagnation streamline.

%
%
\subsubsection{Influence of a $\delta R^4$ perturbation in the vicinity of the wall and the stagnation streamline}\label{sec:dR4_pert_ana_sub}
\begin {figure}[htbp]
  \begin{center}
	  \includegraphics[width=0.5\linewidth]{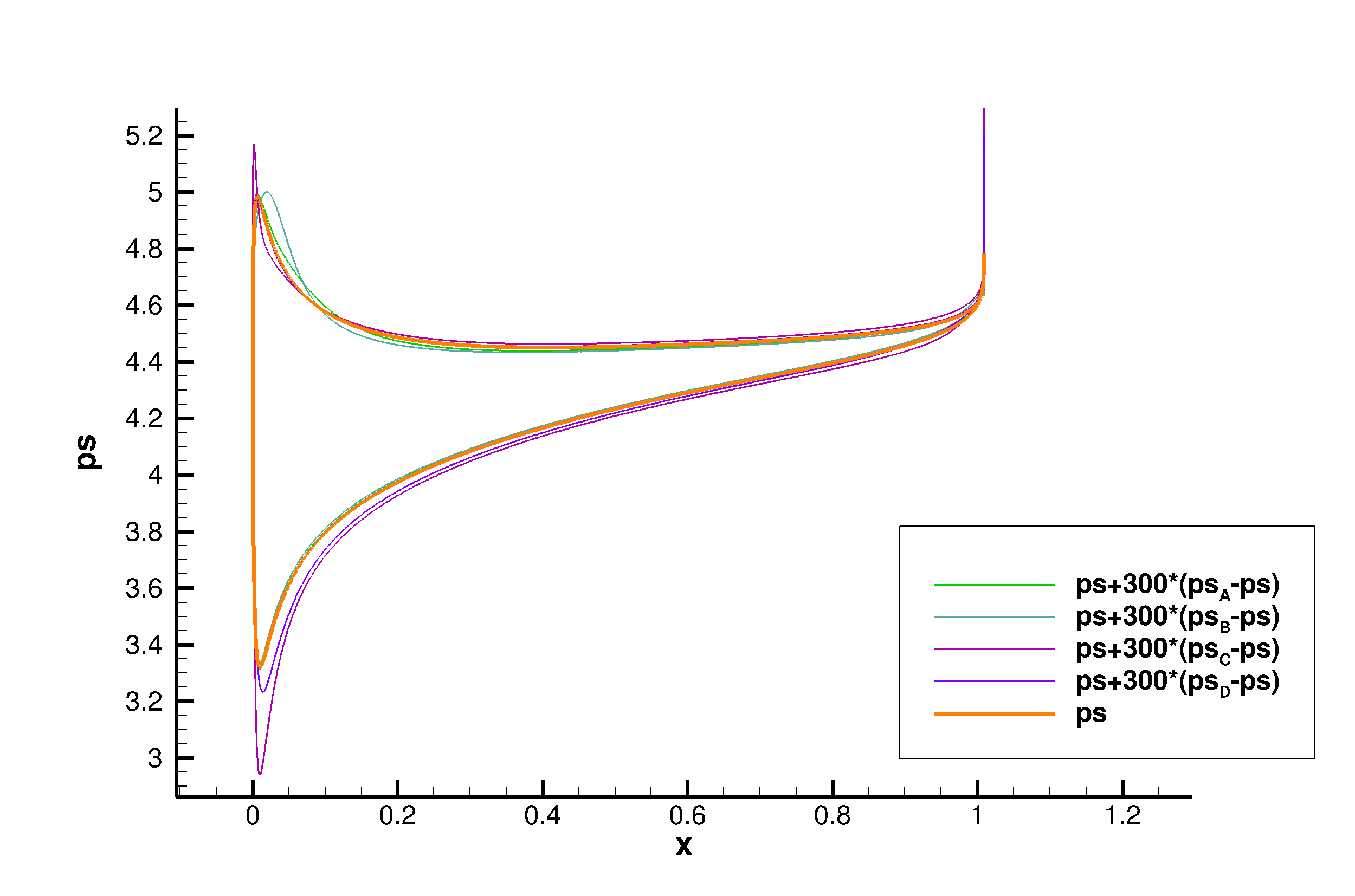}
	\caption{($M_\infty=0.4$, $\alpha=5^o$, 2049$\times$2049 mesh) Pressure at the wall.  Influence of $\delta R^4$ source
    term located at points A, B, C, D, on the static pressure at the wall. Distance of points A,B, C, D to the leading edge $\simeq$ 0.6.
    Respective distance to stagnation line is 0.0400, 0.0187 (trajectory to lower side), 0.0181, 0.397 (trajectory to upper side).}
  \label{f:dR4stream_M040}
  \end{center}
\end {figure}

As in transonic and supersonic flows, a $\delta R^4$ source results in positive increments in the stagnation pressure and
stagnation density and a negative increment in entropy. All three increments are
 convected downstream the \cltw{location of the source} following
 the classical laws for Euler flows with uniform total enthalpy.
 However, the response of the subsonic flow to the local perturbation of these three quantities, is very different from
 the one described in \S~\ref{sec:dR4_pert_ana_tra} for the
 transonic flow: density, velocity, static pressure and all dependent variables exhibit a global
 perturbation (see Fig.~\ref{f:diff_p0sp}). When the source is located close to the 
 stagnation streamline, the perturbation is maximum close to the
 leading edge. If it is located just under (resp. above) the stagnation streamline, the main effect
 on the static pressure at the wall is a decrease all along the pressure (resp. suction) side consistently
 with the local sign of $\delta CLp^4$ (Fig.~\ref{f:dCLp124_M040} right and Fig.~\ref{f:dR4stream_M040}).

As in the transonic case, we examine if $\delta CLp^4 \simeq d^\varphi$ in the vicinity of the wall.
  Once again, a negative $\varphi$ exponent decreasing from 0 to about $-0.40$ when increasing $x$ is derived
  from this assumption (\clon{Not shown here because, as in the transonic test case,
    changing the mesh density, the $F^{JST}_{3/2}$ formula or selecting its adjoint-consistent linearization instead of its exact linearization, 
    significantly modifies this curve although not
   the decreasing trend from a null value.}) 
  
%
%
\section{Conclusion}\label{sec:concl}
In this work, we address open \cltw{questions} in the field of discrete and continuous
 adjoint equations for the compressible inviscid flows. The results are illustrated
through in-depth numerical experiments on lift- and drag-adjoint fields for
lifting flows about the NACA0012 airfoil in supersonic, transonic
and subsonic regimes.

We first investigate the dual consistency of the JST scheme in cell-centered FV formulation for the discretization
of the compressible 2-D Euler equations. Dual consistency of the scheme is \cltw{proven} at interior cells,
 while inconsistency may occur
  at the penultimate cell close to a physical boundary depending on the selected locally modified AD formulas.
 Two classical and robust AD discretizations at the borders \cltw{have been proven} to be dual-inconsistent and
   we propose a slight modification of the exact scheme
 differentiation w.r.t. the penultimate cell
 flow variables to recover consistency. The implementation of this modification has been carried out
 in our code and appears to locally
 improve the regularity of the adjoint field and to speed up their convergence at wall boundaries under grid refinement.
 
Second, a new heuristic method has been proposed to discuss the adjoint consistency of discrete adjoint fields.
It consists of discretizing the continuous equation with the discrete flow and discrete-adjoint field and evaluating
the convergence of the residual as the mesh is refined. Coherently with the demonstrated property of
adjoint consistency, these residuals have been found to vanish except in areas where: (i) a discontinuous adjoint
field is expected (e.g., in the vicinity of the trailing edge for supersonic flows);
(ii) the discrete adjoint field exhibits increasing values and gradients where theoretical
divergence is suspected (stagnation streamline, wall, specific characteristic lines in supersonic areas).

 Third, we derive the general adjoint Rankine-Hugoniot relations across a shock
  and give relations linking the normal
  derivatives across the shock. These relations have been numerically validated for a supersonic and a transonic flow.
  
  We then use these tools for the analysis of the lift and drag adjoint fields about the NACA0012
   at three flow conditions and, in particular, the analysis of
   the zones where increasing values and gradients of the adjoint fields are observed when the mesh is refined
   \cite{GilPie_97,Loz_17,Loz_18,JPLabRen_19}.  The source term approach of Giles and Pierce 
   \cite{GilPie_97} is recalled and after assessing it for our code w.r.t linear results -- Fig. \ref{f:dCLp4lnl} --
   it is exploited to analyse the lift and drag adjoint behavior in the vicinity of the stagnation streamline
   and the wall for the  transonic test case.  The novelty is \clon{for the transonic regime} in the detailed
   description of the flow perturbation resulting of the involved source term : a pertubation along
    streamline passing by the source results in the displacement of the shocks. The mechanism is similar with a 
   source close to the stagnation streamline or close to the wall so that our numerical results suggest that the
   behavior of the adjoint fields in the vicinity of the stagnation streamline (where the authors of \cite{GilPie_97}
    predicted a numerical divergence under simplifying assumptions) and  in the vicinity of the wall are identical.
    Also explained by this analysis (although not illustrated here) is the well-known difference
    of numerical behavior in lift or drag
  adjoint at high transonic Mach numbers (shock-wave feet at trailing edge, no possible shock-foot displacement, no
  adjoint numerical divergence) and lower transonic Mach numbers (more upwind shock wave feet locations,
  shock feet displacement under the influence of $\delta R^4$, adjoint numerical divergence). \clon{In subsonic regime,
    the fourth source term proposed in \cite{GilPie_97} results in a numerically diverging perturbation of the lift,
 also when approaching the wall or the stagnation streamline, but through a global change of the wall pressure.}      
  
Future work will concern the extension of this analysis to laminar flows  considering the usual FV cell-centered discretizations of the viscous flux.

\section*{Acknowledgments}
The authors express their warm gratitude to J.C. Vassberg and A. Jameson for allowing the co-workers of D. Destarac to 
use their hierarchy of O-grids around the NACA0012 airfoil.

This research did not receive any specific grant from funding agencies in the public, commercial or not-for-profit sector 


\section*{Biblography}
\bibliography{biblio-template}{}

\begin{thebibliography}{10}

\bibitem{GilPie_97}
Giles, M. and Pierce, N.
\newblock Adjoint equations in {CFD}: {D}uality, boundary conditions and
  solution behaviour.
\newblock In {\em AIAA Paper Series, Paper 97-1850}.  (1997).

\bibitem{FraShu_92}
Frank, P. and Shubin, G.
\newblock A comparison of optimisation based approaches for a model
  computational aerodynamics design problem.
\newblock {\em Journal of Computational Physics}{ \bf 98}, 74--89 (1992).

\bibitem{pironneau_74}
Pironneau, O.
\newblock On optimum design in fluid mechanics.
\newblock {\em J. Fluid Mech.}{ \bf 64}(1), 97–110 (1974).

\bibitem{Jam_88}
Jameson, A.
\newblock Aerodynamic design via control theory.
\newblock {\em Journal of Scientific Computing}{ \bf 3}(3), 233--260 (1988).

\bibitem{JamMarPie_98b}
Jameson, A., Martinelli, L., and Pierce, N.
\newblock Optimum aerodynamic design using the {N}avier-{S}tokes equations.
\newblock {\em Theoretical and Computational Fluid Dynamics}{ \bf 10}(1),
  213--237 (1998).

\bibitem{CasLozPal_07}
Castro, C., Lozano, C., Palacios, F., and ZuaZua, E.
\newblock Adjoint approach to viscous aerodynamic design on unstructured grids.
\newblock {\em AIAA Journal}{ \bf 45}(9), 2125--2139 (2007).

\bibitem{becker_rannacher_01}
Becker, R. and Rannacher, R.
\newblock An optimal control approach to a posteriori error estimation in
  finite element methods.
\newblock {\em Acta Numerica}{ \bf 10}, 1–102 (2001).

\bibitem{VenDar_02}
Venditti, D. and Darmofal, D.
\newblock Grid adaptation for functional outputs: {A}pplication to
  two-dimensional inviscid flows.
\newblock {\em Journal of Computational Physics}{ \bf 176}, 40--69 (2002).

\bibitem{Dwi_08}
Dwight, R.
\newblock Heuristic {\sl a posteriori} estimation of error due to dissipation
  in finite volume schemes and application to mesh adaptation.
\newblock {\em Journal of Computational Physics}{ \bf 227}, 2845--2863 (2008).

\bibitem{LosDerAla_10}
Loseille, A., Dervieux, A., and Alauzet, F.
\newblock Fully anisotropic mesh adaptation for {3D} steady {Euler} equations.
\newblock {\em Journal of Computational Physics}{ \bf 229}, 2866--2897 (2010).

\bibitem{FidRoe_10}
Fidkowski, K. and Roe, P.
\newblock An entropy approach to mesh refinement.
\newblock {\em SIAM Journal of Scientific Computing}{ \bf 32}(3), 1261--1287
  (2010).

\bibitem{FidDar_11}
Fidkowski, K. and Darmofal, D.
\newblock Aerodynamic design optimization on unstructured meshes using the
  {N}avier-{S}tokes equations.
\newblock {\em AIAA Journal}{ \bf 49}(4), 673--694 (2011).

\bibitem{JPNguTro_12}
Peter, J., Nguyen-Dinh, M., and Trontin, P.
\newblock Goal-oriented mesh adaptation using total derivative of aerodynamic
  functions with respect to mesh coordinates -- with application to {E}uler
  flows.
\newblock {\em Computers and Fluids}{ \bf 66}, 194--214 (2012).

\bibitem{BelAlaDer_19}
Belme, A., Alauzet, F., and Dervieux, A.
\newblock An a priori anisotropic goal-oriented error estimate for viscous
  compressible flow and application to mesh adaptation.
\newblock {\em Journal of Computational Physics}{ \bf 376}, 1051--1088 (2019).

\bibitem{lions_68}
Lions, J.~L.
\newblock {\em Contr\^ole optimal de syst\`emes gouvern\'es par des \'equations
  aux deriv\'ees partielles}.
\newblock Etudes math\'ematiques. Paris: Dunod, Gauthier-Villars,  (1968).

\bibitem{SarMetSip_15}
Sartor, F., Mettot, C., and Sipp, D.
\newblock Stability, receptivity, and sensitivity analyses of buffeting
  transonic flow over a profile.
\newblock {\em AIAA Journal}{ \bf 53}(7), 1980--1993 (2015).

\bibitem{MorMitYlv_93}
Morris, M., Mitchell, T., and Ylvisaker, D.
\newblock Bayesian design and analysis of computer experiments: Use of
  derivatives in surface prediction.
\newblock {\em Technometrics}{ \bf 35}(3), 243--255 (1993).

\bibitem{luchini_bottaro_14}
Luchini, P. and Bottaro, A.
\newblock Adjoint equations in stability analysis.
\newblock {\em Annu. Rev. Fluid Mech.}{ \bf 46}(1), 493--517 (2014).

\bibitem{talagrand_courtier_87}
Talagrand, O. and Courtier, P.
\newblock Variational assimilation of meteorological observations with the
  adjoint vorticity equation. {I}: Theory.
\newblock {\em Q. J. R. Meteorol. Soc.}{ \bf 113}(478), 1311--1328 (1987).

\bibitem{GilPie_00}
Giles, M. and Pierce, N.
\newblock An introduction to the adjoint approach to design.
\newblock {\em Flow, Turbulence, Combustion}{ \bf 65}, 393--415 (2000).

\bibitem{JPDwi_10}
Peter, J. and Dwight, R.
\newblock Numerical sensitivity analysis for aerodynamic optimization: a survey
  of approaches.
\newblock {\em Computers and Fluids}{ \bf 39}, 373--391 (2010).

\bibitem{AndBon_99}
Anderson, W. and Bonhaus, D.
\newblock Airfoil design optimization on unstructured grids for turbulent
  flows.
\newblock {\em AIAA Journal}{ \bf 37}(2), 185--191 (1999).

\bibitem{NieAnd_99}
Nielsen, E. and Anderson, W.
\newblock Aerodynamic design optimization on unstructured meshes using the
  {N}avier-{S}tokes equations.
\newblock {\em AIAA Journal}{ \bf 37}(11), 185--191 (1999).

\bibitem{NemZin_02}
Nemec, N. and Zingg, D.
\newblock {Newton–Krylov} algorithm for aerodynamic design using the
  navier–stokes equations.
\newblock {\em AIAA Journal}{ \bf 40}(6), 1146--1154 (2002).

\bibitem{XuRadMey_15}
Xu.~S., Radford, D., M., M., and J.-D., M.
\newblock Stabilisation of discrete steady adjoint solvers.
\newblock {\em Journal of Computational Physics}{ \bf 299}, 175--195 (2015).

\bibitem{GilDutMul_01}
Giles, M., Duta, M., and M\"{u}ller, J.-D.
\newblock Adjoint code developments using the exact discrete approach.
\newblock In {\em AIAA Paper Series, Paper 2001-2596}.  (2001).

\bibitem{NadJam_00}
Nadarajah, S. and Jameson, A.
\newblock A comparison of the continuous and discrete adjoint approach to
  automatic aerodynamic shape optimization.
\newblock In {\em AIAA Paper Series, Paper 2000-667}.  (2000).

\bibitem{NadJam_01}
Nadarajah, S. and Jameson, A.
\newblock Studies of the continuous and discrete adjoint approaches to viscous
  automatic aerodynamic shape optimization.
\newblock In {\em AIAA Paper Series, Paper 2001-2530}.  (2001).

\bibitem{LuDar_04}
Lu, J. and Darmofal, D.
\newblock Adaptative precision methodology for flow optimisation via
  discretization and iteration error control.
\newblock In {\em AIAA Paper Series, Paper 2004-1996}.  (2004).

\bibitem{Har_07}
Hartmann, R.
\newblock Adjoint consistency analysis of discontinuous {G}alerkin
  discretizations.
\newblock {\em SIAM J. Numer. Anal.}{ \bf 45}(6), 2671--2696 (2007).

\bibitem{shi_wang_2015}
Shi, L. and Wang, Z.
\newblock Adjoint-based error estimation and mesh adaptation for the correction
  procedure via reconstruction method.
\newblock {\em Journal of Computational Physics}{ \bf 295}, 261 -- 284 (2015).

\bibitem{DuiBijKor_05}
Duivesteijn, G., Bijl, H., Koren, B., and van Brummelen, E.
\newblock On the adjoint solution of the {quasi-1D Euler} equations: the effect
  of boundary conditions and the numerical flux function.
\newblock {\em International Journal for Numerical Methods in Fluids}{ \bf 47},
  987--993 (2005).

\bibitem{Loz_16}
Lozano, C.
\newblock A note on the dual consistency of the discrete adjoint quasi-one
  dimensional {Euler} equations with cell-centred and cell-vertex central
  discretization.
\newblock {\em Computers and Fluids}{ \bf 134-135}, 51--60 (2016).

\bibitem{LiuSan_08}
Liu, Z. and Sandu, A.
\newblock On the properties of the discrete adjoints of numerical methods for
  the advection equation.
\newblock {\em International Journal for Numerical Methods in Fluids}{ \bf 56},
  769--803 (2008).

\bibitem{HicZin_14}
Hicken, J. and Zingg, D.
\newblock Dual consistency and functional accuracy: a finite-difference
  perspective.
\newblock {\em Journal of Computational Physics}{ \bf 256}, 161--182 (2014).

\bibitem{Stu_15}
St\"uck, A.
\newblock An adjoint view on flux consistency and strong wall boundary
  conditions to the {Navier-Stokes} equations.
\newblock {\em Journal of Computational Physics}{ \bf 301}, 247--264 (2015).

\bibitem{Stu_17}
St\"uck, A.
\newblock Dual-consistency study for {Green-Gauss} gradient schemes in an
  unstructured {Navier-Stokes} method.
\newblock {\em Journal of Computational Physics}{ \bf 350}, 530--549 (2017).

\bibitem{majda_83}
Majda, A.
\newblock The stability of multidimensional shock fronts.
\newblock In {\em Memoirs of the AMS, 275}. Amer. Math. Soc.,  (1983).

\bibitem{ulbrich_02}
Ulbrich, S.
\newblock A sensitivity and adjoint calculus for discontinuous solutions of
  hyperbolic conservation laws with source terms.
\newblock {\em SIAM J. Control Optim.}{ \bf 41}(3), 740--797 (2002).

\bibitem{giles_pierce_2001}
Giles, M.~B. and Pierce, N.~A.
\newblock Analytic adjoint solutions for the quasi-one-dimensional euler
  equations.
\newblock {\em J. Fluid Mech.}{ \bf 426}, 327–345 (2001).

\bibitem{BaeCasPal_09}
Baeza, A., Castro, C., Palacios, F., and Zuazua, E.
\newblock 2d {Euler} shape design on non-regular flows using adjoint
  {R}ankine-{H}ugoniot relations.
\newblock {\em AIAA Journal}{ \bf 47}(3), 552--562 (2009).

\bibitem{Loz_18}
Lozano, C.
\newblock Singular and discontinuous solutions of the adjoint {Euler}
  equations.
\newblock {\em AIAA Journal}{ \bf 56}(11), 4437--4451 (2018).

\bibitem{Loz_17}
Lozano, C.
\newblock On the properties of the solutions of the 2{D} adjoint {E}uler
  equations.
\newblock In {\em Proceedings of EUROGEN 2017, Madrid},  (2017).

\bibitem{Loz_19}
Lozano, C.
\newblock Watch your adjoints! lack of mesh convergence in inviscid adjoint
  solutions.
\newblock {\em AIAA Journal}{ \bf 56}(11), 4437--4451 (2018).

\bibitem{TodVonBou_16}
Todarello, G., Vonck, F., Bourasseau, S., Peter, J., and D\'esid\'eri, J.-A.
\newblock Finite-volume goal-oriented mesh-adaptation using functional
  derivative with respect to nodal coordinates.
\newblock {\em Journal of Computational Physics}{ \bf 313}, 799--819 (2016).

\bibitem{JamSchTur_81}
Jameson, A., Schmidt, W., and Turkel, E.
\newblock Numerical solutions of the {E}uler equations by finite volume methods
  using {Runge-Kutta} time-stepping schemes.
\newblock In {\em AIAA Paper Series, Paper 1981-1259}.  (1981).

\bibitem{AndVen_99}
Anderson, W. and Venkatakrishnan, V.
\newblock Aerodynamic design optimization on unstructured grids with a
  continuous adjoint formulation.
\newblock {\em Computers and Fluids}{ \bf 28}, 443--480 (1999).

\bibitem{HieEss_99}
Hiernaux, S. and Essers, J.-A.
\newblock An optimal control theory based algorithm to solve {2D} aerodynamic
  shape optimisation problems for inviscid and viscous flows.
\newblock In {\em Proceedings of the RTO-AVT Symposium on Aerodynamic Design
  and Optimisation of Flight Vehicles},  (1999).

\bibitem{HieEss_99b}
Hiernaux, S. and Hessers, J.-A.
\newblock Aerodynamic optimization using {N}avier-{S}tokes equations and
  optimal control theory.
\newblock In {\em AIAA Paper Series, Paper 99-3297}.  (1999).

\bibitem{Gut_06}
Gutknecht, M.~H.
\newblock Block krylov space methods for linear systems with multiple
  right-hand sides: An introduction,  (2006).

\bibitem{PinMon_13}
Pinel, X. and Montagnac, M.
\newblock Block {Krylov} methods to solve adjoint problems in aerodynamic
  design optimization.
\newblock {\em AIAA Journal}{ \bf 51}(9), 2183--2191 (2013).

\bibitem{Hir_07}
Hirsch, C.
\newblock {\em Numerical Computation of Internal and External Flows: The
  Fundamentals of Computational Fluid Dynamics (second edition)}.
\newblock Butterworth -- Heineman. Elsevier,  (2007).

\bibitem{GOODMAN1985336}
Goodman, J. and Majda, A.
\newblock The validity of the modified equation for nonlinear shock waves.
\newblock {\em J. Comput. Phys.}{ \bf 58}(3), 336 -- 348 (1985).

\bibitem{SwaTur_87}
Swanson, R. and Turkel, E.
\newblock Artificial dissipation and central difference schemes for the euler
  and navier-stokes equations.
\newblock Technical Report NAS1-18107, ICASE,  April  (1987).

\bibitem{SwaTur_92}
Swanson, R. and Turkel, E.
\newblock On central-difference and upwinf schemes.
\newblock {\em Journal of Computational Physics}{ \bf 101}, 292--306 (1992).

\bibitem{GilPie_99}
Giles, M. and Pierce, N.
\newblock An introduction to the adjoint approach to design.
\newblock In {\em Proceedings of ERCOFTAC Workshop on Adjoint Methods},
  (1999).

\bibitem{Ngu_14}
Nguyen-Dinh, M.
\newblock {\em Qualification des simulations num\'eriques par adaptation
  anisotropique de maillages}.
\newblock PhD thesis, Universit\'e de Nice-Sophia Antipolis,  March  (2014).

\bibitem{DwiBre_06}
Dwight, R. and Brezillon, J.
\newblock Effect of approximations of the discrete adjoint on gradient-based
  optimization.
\newblock {\em AIAA Journal}{ \bf 44}(12), 3022--3031 (2006).

\bibitem{CamHeiPlo_13}
Cambier, L., Heib, S., and Plot, S.
\newblock The els{A} {CFD} software: input from research and feedback from
  industry.
\newblock {\em Mechanics \& Industry}{ \bf 14}(3), 159--174 (2013).

\bibitem{VasJam_10}
Vassberg, J. and Jameson, A.
\newblock In pursuit of grid convergence for two-dimensional {E}uler solutions.
\newblock {\em Journal of Aircraft}{ \bf 47}(4), 1152--1166 (2010).

\bibitem{Des_03}
Destarac, D.
\newblock Far-field / near-field drag balance and applications of drag
  extraction in cfd,  February  (2003).

\bibitem{JPRenDum_15}
Peter, J., Renac, F., Dumont, A., and M\'eheut, M.
\newblock Discrete adjoint method for shape optimization and mesh adaptation in
  the $els{A}$ code. status and challenges.
\newblock In {\em Proceedings of 50th 3AF Symposium on Applied Aerodynamics,
  Toulouse},  (2015).

\bibitem{And_03}
Anderson, J.
\newblock {\em Modern Compressible Flow (third edition)}.
\newblock McGraw-Hill,  (2003).

\bibitem{JP_20}
Peter, J.
\newblock Contributions to discrete adjoint method in aerodynamics for shape
  optimization and goal-oriented mesh adaptation.
\newblock {University of Nantes. Mémoire pour Habilitation à Diriger des
  Recherches},  (2020).

\bibitem{JPLabRen_19}
Peter, J., Labbé, C., and Renac, F.
\newblock Analysis of discrete adjoint fields for 2d {Euler} flows.
\newblock In {\em Proceedings of eurogen 2019}. ECCOMAS,  (2019).

\end{thebibliography}
\bibliographystyle{nature}


\appendix

\section{Additional figures and arrays}

\begin{table}[htbp]
\begin{center}
\begin{tabular}[t]{|l||c|c|c|c|c|c|c|c|c|c|c|} 
 \hline
 \hline
 reference   & $Bl_1$ & $Bl_2$ & $Bl_3$  & $Bl_4$ & $Bl_5$  & $Bu_5$ & $Bu_4$ & $Bu_3$ & $Bu_2$ & $Bu_1$   \\
 \hline
 \hline
 $10^2~(dCLp/da_i)_{f.d.}$    &   -8.64    &   -4.86     &   -5.34   &  -1.10  &  -4.78   & -1.84   & -1.04 & -1.32 & -2.74 & -1.50 \\
 \hline
 $10^2~(dCLp/da_i)_{adj.}$      &   -8.67    &  -4.87    &   -5.36    &  -1.10  &  -4.78   & -1.84  & -1.04  & -1.32 & -2.74 & -1.50 \\
 \hline
 $10^2~(dCLp/da_i)_{c.adj.}$        &   -7.69    & -4.09    &  -4.80    &  -1.39     &  -4.84   & -1.85  & -1.02  & -1.28 & -2.58 & -0.91 \\
 \hline
 $10^2~(dCLp/da_i)_{adj.fr.}$  &   -7.68    &  -4.36   &  -4.75   &   -0.87   &   -4.59 &  -1.92  & -0.98 & -1.22 & -2.55 &  -1.86 \\
 \hline
 \hline
 $10^3~(dCDp/da_i)_{f.d.}$      &   -6.47    &  -4.15   &   -8.32   &  4.14  &  9.57   &  5.45  & 2.63  & 0.0892 & -1.52 & -3.58\\
 \hline
 $10^3~(dCDp/da_i)_{adj.}$      &   -6.49    &  -4.16   &   -8.33   &  4.15  &  9.57   &  5.45  & 2.63  & 0.0867& -1.52 & -3.58 \\
 \hline
 $10^3~(dCDp/da_i)_{c.adj.}$    &   -5.70    &   -3.53  &   -7.88   &  3.91  &  9.57   & 5.45  & 2.65   & 0.0125 & -1.39 & -3.10 \\
 \hline
 $10^3~(dCDp/da_i)_{adj.fr.}$    &  -5.57   &   -3.69   &   -7.78   &  4.33  &  9.72   & 5.37   & 2.67  & 0.0165 & -1.35 & -3.90\\
 \hline
 \hline
\end{tabular}
\end{center}
\caption{($M_\infty=0.85$, $\alpha=2^o$, 513$\times$513 mesh) lift and drag gradient w.r.t. parameters driving
 the amplitudes of ten bumps. Finite-differences $(f.d.)$, exact discrete adjoint $(adj.)$, consistent
 adjoint $(c.adj.)$, discrete adjoint with frozen $\kappa$ and $\nu$ $(adj.fr.)$}
\label{tb:gradbumps}
\end{table}

\begin {figure}[htbp]
  \begin{center}
	  \includegraphics[width=0.9\linewidth]{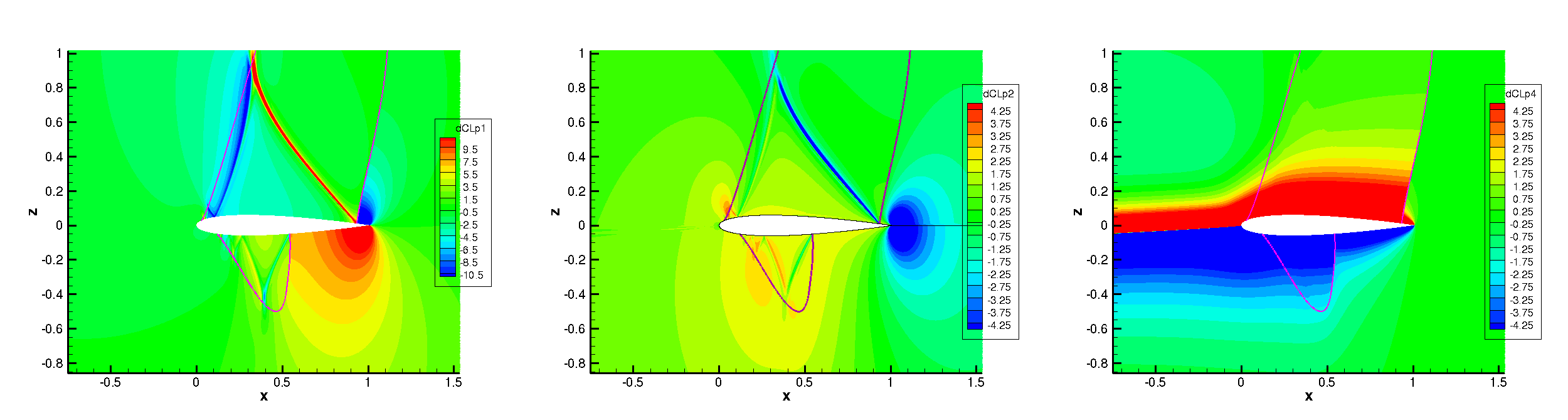}
	\caption{($M_\infty=0.85$, $\alpha=2^o$, 2049$\times$2049 mesh) Contours of $\delta CL_p^1$, $\delta CL_p^2$, $\delta CL_p^4$
 (equations (\ref{e:deltaR}) and (\ref{e:physpre}) with $\epsilon$=1).  $\delta CL_p^3$ is zero.}
  \label{f:dCLp124_M085}
  \end{center}
\end {figure}  
\begin {figure}[htbp]
  \begin{center}
	  \includegraphics[width=0.9\linewidth]{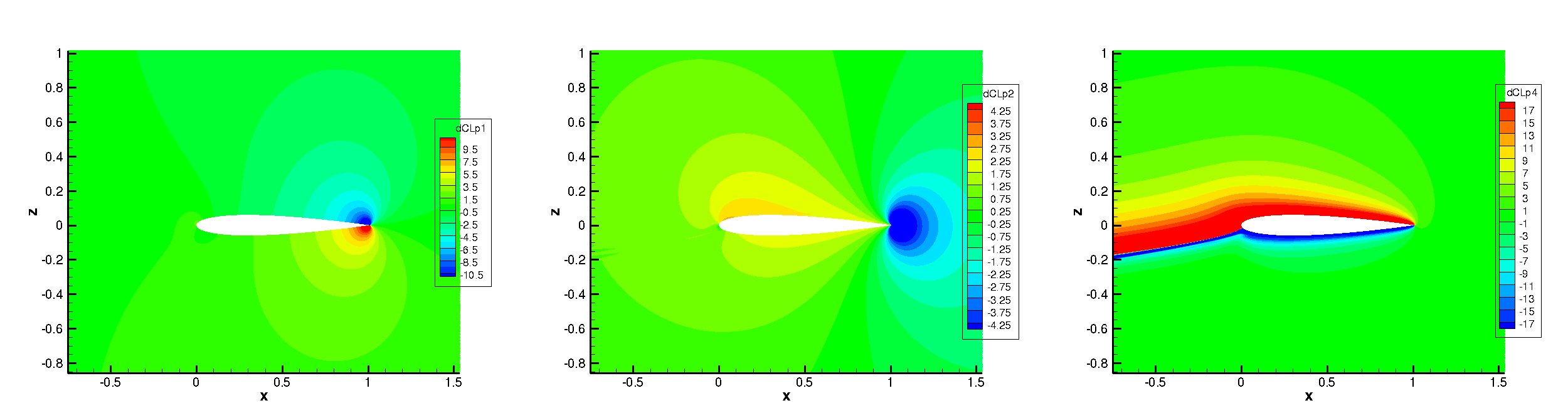}
	\caption{($M_\infty=0.4$, $\alpha=5^o$, 2049$\times$2049 mesh) Contours of $\delta CL_p^1$, $\delta CL_p^2$, $\delta CL_p^3$, $\delta CL_p^4$
 (equations (\ref{e:deltaR}) and (\ref{e:physpre}) with $\epsilon$=1). $\delta CL_p^3$ is zero.}
  \label{f:dCLp124_M040}
  \end{center}
\end{figure}


\begin {figure}[htbp]
  \begin{center}
	  \includegraphics[width=0.48\linewidth]{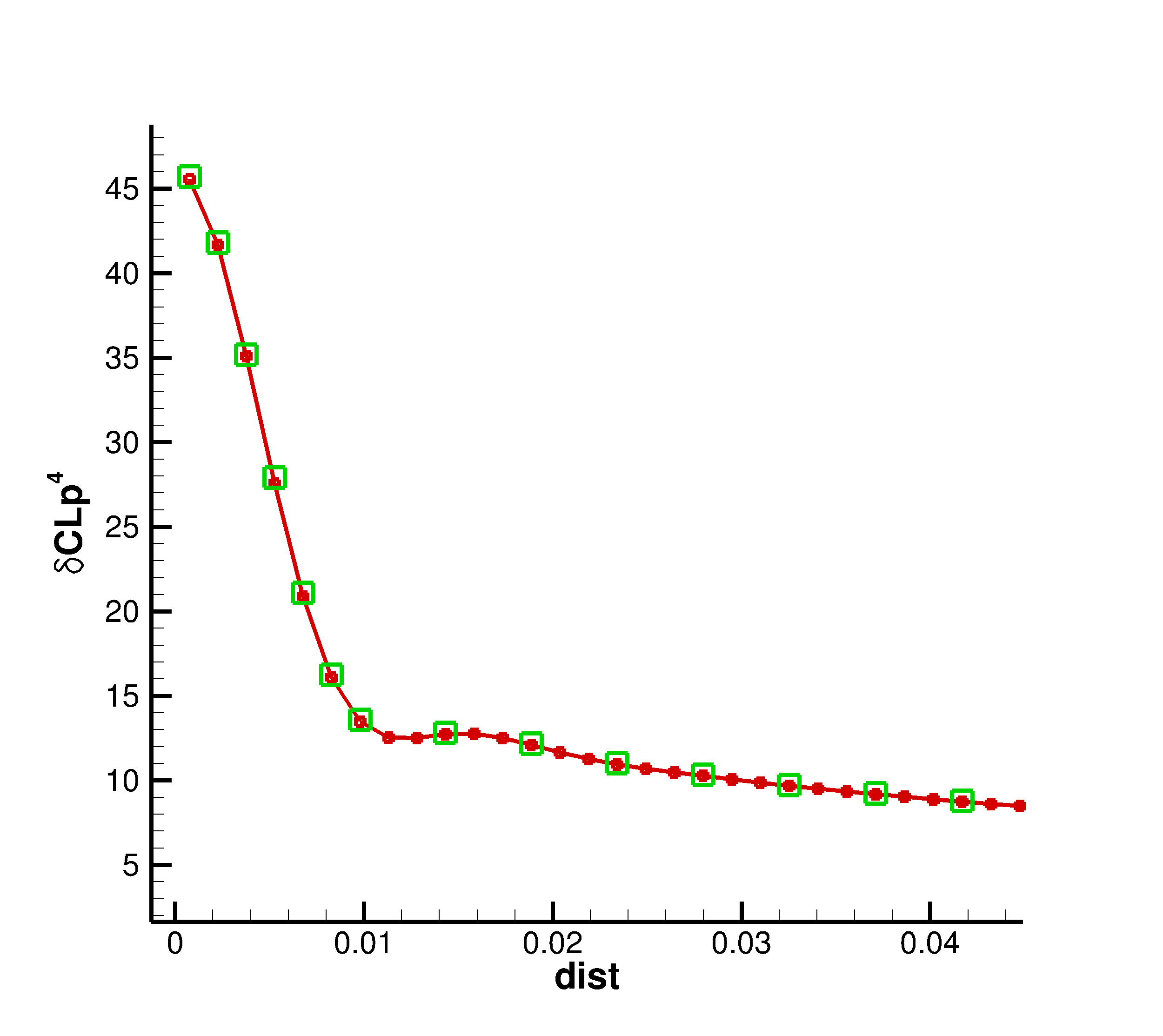}
	  \includegraphics[width=0.48\linewidth]{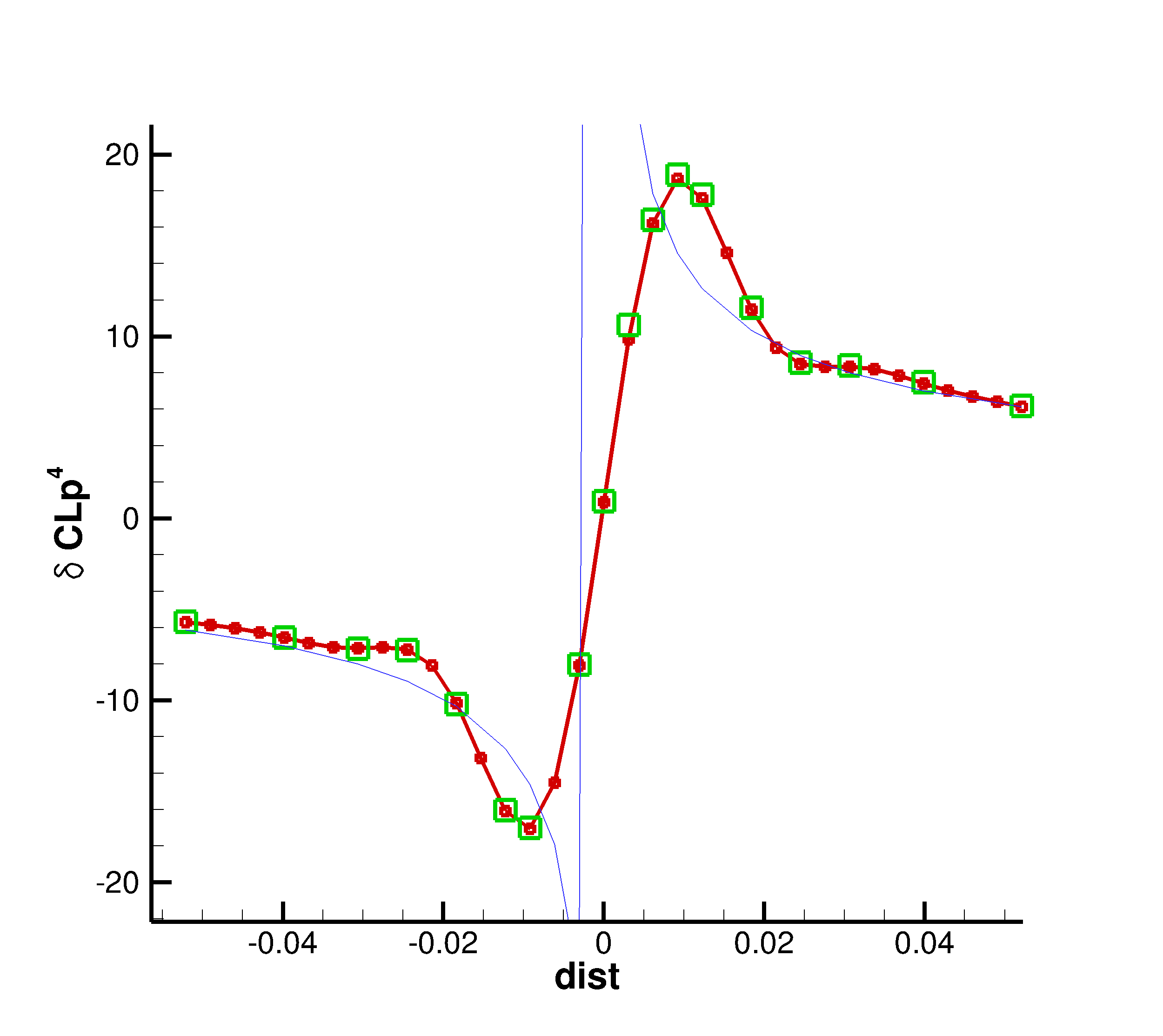}
	  \caption{($2049\times2049$ mesh) $M_\infty=0.85$, $\alpha=2^o$. Linear (red curve) and non-linear (green squares)
      estimations of $\delta CL_p^4$. Left: upper side at the wall $x\simeq 0.5$. Right: in the vicinity
      of the stagnation streamline  $x\simeq - 0.6$ (dist being a signed distance to it). The blue curve is a  tentative
     fit with $K sgn(dist)/\sqrt{|dist|}$ (\cite{GilPie_97}). The extraction lines are drawn in Fig.  \ref{f:isoM_M085_M040} left.} 
  \label{f:dCLp4lnl}
  \end{center}
\end{figure}

\begin {figure}[htbp]
  \begin{center}
	  \includegraphics[width=0.9\linewidth]{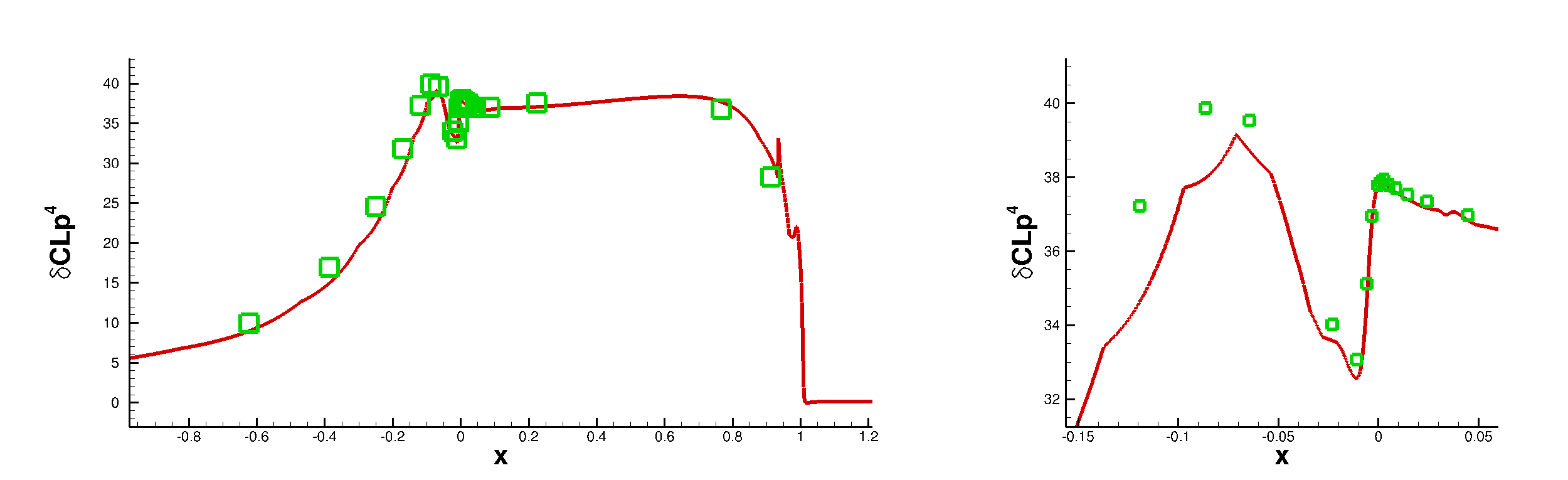}
	  \caption{($2049\times2049$ mesh) $M_\infty=0.85$, $\alpha=2^o$. Linear (red curve) and non-linear (green squares)
      estimations of $\delta CL_p^4$ along the streamline passing by (-0.6224,-0.0383). The mean distance of this streamline
      to the stagnation streamline is about 0.02c. Its distance to the upperside wall is about 0.015c.
      The shock foot abscissa is $x \simeq
       0.935$ (upwards oscillation right of the last green square).  The streamline is drawn in Fig.  \ref{f:isoM_M085_M040} left.
    } 
  \label{f:dCLp4lnltr}
  \end{center}
\end{figure}

%
\begin {figure}[htbp]
  \begin{center}%
	  \includegraphics[width=0.9\linewidth]{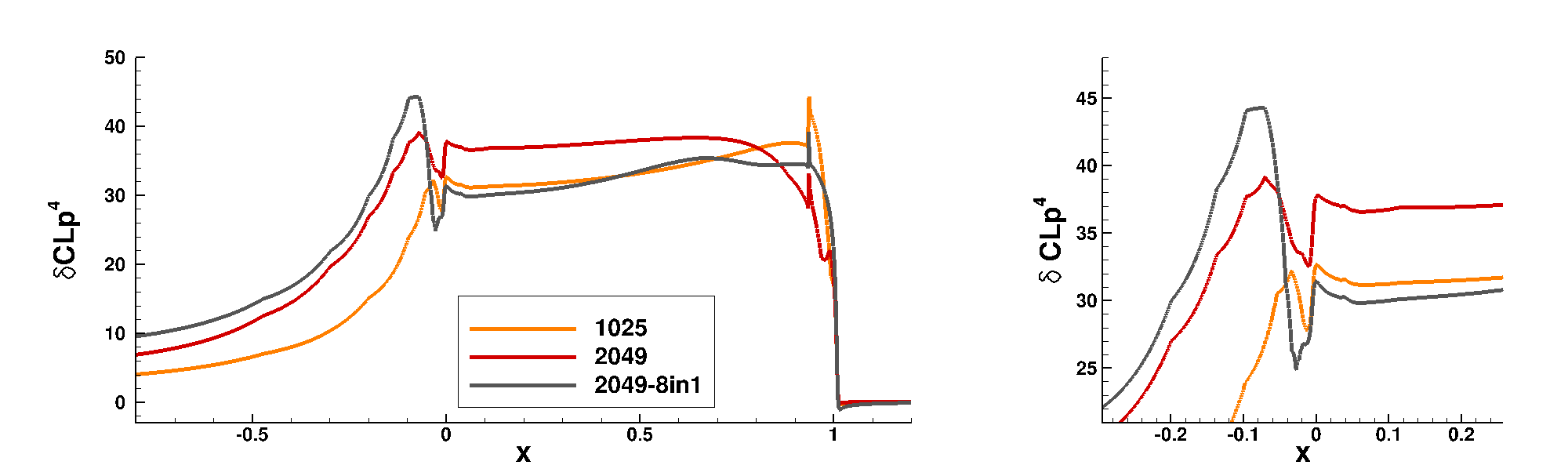}
	  \caption{  \clon{$M_\infty=0.85$, $\alpha=2^o$. Exact discrete adjoint $\delta CL_p^4$
      along the streamline passing by (-0.6224,-0.0383) for the finest meshes. The mesh denoted 2049-8in1 is a $2049\times 2049$ mesh built by regular linear interpolation inside the standard  $2049\times 2049$ mesh. It is stretched at the wall, the aspect ratio of the cells adjacent to the wall being 1 to 8. (Not shown for the 4097$\times$4097 mesh for which the trajectory is
  a bit different and higher values are obtained.)}} 
  \label{f:3streamtr}
  \end{center}
\end{figure}
\begin {figure}[htbp]
  \begin{center}
	  \includegraphics[width=0.45\linewidth]{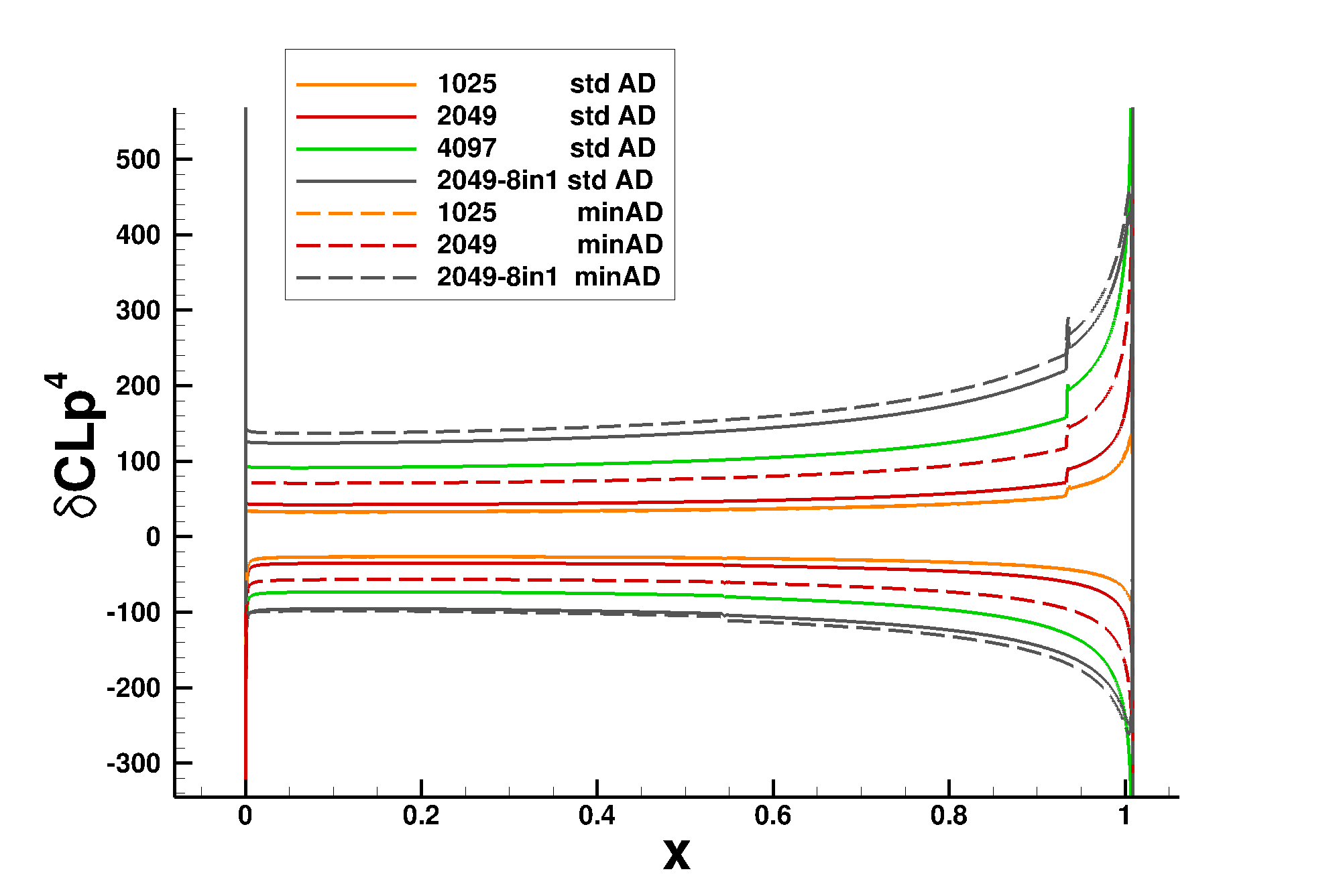}
	  \includegraphics[width=0.45\linewidth]{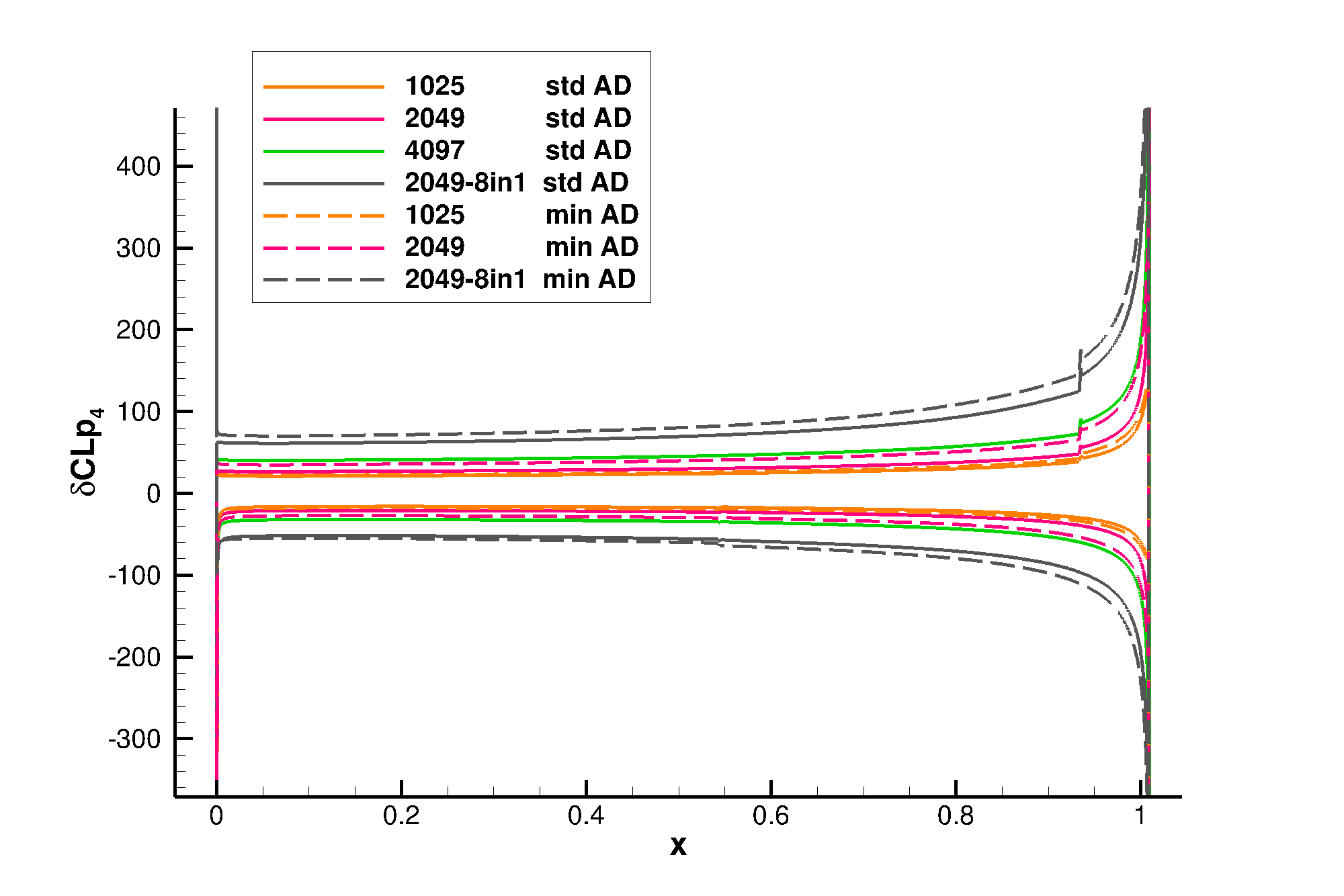}
	  \caption{(finer meshes)  $M_\infty=0.85$, $\alpha=2^o$. $\delta CLp4$ in the cells adjacent to the wall
      (equations (\ref{e:deltaR}) and (\ref{e:physpre}) with $\epsilon$=1). 
      Left: exact discrete adjoint. Right: consistent adjoint. \clon{The solid line curves correspond to the calculations
      presented elsewhere (``standard AD'' coefficient $k^4=0.032$). The dashed line curves correspond to calculations with halved artificial dissipation (``minimum AD''
      coefficient k4=0.016). This small coefficient does not allow a proper flow convergence with the 4097$\times$4097 mesh.}} 
  \label{f:finewalladj}
  \end{center}
\end{figure}


\begin {figure}[htbp]
  \begin{center}
	  \includegraphics[width=0.9\linewidth]{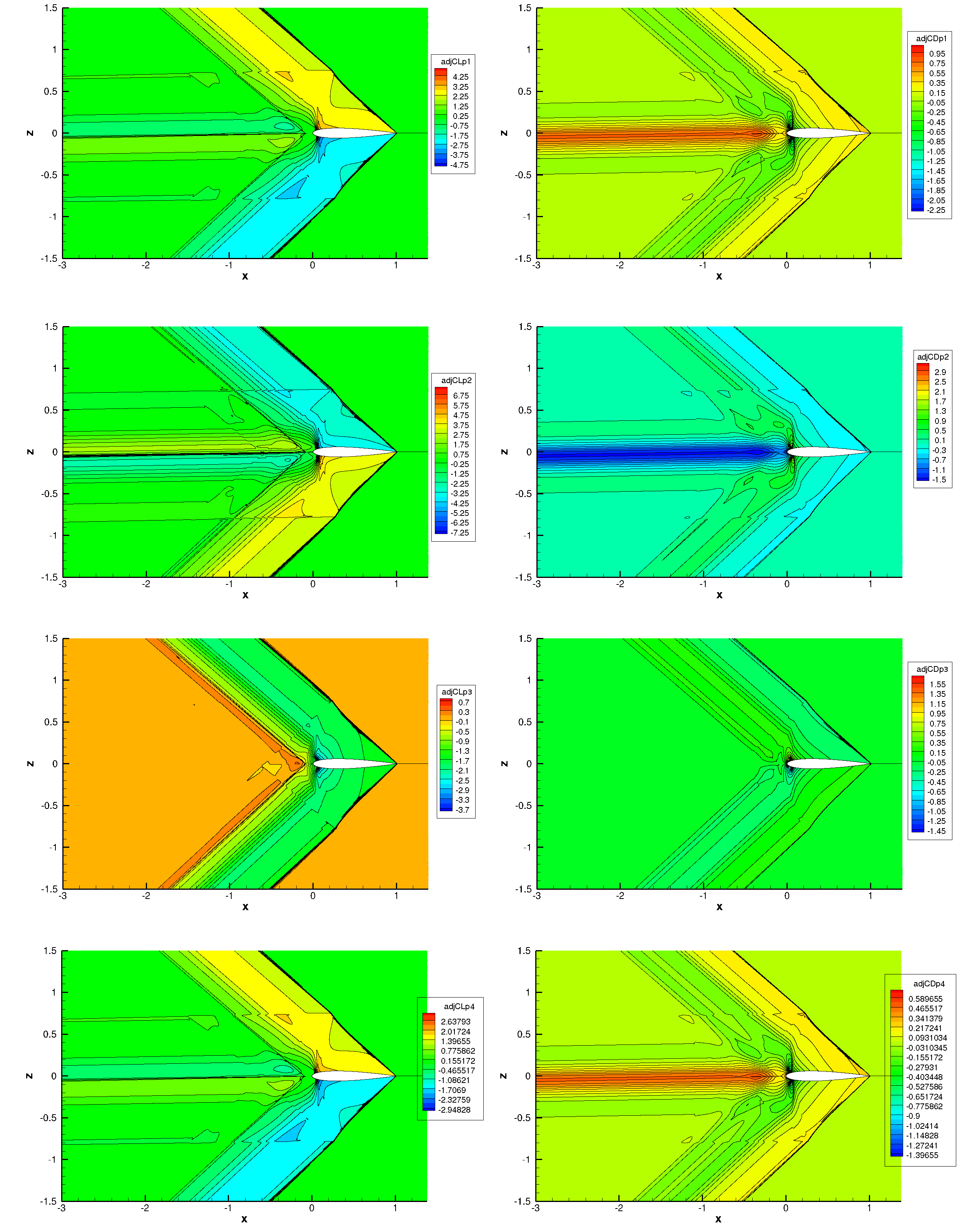}
	\caption{($M_\infty=1.5$, $\alpha=1^o$, 4097$\times$4097 mesh) Left: lift adjoint, right: drag adjoint.}
  \label{f:g_M150}
  \end{center}
\end {figure} 

\begin {figure}[htbp]
  \begin{center}
	  \includegraphics[width=0.9\linewidth]{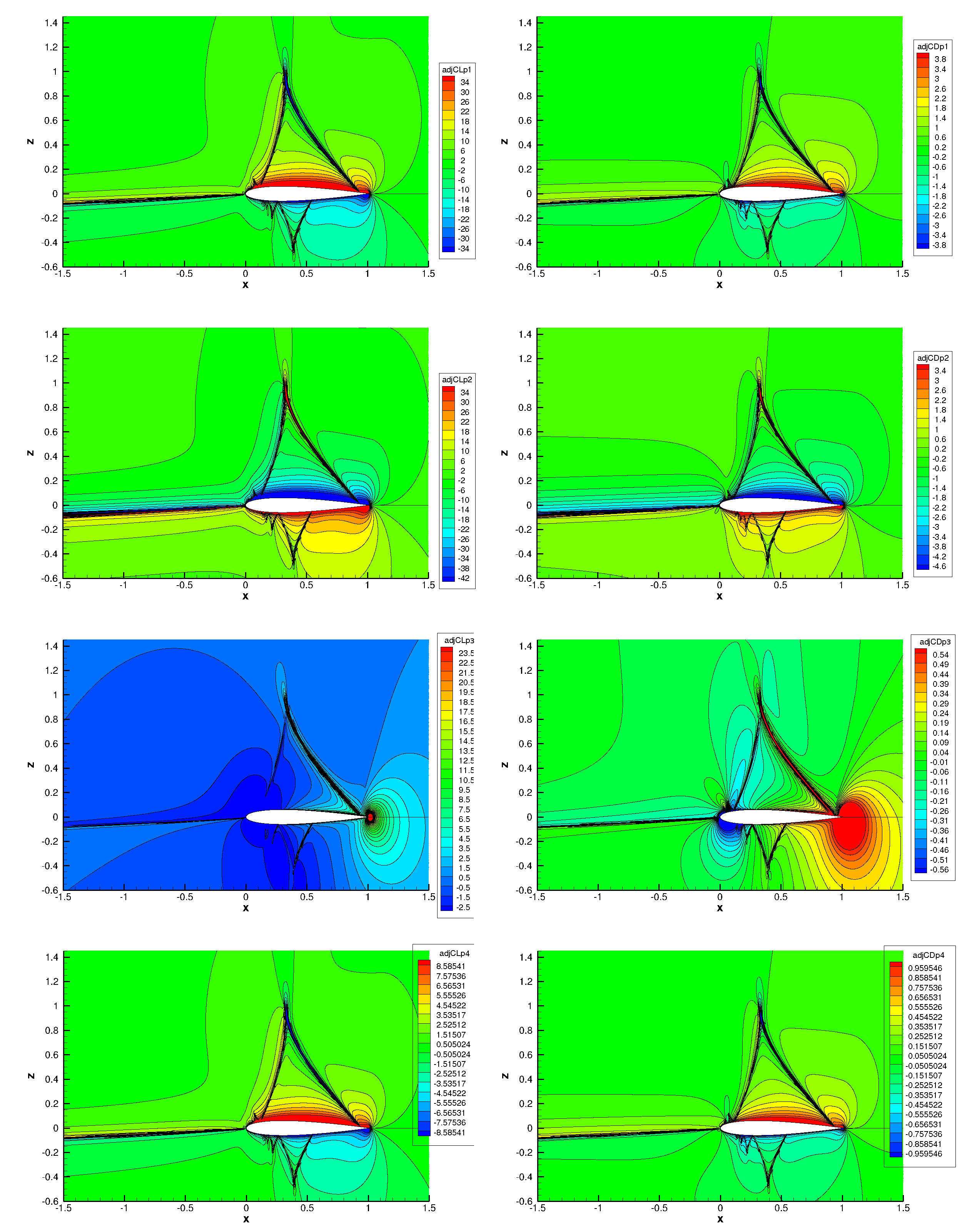}
	\caption{($M_\infty=0.85$, $\alpha=2^o$, 4097$\times$4097 mesh) Left: lift adjoint, right: drag adjoint.}
  \label{f:g_M085}
  \end{center}
\end {figure} 

\begin {figure}[htbp]
  \begin{center}%
	  \includegraphics[width=0.9\linewidth]{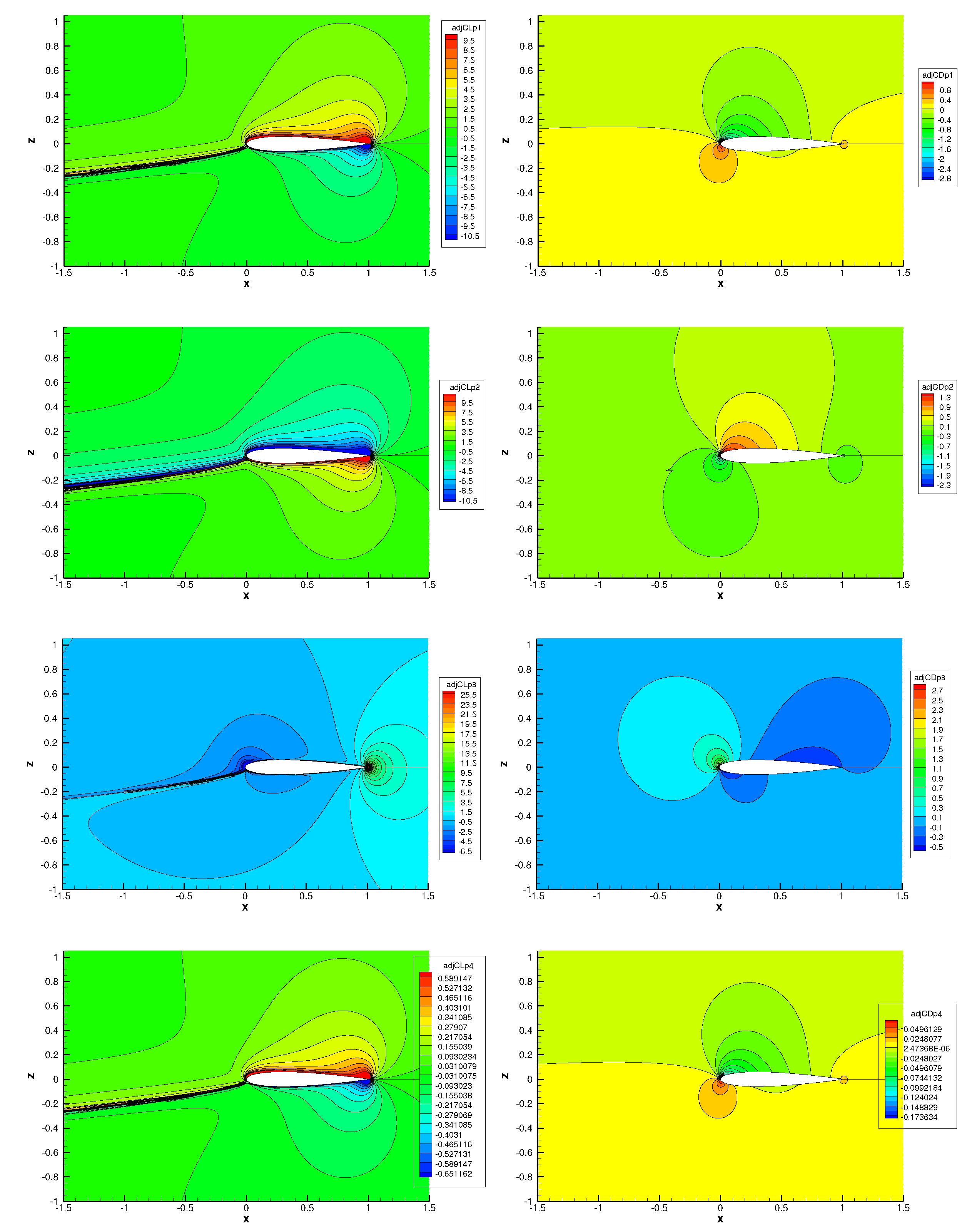}
	\caption{($M_\infty=0.4$, $\alpha=5^o$, 4097$\times$4097 mesh) Left: lift adjoint, right: drag adjoint.}
  \label{f:g_M040}
  \end{center}
\end {figure}


\end{document}